\documentclass[aps,onecolumn,superscriptaddress,nofootinbib,11pt,a4]{revtex4-1}

\RequirePackage{fix-cm}
\usepackage{graphicx}
\usepackage{subfig}
\usepackage{color}
\usepackage[colorlinks=true,linkcolor=blue]{hyperref}
\usepackage{url}
\usepackage[normalem]{ulem}
\usepackage{cancel}
\usepackage{caption}
\usepackage{mathrsfs}
\usepackage{bbm}
\usepackage{stmaryrd}
\usepackage{bm}
\usepackage{amsfonts}
\usepackage{amssymb}

\newcommand{\rmb}{{\mathrm{b}}}
\newcommand{\rmpp}{{\mathrm{p}}}
\newcommand{\rmw}{{\mathrm{w}}}
\newcommand{\rmD}{{\mathrm{D}}}

\newcommand{\rmP}{{\mathrm{P}}}
\newcommand{\rmT}{{\mathrm{T}}}
\newcommand{\gamL}{{\gamma_{\mathrm{L}}}}

\newcommand{\bfa}{\mathbf{a}}

\newcommand{\bfk}{\mathbf{k}}
\newcommand{\bfl}{\mathbf{l}}
\newcommand{\bfm}{\mathbf{m}}
\newcommand{\bfn}{\mathbf{n}}
\newcommand{\bfp}{\mathbf{p}}
\newcommand{\bfr}{\mathbf{r}}
\newcommand{\bfu}{\mathbf{u}}
\newcommand{\bfv}{\mathbf{v}}
\newcommand{\bfw}{\mathbf{w}}

\newcommand{\bfX}{\mathbf{X}}
\newcommand{\bfM}{\mathbf{M}}

\newcommand{\rmd}{{\mathrm{d}}}
\newcommand{\rme}{{\mathrm{e}}}
\newcommand{\rmi}{{\mathrm{i}}}
\newcommand{\rmr}{{\mathrm{r}}}

\newcommand{\rmcc}{{\mathrm{c.c.}}}

\newcommand{\lamD}{{\lambda_{\mathrm{D}}}}
\newcommand{\lma}{{\lambda_{\mathrm{ca}}}}
\newcommand{\lid}{{\lambda_{\mathrm{id}}}}

\newcommand\norm[1]{{\| #1 \|}}

\newcommand{\bsr}{{\boldsymbol{r}}}

\begin{document}

\title{Basic microscopic plasma physics from $N$-body mechanics \\
A tribute to Pierre-Simon de Laplace}


\def\Marseille{Aix-Marseille Universit\'{e}, CNRS, PIIM, Marseille, France}
\def\Arpajon{CEA, DAM, DIF F-91297 Arpajon, France}



\author{D.F. Escande}
              \email{dominique.escande@univ-amu.fr}        
              \affiliation\Marseille
\author{D. B\'enisti}
              \email{didier.benisti@cea.fr}
              \affiliation\Arpajon
\author{Y. Elskens}
              \email{yves.elskens@univ-amu.fr}
              \affiliation\Marseille
\author{D. Zarzoso}
              \email{david.zarzoso-fernandez@univ-amu.fr}
              \affiliation\Marseille
\author{F. Doveil}
              \email{fabrice.doveil@univ-amu.fr}
              \affiliation\Marseille


\begin{abstract}

Computing is not understanding. This is exemplified by the multiple and discordant interpretations of Landau damping still present after seventy years. For long deemed impossible, the mechanical $N$-body description of this damping, not only enables its rigorous and simple calculation, but makes unequivocal and intuitive its interpretation as the synchronization of almost resonant passing particles. This synchronization justifies mechanically why a single formula applies to both Landau growth and damping. As to the electrostatic potential, the phase mixing of many beam modes produces Landau damping, but it is unexpectedly essential for Landau growth too. Moreover, collisions play an essential role in collisionless plasmas. In particular, Debye shielding results from a cooperative dynamical self-organization process, where ``collisional" deflections due to a given electron diminish the apparent number of charges about it. The finite value of exponentiation rates due to collisions is crucial for the equivalent of the van Kampen phase mixing to occur in the $N$-body system. The $N$-body approach incorporates spontaneous emission naturally, whose compound effect with Landau damping drives a thermalization of Langmuir waves. O'Neil's damping with trapping typical of initially large enough Langmuir waves results from a phase transition. As to Coulomb scattering, there is a smooth connection between impact parameters where the two-body Rutherford picture is correct, and those where a collective description is mandatory. The $N$-body approach reveals two important features of the Vlasovian limit: it is singular and it corresponds to a renormalized description of the actual $N$-body dynamics.




\begin{tabular}{l  l}
    \emph{PACS 2010} & \\
    45.50.Jf     &   Few- and many-body systems \\
    45.50.Tn   &   Collisions     \\
    52.35.Fp   &   Electrostatic waves and oscillations (e.g., ion-acoustic waves)    \\
    52.20.Fs   &   Electron collisions    \\
\end{tabular}

\end{abstract}

\maketitle

\section{Introduction}
\label{intro}

This review deals with the microscopic physics of plasmas, mainly collisionless ones. Its main purpose is to improve the foundations of this physics by laying out, in a pedagogic and elementary manner, a systematic exposition of its approaches using $N$-body classical mechanics. These approaches enable in particular: (i) a short, yet explicit and rigorous, derivation of Landau damping and growth, (ii) an associated intuitive, yet rigorous, interpretation of these phenomena, (iii) another derivation showing Landau damping to result from phase mixing, as originally proved by van Kampen in a Vlasovian setting, (iv) a third derivation recovering the usual Vlasovian dielectric function and unifying Landau damping and Debye shielding (or screening), (v) unveiling how the microscopic mechanism of Debye shielding is intimately connected to collisions, (vi) a unification of the derivations of spontaneous emission and of Landau damping, (vii) proving the depletion of nonlinearity when there is a plateau in the tail distribution function, (viii) proving that damping with trapping results from a phase transition, (ix) a calculation of Coulomb scattering describing for the first time correctly all impact parameters with no ad hoc cut-off.

Item (ii) is important, since the lack of mechanical interpretation prevented the plasma community from accepting the reality of Landau damping from its publication in 1946 \cite{lan46} till 1964 when Malmberg and Wharton's celebrated experiment proved its existence (see e.g.\ \cite{Ryutov} for a historical sketch). Unfortunately, more than seven decades after Landau's publication, textbooks still propose discordant physical interpretations of the effect, invoking trapping, surfing, or synchronization of almost resonant passing particles. As to item (v), the new approach elucidates a longstanding mystery: how can a given particle be shielded by all other particles, while contributing to their individual shieldings? Finally, we stress that the mechanical approach of this review does not aim at challenging the Vlasovian one, as far as efficiency or convenience is concerned to perform kinetic calculations in plasma physics.

For macroscopic classical systems, the $N$-body description by classical mechanics was deemed impossible. This led to the development of thermodynamics, of fluid mechanics, and of kinetic equations to describe various macroscopic systems made up of particles like electrons, gas atoms or molecules, stars, or microorganisms. When plasma physicists had to address the microscopic description of their state(s) of matter, they did not consider using $N$-body classical mechanics, but directly derived kinetic analogues of the Boltzmann and Liouville equations, in particular Vlasov with his celebrated equation. This trend has been dominant till nowadays. However, the theoretical approach presented in this review shows that, actually, $N$-body classical mechanics is a practical tool both for solving important linear and nonlinear problems in microscopic plasma physics: Laplace's dream\footnote{The sentence defining what was called afterwards Laplace's demon is well known : ``Une intelligence qui, pour un instant donn\'e, conna\^itrait toutes les forces dont la nature est anim\'ee et la situation respective des \^etres qui la composent, si d'ailleurs elle \'etait assez vaste pour soumettre ces donn\'ees \`a l'analyse, embrasserait dans la m\^eme formule les mouvements des plus grands corps de l'univers et ceux du plus l\'eger atome : rien ne serait incertain pour elle, et l'avenir, comme le pass\'e, serait pr\'esent \`a ses yeux." [Essai philosophique sur les probabilit\'es (1814) \cite{laplace1829essai}] \emph{English translation} : ``An intellect, which at a certain moment would know all forces that set nature in motion, and all positions of all items of which nature is composed, if this intellect were also vast enough to submit these data to analysis, it would embrace in a single formula the movements of the greatest bodies of the universe and those of the tiniest atom; for such an intellect nothing would be uncertain and the future just like the past would be present before its eyes." However, the genuine Laplace's dream is reasonable, since a few sentences later he states: ``Tous ses efforts dans la recherche de la v\'erit\'e tendent \`a rapprocher [l'esprit humain] sans cesse de l'intelligence que nous venons de concevoir, mais dont il restera toujours infiniment \'eloign\'e." \emph{English translation} : ``All its efforts in the quest of truth tend at moving [the human spirit] closer to the intelligence we have just conceived, but from which it will always stay infinitely distant".} was not a mere utopia, since the calculation of classical orbits starting from prescribed initial conditions can genuinely describe and explain non-trivial aspects of the macroscopic dynamics of a many-body system. This is the reason why this review is ``A tribute to Pierre-Simon de Laplace". A second reason is the important role of the Laplace transform in the present $N$-body approach.

While deemed impossible to use, $N$-body classical mechanics is generally considered as the ultimate reference for the description of the microscopic physics of classical plasmas. However, if attempted, the corresponding ultimate description may a priori run into difficulties. Indeed, as shown in Chibbaro, Rondoni, and Vulpiani's book ``Reductionism, Emergence and Levels of Reality" \cite{vulpiani} with various examples in statistical mechanics, in chaotic systems, and in chemistry, the relations between different levels of description of physical phenomena are not simple. As stated by Michael Berry in the foreword to this book, ``It is far from straightforward to derive the formula relating the object and image of a simple lens by starting from the field operators of quantum optics supposedly the deepest of our current pictures of light. [...] The resolution of these difficulties starts from the observation that the theories of physics are mathematical, and relations between them involve limits as some parameter vanishes: wave optics `reduces to' geometrical optics when the wavelength is negligibly small, quantum physics `reduces' to classical physics when Planck's constant can be neglected, etc. Therefore understanding relations between levels must involve the study of limits, that is, mathematical asymptotics. And the central reason why `reduces to' is so problematic is the fact that the limits involved are usually singular." Plasma physics would be exceptional if it would escape this problem. It does not.

\newpage

Indeed, this issue is already present in the derivation of the kinetic equations of plasmas. Deriving the Vlasov equation by the BBGKY hierarchy assumes the two-point correlation function to be small. Because of the singularity of the Coulomb force at vanishing distances, this is not true when two particles come close to each other, and the singularity must be cured by imposing a short-range cut-off, but this works only for a uniform plasma, without waves in particular (see section 5.2 of \cite{Nicholson}). The usual way to obtain a vanishing correlation function is the ``pulverization procedure" : one cuts each particle into $M$ equal pieces, and one lets $M$ and the number of particles in the Debye sphere (defined in section \ref{Bs}) go to infinity (see section 4.3 of \cite{Nicholson}). However, this entails the loss of the actual correlations provided by the genuine two-point correlation function of the plasma of interest. Furthermore, the issue of the short-range cut-off is important to evaluate Coulomb scattering. The general practice is to take this cut-off equal to the classical distance of closest approach for the lower range of temperatures, and the quantum uncertainty on the electron's position for temperatures high enough for the latter to be larger than the former.

The singularity of the Coulomb force at vanishing distances is also a problem (see \cite{Spohn} and references therein for a discussion) in the mean-field derivation of the Vlasov equation \cite{NeunzertWick,Neunzert84,Dobru,BraunHepp,Spohn,ElsVla,boers,jabin}. This derivation requires a smoothing of this singularity at a length scale about $N^{-\alpha}$, for some $\alpha>0$ \cite{Kie14,HJ15}. However, even so, the equation is proved to be accurate only over a time of the order of the inverse of the largest Lyapunov exponent of the $N$-body system. This derivation of Vlasov equation deals with a distribution describing a single realization of the plasma, while the BBGKY derivation deals with a function describing an ensemble of plasmas. These two cases indicate that the issue of singular limits is likely to come about in approaches using $N$-body classical mechanics too. It does, as we will see.

This paper was initially meant as a mere review of results already published on the description of microscopic plasma physics by $N$-body mechanics. However, this endeavour rapidly ran into a difficulty: a part of these results were derived for three-dimensional plasmas described as sets of particles coupled by the Coulomb force, while another part came from a one-dimensional wave-particle description. This induced an artificial separation in the presentation of concepts and opposed logical continuity. This was an incentive to reformulate some of the one-dimensional results in three dimensions. When doing so for the proof of the average synchronization of particles at work in Landau damping, it proved useful to extend somewhat the calculation with respect to the one-dimensional case, to make it more intuitive. Unexpectedly, this eventually led to a very simple and rigorous description of Langmuir waves analogous to that published by Kaufman in a Vlasovian setting in 1972. Because of its simplicity, this new description is the first presented in this review\footnote{This rigorous derivation is accessible to students knowing Newton's second law of motion and the Fourier transform, but neither analytic functions, nor the Laplace transform. It also provides a correction to the lowest order expression of Landau damping. In a second step, the calculation is extended to the corresponding particle dynamics, showing that it produces an average synchronization of almost resonant passing particles with the wave.}. Paradoxically, going to the more realistic three dimensions brought a simplified theory! The new description is followed by a reformulation in three dimensions of a previous van Kampen-Dawson-like description of Langmuir waves.

This review is organized as follows. Section \ref{sec:1} introduces the basic scales, the equations of motion of the present $N$-body approach, and a class of granular distributions close to being spatially uniform. Then, the first half of this review provides three different derivations of Landau damping. First, section \ref{LwaK} deals with Langmuir waves by using a technique introduced by Kaufman in a Vlasovian setting, which provides a very short, though rigorous, derivation of Landau damping together with the interpretation of this effect as an average synchronization of particles. Then, section \ref{Feqpot} introduces a fundamental equation for the electrostatic potential which is reminiscent of the one obtained by the calculation \`a la Landau starting with the Vlasov equation. This equation is the basis of the next two derivations of Landau damping. First, section \ref{LwKD} deals with Langmuir waves in a way reminiscent of both van Kampen's and Dawson's works: phase mixing is at work in Landau damping. Second, section \ref{LwDsL} takes a singular limit of the fundamental equation for the electrostatic potential; it recovers simultaneously Debye shielding and the classical expression for Langmuir waves obtained by dealing \`a la Landau with Vlasov equation. Section \ref{Vlso} shows the Vlasovian limit to be a singular one providing a renormalized description of the plasma.

The second half of this paper reviews a series of previously published results. In section \ref{MIIDS} a calculation using Picard technique shows that the acceleration of a particle due to another one is mediated by all other particles. This unveils the microscopic mechanism of Debye shielding to be intimately connected to collisions, and explains how a given particle can be shielded by all other particles, while contributing to their individual shieldings. Section \ref{Wpilnle} is devoted to wave-particle interaction. First one derives from the $N$-body dynamics a Hamiltonian describing the self-consistent motion of $M$ waves with $N'$ particles in the tail of the velocity distribution. On this basis, a simple statistical calculation derives both spontaneous emission and Landau damping, and the corresponding quasilinear friction and diffusion coefficients. Then, the saturation of the weak warm beam instability and the dynamics with a single wave are discussed. Finally section \ref{Coltrsp} deals with a problem completely out of reach of a Vlasovian description: it provides a calculation of Coulomb scattering describing correctly all impact parameters $b$, with a convergent expression reducing to Rutherford scattering for small $b$. Nonlinear issues are dealt with in sections \ref{Wpilnle} and \ref{Coltrsp}. Finally, section \ref{Disc} discusses in a self-contained way \emph{the new physical picture of basic microscopic plasma physics provided by the $N$-body approach} and the new insight into the Vlasovian limit resulting from this approach. The reader might benefit from reading this last section first.

\section{One Component Plasma model}
\label{sec:1}

This section defines the basic scales and equations of the One Component Plasma model used up to and including section \ref{Wpilnle}, and introduces a class of granular distributions close to being spatially uniform. Since each new charged species brings new elements of complexity to the system, for the sake of simplicity we focus on the dynamics due to electrons, while neglecting that of ions because of their higher inertia. We also neglect any magnetic effect. More specifically, we deal with the One Component Plasma (OCP) model \cite{Salp,Abe,BH}, which considers the plasma as infinite with spatial periodicity $L$ in three orthogonal directions with coordinates $(x,y,z)$, and made up of $N$ electrons in each elementary cube with volume $L^3$. Ions are present only as a uniform neutralizing background, enabling periodic boundary conditions.

\subsection{Basic scales}
\label{Bs}

The number density is $n = N/L^3$. It enables the definition of a fundamental length scale of the system,
\begin{equation}
\lid = n^{-1/3},
\label{lid}
\end{equation}
which is the average inter-particle distance. At this scale, the plasma looks granular. The number density also enables the definition of a fundamental frequency of the system, the plasma (angular) frequency
\begin{equation}
\omega_{\rmpp} = \left[\frac{e^2 n }{\varepsilon_0 m }\right]^{1/2},
\label{omegap}
\end{equation}
where $m$ is the electron mass, $e$ is the proton charge, and $\varepsilon_0$ is the vacuum permittivity. In solid state physics, since $\hbar\omega_{\mathrm p} \gg T$, where $T$ is the plasma temperature ($T$ stands for $k_{\rm{B}} T$, where $k_{\rm{B}}$ is the Boltzmann constant), it is very hard to excite oscillations at the plasma frequency; these oscillations are quantized and are called plasmons. For the plasmas of interest here, the density and temperature are such that the opposite ordering holds. Therefore, there are many plasmons already at the thermal level, and they may be described through their coherent states, i.e.\ classical electrostatic waves.

If the plasma has a temperature $T$ measured in units of energy, two new fundamental length scales may be defined. First, the classical distance of closest approach
\begin{equation}
\lma = \frac{e^2}{4 \pi \, \varepsilon_0 T}
\label{lma}
\end{equation}
is the minimum distance of two electrons in a Rutherford collision, since $ {e^2}/(4 \pi \, \varepsilon_0 \lma) = m v_{\rmT}^2$, where $v_{\rmT} = [ T / m]^{1/2}$ is their thermal velocity\footnote{Many textbooks multiply this expression by $\sqrt 2$.}.

The second length scale involves both the density and the temperature, and is the Debye length
\begin{equation}
 \lambda_\rmD = \frac{v_{\rmT}}{\omega_{\rmpp}} = \left[\frac{\varepsilon_0 T }{n e^2}\right]^{1/2}.
\label{LDeb}
\end{equation}
This length will turn out to be the typical distance where the Coulomb potential of a point charge is shielded out (see section \ref{LwDsL}). For the plasma to be quasineutral, we impose $L \gg \lambda_\rmD$.

\begin{figure}
\begin{center}
\includegraphics[width=1\textwidth]{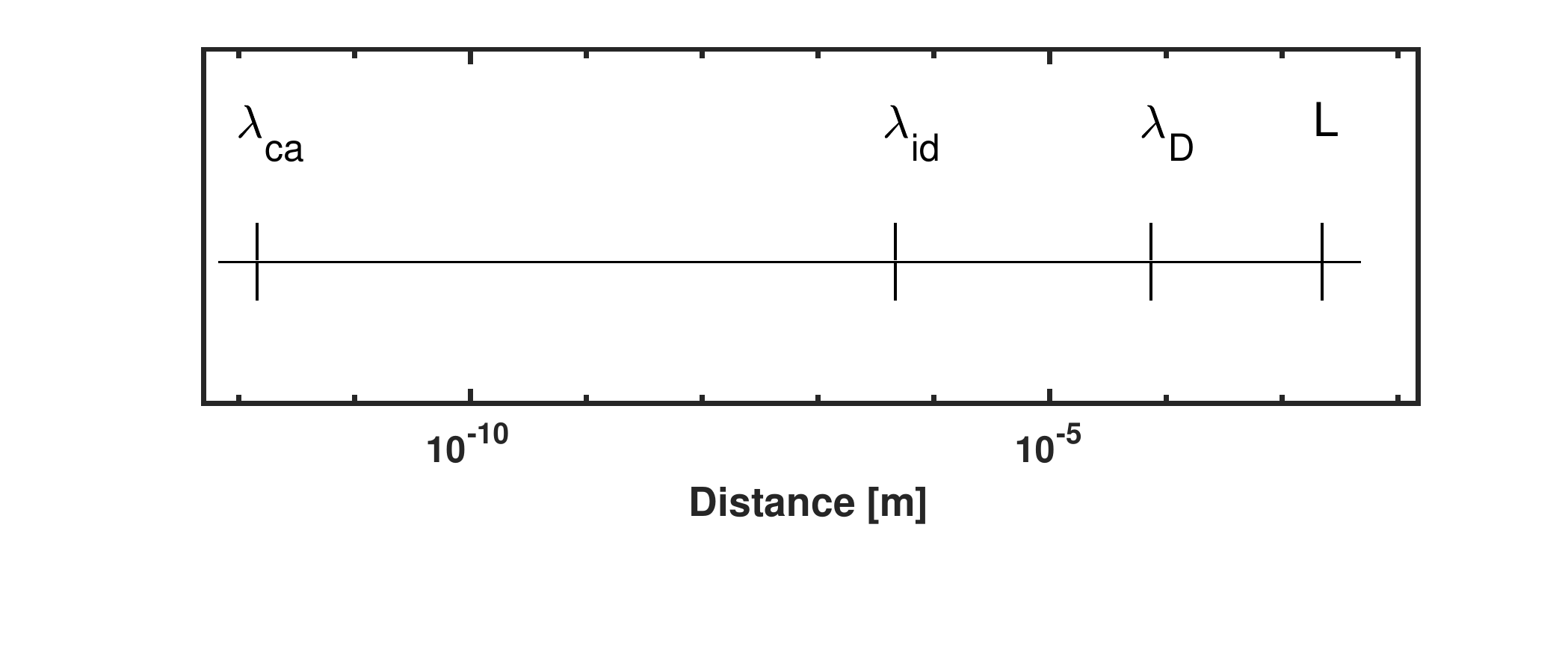}
\caption{View in logarithmic scale of the different scale lengths discussed in this section.}
\label{fig:length_scales}
\end{center}
\end{figure}

We define the plasma parameter
 \begin{equation}
 \Lambda = n \lambda_\rmD^3 = \left[\frac{\lambda_\rmD }{\lid}\right]^{3}
 = \left[\frac{\lid}{4 \pi \lma}\right]^{3/2},
\label{Lambda}
\end{equation}
which is about one fourth of the number of particles in the Debye sphere, a sphere of radius $\lambda_\rmD $. A large value of $\Lambda$ implies $\lma \ll \lid \ll \lambda_\rmD$.
This ordering is represented in figure \ref{fig:length_scales} using a logarithmic scale,
for a plasma with a density $n=10^{19}\,{\rm m}^{-3}$ and a temperature $T=1$ keV. It rules out strongly coupled and/or degenerate plasmas.

By definition, $\lma$ is the typical distance where the electrostatic potential of a pair of electrons balances their kinetic energy. Since for $\Lambda \gg 1$, $\lma \ll \lid$, particles typically interact weakly with one another, which corresponds to weakly coupled plasmas.

\subsection{Equations of motion}
\label{Em}

Because of the spatial periodicity of our One Component Plasma model, we write the potential created by the $N$ particles as a Fourier series
\begin{equation}
\varphi(\bfr) = \frac{1}{L^3}\sum_{\bfm, \, k_{\bfm} b_{\mathrm{smooth}} \leq 1} \tilde{\varphi} (\bfm) \exp(\rmi \bfk_{\bfm} \cdot \bfr),
\label{phiInv}
\end{equation}
where $\bfk_{\bfm} = \frac{2 \pi}{L} \, \bfm$, and where we restrict the Fourier expansion to $k_{\bfm}$'s such that $k_{\bfm} b_{\mathrm{smooth}} \leq 1$, where
\begin{equation}
b_{\mathrm{smooth}} \ll \lambda_\rmD,
\label{bsmlD}
\end{equation}
in order to avoid the $1/r$ singularity of the Coulomb potential\footnote{This smoothing is similar to the one performed in the mean-field derivation of the Vlasov equation, which was recalled in the introduction.}. The coefficients of the series are $\tilde{\varphi}(\bfm)
= \int \varphi(\bfr) \exp(- \rmi \bfk_{\bfm} \cdot \bfr) \, \rmd^3 \bfr$. We choose $\tilde{\varphi}(\mathbf{0}) = 0$. For $\bfm \neq \mathbf{0}$, they are readily obtained from the Poisson equation $\triangle \varphi = \frac{e}{\varepsilon_0} \sum_{j = 1}^N
     \delta[\bfr - \bfr_j(t)]$ with periodic boundary conditions, and are given by
\begin{equation}
  \tilde{\varphi}(\bfm)
  = -\frac{e}{\varepsilon_0 k_{\bfm}^2} \sum_{j = 1}^N
     \exp[- \rmi \bfk_{\bfm} \cdot \bfr_j(t)],
\label{phitildetotM}
\end{equation}
where $\bfr_j(t)$ is the position at time $t$ of particle $j$ acting as a source, and $k_{\bfm} = \|\bfk_{\bfm}\|$. Since the self-field due to the smoothed Coulomb potential vanishes, it is not necessary to exclude self-interaction in the equations of motion.

Newton's second law of motion for particle $j$ reads
\begin{equation}
  \ddot{\bfr}_j
  = \frac{e}{m} \nabla \varphi (\bfr_j).
\label{rsectot}
\end{equation}
The $N$-body dynamics is defined by equations (\ref{phiInv}) and (\ref{rsectot}) written for the $N$ particles.

\subsection{Spatially uniform granular distribution of particles}
\label{Aud}

\begin{figure}
\begin{center}
\includegraphics[width=1\textwidth]{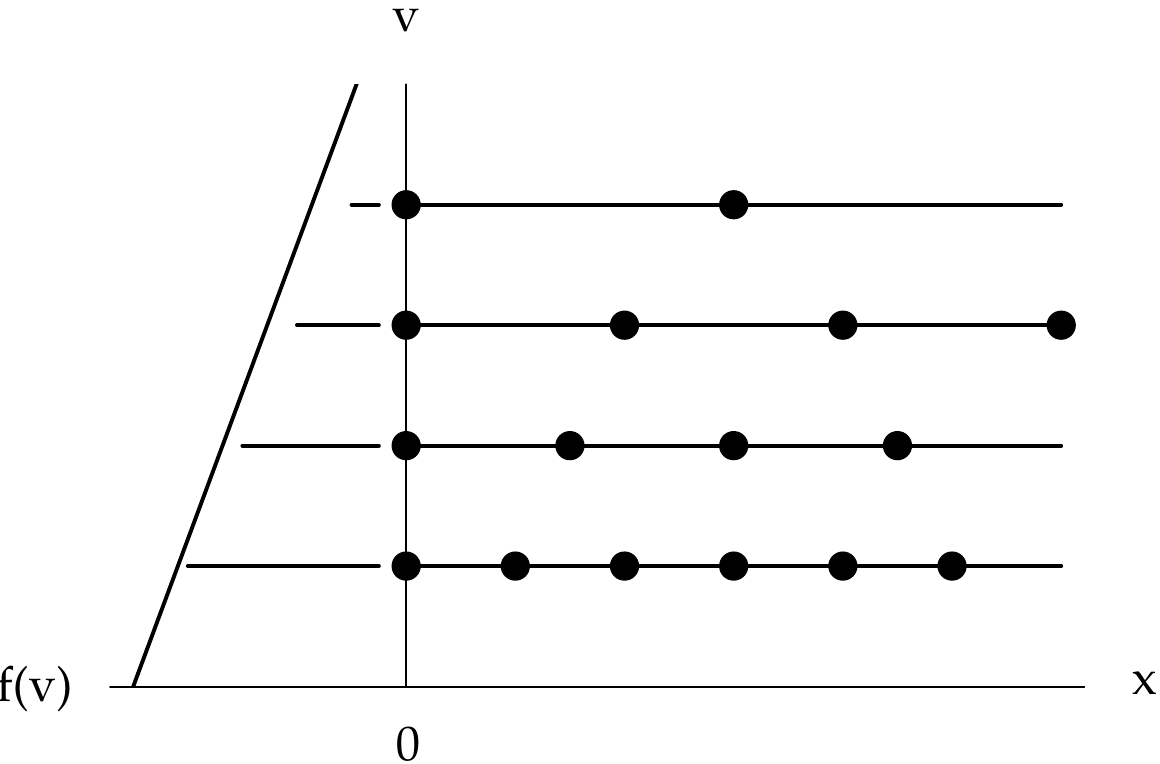}
\caption{One-dimensional multi-beam-multi-array.}
\label{fig:mbma}
\end{center}
\end{figure}

We introduce a class of spatially uniform granular distributions, which will be perturbed in the next sections.
In the case of a cold plasma, such a distribution can be obtained by setting particles with a vanishing velocity over a cubic array.
For a multi-velocity distribution, we take a set of monokinetic beams where each beam is
a simple cubic array of particles whose elementary cube has its edges along the three orthogonal directions with coordinates $(x,y,z)$. Such a set is called \emph{multi-beam-multi-array}. Figure \ref{fig:mbma} displays such a distribution of particles for a one-dimensional plasma.

We first focus on a given beam. It is convenient to consider the index $j$ of its particles as a three-dimensional vector $\bfp$
whose each integer components run from 1 to $n_{\mathrm{edge}}$. Since its particles have the same velocity $\bfv_{j}$, in equation (\ref{rsectot}) combined with equation (\ref{phitildetotM}), the part of the sum due to the particles of this beam bears on $\exp[\rmi \bfk_{\bfm} \cdot \bfr_{j0}]$ only, where $\bfr_{j0} = \frac{L}{n_{\mathrm{edge}}}\bfp$ is the initial position of particle $j$ ($L/n_{\mathrm{edge}}$ is the edge length of the elementary cube). Due to the periodicity of the $\bfr_{j0}$'s, the corresponding sum vanishes unless the three components of $\bfm$
are on the simple cubic lattice $(n_{\mathrm{edge}} {\mathbb{Z}})^3$ with mesh length $n_{\mathrm{edge}}$. Therefore, the sum vanishes for $\bfk_{\bfm} \neq \mathbf{0}$, if we assume
\begin{equation}
\pi b_{\mathrm{smooth}} n_{\mathrm{edge}}/L > 1.
\label{bsmnedge}
\end{equation}
This condition can be satisfied provided the inter-particle distance is such that it fulfills the condition
\begin{equation}
\lid \ll b_{\mathrm{smooth}}.
\label{dllbsmoo}
\end{equation}
This and condition (\ref{bsmlD}) imply $\lid \ll \lambda_\rmD$, which implies a large value of $\Lambda$ as a consequence of equation (\ref{Lambda}). As the contribution of each beam vanishes, the total Coulomb force for this set of beams vanishes identically. Therefore, the system of beams is force-free, and their distribution is invariant in time. We notice that, for this set of beams, both collisions and Debye shielding do not work. As is shown in section \ref{MIIDS}, the fact that these two mechanisms fail simultaneously is natural, because the former produces the latter. For kinetic/collisionless plasmas, the position distribution corresponding to this set of beams is very atypical, and Debye shielding
can generically set in. In contrast, cold plasmas take on a crystalline structure where particles vibrate about positions corresponding to the nodes of perfect lattices, and do not experience shielding.

In the following, it will be useful to consider continuous limits of multi-beam-multi-arrays with bounded velocities related to very large numbers of particles in the Debye sphere ($\Lambda \gg 1$).  We consider the beam velocity distributions to be a granular approximation of a spatially uniform smooth velocity distribution $f_0(\bfv)$. One can deal in this limit with a large number of beams whose velocities are on a three-dimensional grid becoming tighter and tighter in this limit. These tight grids may be taken as cubic arrays with an orientation arbitrary with respect to the spatial frame.

When dealing with wavevector $\bfk_{\bfm}$, the calculations are simpler when the velocity cubic array has an edge parallel to $\bfk_{\bfm}$ and one of its points is at the origin of velocities. Then, we consider the beam velocity distribution to be a granular approximation of a smooth velocity distribution $f_0(\bfv)$ whose integral perpendicular to $\bfk_{\bfm}$ is $g(v)$ (we omit the index $\bfm$ to simplify notations). We assume $\int f_0(\bfv)\rmd^3 \bfv = \int g(v) \rmd v = 1$. When $\Lambda$ increases, $f_0(\bfv)$ and $g(v)$ are split over a growing number of beams whose velocities lie on grids with a mesh size going to 0.

\section{Langmuir waves \`a la Kaufman}
\label{LwaK}

This section deals with Langmuir waves by using a technique introduced by Kaufman in a Vlasovian setting \cite{Kaufman}. The calculation provides a very short and rigorous derivation of Landau damping together with the intuitive interpretation of this effect\footnote{In the similar spirit of considering a wave with a slowly varying amplitude, but by using smooth distribution functions, references \cite{benisti2007,benisti2009} derive not only Landau damping, but also its equivalent in the nonlinear regime.}.

\subsection{Landau damping}
\label{LandamK}

We now consider the case where all particles have their initial positions slightly perturbed with respect to those of a multi-beam-multi-array; this choice will provide useful cancellations in the following. Let $\bfr_{j0}$ be the initial position of the unperturbed beam particle with index $j$, and $\bfv_{j}$ be its velocity. Let $\delta \bfr_{ j}$ be the smooth amplitude of modulation at $t=0$ of the positions of the particles belonging to the same beam as particle $j$, and $\Delta {\bfr}_j(t) = \bfr_j(t) - \bfr_{j0} - \bfv_j t$ be the mismatch of particle $j$ with respect to its ballistic position. 
We now assume that the beam particles have their initial positions sinusoidally modulated with the wavevector $\bfk_{\bfm}$ according to $\Delta {\bfr}_j(0) = \delta \bfr_{ j} \sin(\bfk_{\bfm} \cdot \bfr_{j0})$, while keeping the same velocity\footnote{An arbitrary phase should be present in the sine, but it is set to 0, since it is not important for the derivation. The case with a concomitant small velocity modulation can be dealt with in a similar manner, but with longer expressions. Then a $\Delta \bfv_{ j} \, t \, \sin(\bfk_{\bfm} \cdot \bfr_{j0} + \psi_j)$ contribution must be added to $\Delta {\bfr}_j(t)$, where $\psi_j$ and $\Delta \bfv_{ j}$ are respectively the phase and the amplitude of modulation of the velocities of the particles belonging to the same beam as particle $j$.}. 
Setting these positions in the expression of $\tilde{\varphi} (\bfl)$ provided by equation (\ref{phitildetotM}), shows that if the $\delta \bfr_{ j}$'s are of order $\varepsilon$, because of condition (\ref{bsmnedge}), $\tilde{\varphi} (\bfl)$ is of order $\varepsilon^2$ or higher for $\bfl \neq \pm \bfm$. In contrast, $\tilde{\varphi} (\bfm)$ is\footnote{The same property would hold as well for other types of interactions than Coulombian ones.} of order $\varepsilon$. This limits the number of terms to be taken into account when computing $\tilde{\varphi} (\bfm,t)$ to this order. However, the excitation of other wavevectors at order $\varepsilon^2$ is important physically, since it corresponds to \emph{spontaneous emission} enabling the excitation of a wave with an initially vanishing amplitude. This emission will be described in a statistical setting in section \ref{SpontEmission}.

We consider the linearized motion of the OCP about the multi-beam-multi-array distribution, and we look for a solution of the type $\tilde{\varphi} (\bfm,t) = A(t) \exp (- \rmi \omega t)$, where $\omega$ is real and $A(t)$ is a complex amplitude of order $\varepsilon$ varying on the scale $\tau_A \gg 1/|\omega|$ (its dependence on $\bfm$ is not shown to simplify the notations). Retaining only the contribution of the wave of interest, the equation of motion of particles is
\begin{equation}
\ddot{\bfr}_j
  = \alpha \bfk_{\bfm} A(t) \exp [\rmi (\bfk_{\bfm} \cdot \bfr_j(t) - \omega t)] + \rmcc,
\label{Acc1w}
\end{equation}
where ``c.c.'' means complex conjugate, and
\begin{equation}
\alpha = \rmi e/(m L^3).
\label{alpha}
\end{equation}
Integrating formally twice in time the equation of motion for particle $j$ up to first order in $\varepsilon$, and changing the order of integrations yields
\begin{equation}
\Delta \bfr_{j1}(t)
  = \delta \bfr_{ j} \sin(\bfk_{\bfm} \cdot \bfr_{j0})
  + \alpha \bfk_{\bfm} \int_0^{t} \tau A(t-\tau) \exp [\rmi (\Omega_j (t-\tau) + \bfk_{\bfm} \cdot \bfr_{j0})] \rmd \tau + \rmcc \,,
\label{delrj}
\end{equation}
where $\Omega_j = \bfk_{\bfm} \cdot \bfv_j - \omega$, and both contributions of $\tilde{\varphi} (\bfm,t)$ and of its complex conjugate $\tilde{\varphi} (- \bfm,t)$ are taken into account; we stress that the resonant
component of the particle position (and velocity) is along the considered wavevector. Setting this into the definition (\ref{phitildetotM}) of $\tilde{\varphi} (\bfm)$, and approximating $\exp[- \rmi \bfk_{\bfm} \cdot \Delta \bfr_j(t)]$ by $1- \rmi \bfk_{\bfm} \cdot \Delta \bfr_j(t)$ yields to order $\varepsilon$
\begin{eqnarray}
A(t) &=& \frac{e}{2 \varepsilon_0 k_{\bfm}^2 }\sum_{j=1}^N \Bigl[ \bfk_{\bfm} \cdot \delta \bfr_{ j} \exp (- \rmi \Omega_j t) \Bigr.
\nonumber \\ && \qquad \qquad \qquad \Bigl.
- 2 \alpha k_{\bfm}^2 \frac{\partial}{\partial \Omega_j} \int_0^{t} A(t-\tau) \exp (- \rmi \Omega_j \tau) \rmd \tau \Bigr] ,
\label{At}
\end{eqnarray}
where we took advantage of cancellations resulting from condition (\ref{bsmnedge}) and from the fact that beam particles sit on arrays. In the spirit of the continuous limits of multi-beam-multi-arrays introduced in section \ref{Aud}, we now substitute the discrete sum over particles with the integral over a smooth distribution function $f(\bfv)$ whose integral perpendicular to the $\bfk_{\bfm}$ axis is a function $g(v)$ of the type introduced in section \ref{Aud} (this substitution is justified below equation (\ref{gammaL})). This yields
\begin{eqnarray}
A(t) & = & \int f(\bfv)  \frac{Ne}{2\varepsilon_0 k_{\bfm}^2 }\, \bfk_{\bfm} \cdot \delta \bfr(\bfv)
                         \exp [- \rmi (\bfk_{\bfm} \cdot \bfv - \omega)  t] \, \rmd^3 \bfv
 \nonumber \\
&& \qquad  - \rmi \, \omega_{\rmpp}^2
\int g(v)
\frac{\partial}{\partial \Omega} \int_0^{t} A(t-\tau) \exp (- \rmi \Omega \tau) \rmd \tau \rmd v ,
\label{At1}
\end{eqnarray}
where $\delta \bfr(\bfv)$ is the equivalent of $\delta \bfr_{ j} $ in the continuous limit, 
$\Omega = k_{\bfm} v - \omega$ with $k_{\bfm} =\| \bfk_{\bfm}\|$, 
$g(v)= \int f(\bfv) \rmd^2 \bfv_\bot$, with $v$ and $\bfv_\bot$ the components of $\bfv$ respectively parallel and perpendicular to $\bfk_{\bfm}$. Then, a Taylor expansion of $A(t-\tau)$ to first order in $\tau$ yields
\begin{eqnarray}
A(t) =&& \int \left[\frac{Ne}{2\varepsilon_0 k_{\bfm}^2 }\, h(v)  \exp (- \rmi \Omega t) + \omega_{\rmpp}^2 A(t) g(v) \frac{\partial}{\partial \Omega}  \frac{\exp (- \rmi \Omega t) -1}{\Omega} \right] \rmd v
 \nonumber \\
&&- \, \rmi \, \omega_{\rmpp}^2
\dot{A}(t)  \int g(v) \frac{\partial^2}{\partial \Omega^2}  \frac{\exp (- \rmi \Omega t) -1}{\Omega} \,  \rmd v,
\label{AtZ}
\end{eqnarray}
where $h(v)= \int f(\bfv) \, \bfk_{\bfm} \cdot \delta \bfr(\bfv) \, \rmd^2 \bfv_\bot$. This truncated Taylor expansion makes sense since, as shown in \ref{fdfe}, $A(t)$ is an entire function, and the successive contributions in the expansion decay like powers of $(\tau_A k_{\bfm} v_{\rmT})^{-1}$. After integrating by parts, this becomes
\begin{eqnarray}
A(t) =&& \int \left[\frac{Ne}{2\varepsilon_0 k_{\bfm}^2 }\, h(v)  \exp (- \rmi \Omega t) + \frac{\omega_{\rmpp}^2}{k_{\bfm}} A(t) g'(v) \frac{1-\cos (\Omega t)+ \rmi \sin (\Omega t)}{\Omega} \right] \rmd v
 \nonumber \\
&&- \, \rmi \frac{\omega_{\rmpp}^2}{k_{\bfm}^2}
\dot{A}(t)  \int g^{\prime \prime}(v) \frac{\cos (\Omega t)- \rmi \sin (\Omega t) -1}{\Omega} \rmd v,
\label{Atint}
\end{eqnarray}
where $g'(v)$ and $g^{\prime \prime}(v)$ are respectively the first and the second derivative of $g(v)$.

For $t=0$, this equation defines $A(0)$. When $t$ grows, the first exponential and the cosine terms in equation (\ref{Atint}) produce a phase mixing\footnote{Phase mixing is a classical concept in the theory of kinetic plasma waves, and especially in the van Kampen-Case approach to Landau damping \cite{vanKampen,vanKampen2,Case}. Intuitively, it corresponds to the idea that the integral of a rapidly oscillating function is close to zero. Mathematically, it is grounded on the fact that $\int \rmd \nu F(\nu)  \exp (- \rmi \nu t)$ is the Fourier transform of $F(\nu)$, which decays for large $t$'s in various cases. This occurs in particular, at least on average, if $F(\nu)$ is an $L^2$ function; also if $F(\nu)$ has an integrable derivative of order at least one. Having one of such properties is natural for $h(v)$, $g'(v)$, and $g^{\prime \prime}(v)$, especially if $g(v)$ is analytic, as assumed in Landau's derivation of Landau damping \cite{LL}.} leading asymptotically to the expression\footnote{Phase mixing works here in the following way: because of the integration over $v$, the first exponential and the cosine terms in equation (\ref{Atint}) produce the Fourier transforms of $h(v)$, $g'(v)$ and $g^{\prime \prime}(v)$. For $t$ large, the vanishing values of the tails of these transforms are involved, and thus neglected. For $t$ large, in the integrals involving $\sin(\Omega t)/\Omega$, this factor has non vanishing values over a vanishing domain in $v$. This enables extracting $g'(\frac{\omega}{k_{\bfm}})$ out of the integral. The similar contribution involving $g^{\prime \prime}(v)$ is neglected, since it is of higher order in $(\tau_A k_{\bfm} v_{\rmT})^{-1}$. We stress that we do not use Plemelj formula, in contrast with what was done in reference \cite{Kaufman}.}
\begin{equation}
A(t) = \frac{\omega_{\rmpp}^2}{k_{\bfm}} A(t) \rmP \! \! \! \int g'(v) \frac{1}{\Omega} \, \rmd v
+ \rmi \frac{\pi \omega_{\rmpp}^2}{k_{\bfm}^2}  g'(\frac{\omega}{k_{\bfm}}) A(t)
+ \rmi \frac{\omega_{\rmpp}^2}{k_{\bfm}^2} \dot{A}(t) \rmP \! \! \! \int g^{\prime \prime}(v) \frac{1}{\Omega} \, \rmd v,
\label{Atintasym}
\end{equation}
where $\rmP$ stands for the Cauchy principal value, and only the dominant real and imaginary terms are kept; the other ones are at most of order $(k_{\bfm} v_{\rmT} \tau_A)^{-1}$, where $\tau_A \gg 1/|\omega|$ is the time scale of variation of $A(t)$. The typical time scale for phase mixing to become strong is $\tau_{\rm{mix}} = (k_{\bfm} v_{\rmT})^{-1}$.

Equation (\ref{Atintasym}) is a homogeneous, linear ordinary differential equation with constant coefficients.
Its general solution thus reads $A(t) = A_0 \exp(\gamma t)$, where $\gamma$ is the solution to the algebraic equation
\begin{equation}
   \epsilon_{\rm{r}}(\bfm,\omega) - \rmi \frac{\pi \omega_{\rmpp}^2}{k_{\bfm}^2}  g'(\frac{\omega}{k_{\bfm}})
   =
   \gamma \, \rmi \frac{\omega_{\rmpp}^2}{k_{\bfm}^2} \, \rmP \! \! \! \int g^{\prime \prime}(v) \frac{1}{\Omega} \, \rmd v
\label{gammaAasym}
\end{equation}
where we define
\begin{equation}
   \epsilon_{\rm{r}}(\bfm,\omega)
   = 1 - \frac{\omega_{\rmpp}^2}{k_{\bfm}} \, \rmP \! \! \! \int g'(v) \frac{1}{\Omega} \, \rmd v.
\label{epsr}
\end{equation}
The unique solution of the problem is obtained by requiring $\epsilon_{\rm{r}}(\bfm,\omega)=0$, which provides the usual definition of $\omega_{\bfm \rmr}$, the real part of the frequency in the Landau calculation of Langmuir waves.

Assuming $g$ is regular enough (e.g.\  with a uniformly continuous second derivative),
integrating by parts under the principal value shows that the right-hand side in equation (\ref{gammaAasym}) is
$- \rmi \gamma \, {\partial_\omega \epsilon_{\rm{r}}}$, and equation (\ref{gammaAasym}) reduces to
$\gamma = \gamma_{\rm L}(\bfk_{\bfm})$ where
\begin{equation}
\gamma_{\rm L}(\bfk_{\bfm}) = \frac{\pi \omega_{\rmpp}^2}{k_{\bfm}^2 \frac{\partial \epsilon_{\rm{r}}}{\partial \omega}} g' (\frac{\omega_{\bfm \rmr}}{k_{\bfm}})
\label{gammaL}
\end{equation}
\newpage\noindent
is the Landau damping/growth rate to lowest order in $\gamma_{\rm L}(\bfk_{\bfm}) (k_{\bfm} v_{\rmT})^{-1}$, whose sign is\footnote{At this point, we notice that the derivation of Landau damping in the mechanical $N$-body setting is more accessible to students than Landau's derivation: the number of pages is divided by three (see for instance sections 6.3 to 6.5 of \cite{Nicholson}), the mathematics is elementary, and there is no need to introduce Vlasov equation.} the one of $g' (\frac{\omega_{\bfm \rmr}}{k_{\bfm}})$. The above calculation is performed to all orders in \ref{fdfe}, and yields exactly Landau's complete formula if $g(v)$ is analytic. The latter property is not necessary for the present calculation.

In practice, for the Landau effect to be observable, $\frac{\omega_{\bfm \rmr}}{k_{\bfm}}$ must stay close to $v_{\rmT}$. 
Then the relevant mixing time is close to a plasma period. If the considered multi-beam-multi-array corresponds in $v$ to a series of adjacent equidistant monokinetic beams with a mismatch $\delta$ in velocity, the passage to a smooth distribution function in equation (\ref{At1}) makes sense provided $\bfk_{\bfm}\, \delta t \ll 1$. 
Therefore, to be observable, Landau damping requires\footnote{Because of the Floquet exponents introduced in \ref{fdfe} and computed in \ref{App:InfNbBeams}, the actual condition (\ref{CondL}) is somewhat stronger.} $k_{\bfm} \delta \, |\gamma_{\rm L}(\bfk_{\bfm})|^{-1} \ll 1$. 
Also $\gamma_{\rm L}(\bfk_{\bfm}) \ll k_{\bfm} v_{\rmT}$ for phase mixing to occur faster than the damping time. This condition is well satisfied in a Maxwellian plasma, when using the Bohm-Gross dispersion relation\footnote{The Bohm-Gross dispersion relation is $\omega^2 = \omega_{\rmpp}^2 + 3 k^2 v_{\rmT}^2$.}. We notice that, in the present derivation of Landau damping, phase mixing corresponds to the transients related to the terms complementing those of Landau poles when inverting the Laplace transform in the usual Vlasovian derivation. The relevance of transients, depending on the way the self-consistent electrostatic field has been generated, is further discussed in \ref{RtbLd}.

Finally, since $A(t)$ is entire, the above Taylor expansion is converging for all $\tau$'s. Therefore we may compute the real part of the dispersion relation and the damping rate to arbitrary order in $\gamma_{\rm L}(\bfk_{\bfm}) (k_{\bfm} v_{\rmT})^{-1}$. This only requires $g(v)$ to be infinitely differentiable, not analytic. The next order terms in equation (\ref{Atintasym}) give perturbatively\footnote{Using again phase mixing, but going to next order in $(k_{\bfm} v_{\rmT} \tau_A)^{-1}$ in equation (\ref{Atintasym}), adds $ \frac{\pi \omega_{\rmpp}^2}{k_{\bfm}^3}  g^{\prime \prime} (\frac{\omega}{k_{\bfm}}) \dot{A}(t) - \rmi \frac{\pi \omega_{\rmpp}^2}{2 k_{\bfm}^4}  g^{\prime \prime \prime} (\frac{\omega}{k_{\bfm}}) \ddot{A}(t)$ in the right hand side of this equation. The next orders can be obtained from equations (\ref{D3})-(\ref{D4}).} a modified $\epsilon_{\rm{r}}(\bfm,\omega)$
\begin{equation}
\epsilon_{\rm{r}2}(\bfm,\omega)= \epsilon_{\rm{r}}(\bfm,\omega) +
\frac{\pi \gamma_{\rm L}(\bfk_{\bfm}) \omega_{\rmpp}^2}{k_{\bfm}^3} g^{\prime \prime} (\frac{\omega_{\bfm \rmr}}{k_{\bfm}}),
\label{epsr2}
\end{equation}
and a more precise expression of Landau damping
\begin{equation}
\gamma_{\rm L2}(\bfk_{\bfm}) = \gamma_{\rm L}(\bfk_{\bfm}) [1 -\frac{\gamma_{\rm L}^2(\bfk_{\bfm})}{2 k_{\bfm}^2} \frac{g^{\prime \prime \prime} (\frac{\omega_{\bfm \rmr}}{k_{\bfm}})}{g' (\frac{\omega_{\bfm \rmr}}{k_{\bfm}})}].
\label{gammaL2}
\end{equation}
If $g(v)$ is a Maxwellian, this becomes
\begin{equation}
\gamma_{\rm L2}(\bfk_{\bfm}) = \gamma_{\rm L}(\bfk_{\bfm}) [1 +\frac{\gamma_{\rm L}^2(\bfk_{\bfm})}{2 (k_{\bfm} v_{\rmT})^2} (3 - \frac{\omega_{\bfm \rmr}^2}{(k_{\bfm} v_{\rmT})^2})]
\label{gammaL3}
\end{equation}
\newpage\noindent
and
\begin{equation}
\gamma_{\rm L2}(\bfk_{\bfm}) = \gamma_{\rm L}(\bfk_{\bfm}) [1 -\frac{(\gamma_{\rm L}(\bfk_{\bfm})\, \omega_{\rmpp})^2}{2 (k_{\bfm} v_{\rmT})^4}]
\label{gammaL4}
\end{equation}
by using the Bohm-Gross dispersion relation.

\subsection{Average synchronization of particles with waves}
\label{Aspw}

After introducing Landau damping, we now consider the corresponding dynamics of particles. To this end, we compute the sum of the $\ddot{\bfr}_j$'s for the particles considered at the beginning of the previous subsection whose position is slightly perturbed with respect to monokinetic beams. This sum vanishes, since the system is isolated (multiplying this sum by $m$ corresponds to the time derivative of the total momentum of the system, which is conserved). Using equation (\ref{delrj}) to express $\bfr_j(t)$ in equation (\ref{Acc1w}) to second order in $A$, and changing the order of integrations, we obtain
\begin{eqnarray}
\mathbf{0} =&& -\alpha  A(t) \bfk_{\bfm} \sum_{j=1}^N
 \bfk_{\bfm} \cdot \delta \bfr_j \exp(\rmi \Omega_j t) + \rmcc
 \nonumber \\
  &&+ \, \sum_{j=1}^N 2 \rmi |\alpha|^2 k_{\bfm}^2 A(t) \bfk_{\bfm} \int_0^{t} \tau A^{*}(t-\tau) \exp(\rmi \Omega_j \tau) \, \rmd \tau  + \rmcc \,,
\label{Accav}
\end{eqnarray}
where $ A^{*}(t)$ is the complex conjugate of $ A(t)$, $\alpha$ is defined in equation (\ref{alpha}), and where we took again advantage of cancellations resulting from the fact that beam particles sit on arrays and from condition (\ref{bsmnedge}). For $t \ll \tau_A$, the integral term and its complex conjugate combine to produce a term scaling like $|A(t)|^2 \int_0^{t} \tau \sin (\Omega_j \tau) \rmd \tau$. For $|\Omega_j|t < \pi$, it has the sign of $\Omega_j$, which implies an acceleration with the opposite sign. This corresponds to an \emph{average synchronization of the particles with the wave}. Such a synchronization was observed experimentally for the particles of a monokinetic beam in a travelling wave tube \cite{DovEsMa}. Equation (\ref{Accav}) may also be written
\begin{eqnarray}
\mathbf{0} =&& -\alpha  A(t) \bfk_{\bfm} \sum_{j=1}^N
 \bfk_{\bfm} \cdot \delta \bfr_j \exp(\rmi \Omega_j t) + \rmcc
 \nonumber \\
  &&+ \,
 \sum_{j=1}^N 2 |\alpha|^2 k_{\bfm}^2 A(t) \bfk_{\bfm} \frac{\partial}{\partial \Omega_j}
  \int_0^{t} A^{*}(t-\tau) \exp(\rmi \Omega_j \tau) \, \rmd \tau + \rmcc \,.
\label{Accav2}
\end{eqnarray}

Using again Taylor expansion in $\tau$, and following the procedure leading to equation (\ref{AtZ}), equation (\ref{Accav2}) becomes
\begin{eqnarray}
\mathbf{0}
  =&&
 -\alpha  A(t) \bfk_{\bfm} \sum_{j=1}^N
 \bfk_{\bfm} \cdot \delta \bfr_j \exp(\rmi \Omega_j t) + \rmcc
 \nonumber \\
 &&+ \, \sum_{j=1}^N  4 |\alpha|^2 k_{\bfm}^2 \bfk_{\bfm} \left[|A(t)|^2 \frac{\partial}{\partial \Omega_j}  \frac{\sin (\Omega_j t)}{\Omega_j}
\right.
 \nonumber \\
&&\qquad + \left.
A(t) \dot{A}^{*}(t)  \frac{\partial^2}{\partial \Omega_j^2}  \frac{\exp(\rmi \Omega_j t) -1}{2 \Omega_j} + \rmcc \,\right].
\label{Accavt}
\end{eqnarray}
As in the previous subsection, we use the continuous limit of multi-beam-multi-arrays introduced in section \ref{Aud} to introduce a smooth distribution function, and we integrate again by parts, which yields
\begin{eqnarray}
 \mathbf{0} =&& - N \alpha  A(t) \bfk_{\bfm} \int  h(v) \exp[\rmi \Omega_j t] \, \rmd v  + \rmcc
  \nonumber \\
&&
 - 4 N |\alpha|^2 k_{\bfm}^2 \bfk_{\bfm} \left( \frac{|A(t)|^2}{k_{\bfm}}
 \int  g'(v) \frac{\sin (\Omega t)}{\Omega} \, \rmd v \right.
 \nonumber \\
&& \quad \left. - \frac{A(t) \dot{A}^{*}(t)}{k^2_{\bfm}}
 \int g^{\prime \prime}(v) \frac{\cos(\Omega t)-1 + \rmi \sin(\Omega t) }{2 \Omega} \,  \rmd v + \rmcc \, \right) .
\label{Accavttot}
\end{eqnarray}
For $t=0$, the three contributions vanish identically. Moreover, since $\frac{\cos (\Omega t) -1}{\Omega}$ vanishes when $\Omega$ goes to 0, this factor cannot provide a contribution of nearly-resonant particles (those with $\Omega \simeq 0$). On the contrary, since $\frac{\sin (\Omega t)}{\Omega}$ does not vanish when $\Omega$ goes to 0, this factor provides a contribution of nearly-resonant particles. However, this factor eventually provides a vanishing contribution of the third integral, because it involves the factor $A(t) \dot{A}^{*}(t) - A^{*}(t) \dot{A}(t)$, which vanishes since $\gamma$ was found to be real in the previous subsection. Therefore, when $t$ grows, equation (\ref{Accavttot}) becomes asymptotically
\begin{equation}
 \mathbf{0} = - 4 \pi N |\alpha|^2 \bfk_{\bfm}
 g'(\frac{\omega}{k_{\bfm}}) |A(t)|^2
 + 2 N |\alpha|^2 \bfk_{\bfm} \frac{k_{\bfm}^2}{\omega_{\rmpp}^2} \frac{\partial \epsilon_{\rm{r}}}{\partial \omega} \frac{\rmd |A(t)|^2}{\rmd t},
\label{Accavttotasym}
\end{equation}
where the first term corresponds to the contribution of particles nearly-resonant with the wave, while the second one is the contribution of non-resonant particles. Multiplying by the electron mass $m$ yields
\begin{equation}
\frac{\rmd P_{\rm{res}}}{\rmd t} + \frac{\rmd P_{\rm{wave}}}{\rmd t} = 0,
\label{Ptot}
\end{equation}
where
\begin{equation}
\frac{\rmd P_{\rm{res}}}{\rmd t}
  =
  - \frac{4 \pi \varepsilon_0 \omega_{\rmpp}^2}{L^3} g'(\frac{\omega}{k_{\bfm}}) |A(t)|^2 \bfk_{\bfm}
\label{Accav3}
\end{equation}
may be interpreted as the derivative of the momentum of particles nearly-resonant with the wave, and
\begin{equation}
P_{\rm{wave}} =  \frac{2 \varepsilon_0 k_{\bfm}^2}{L^3} \frac{\partial \epsilon_{\rm{r}}(\bfm,\omega)}{\partial \omega} |A(t)|^2 \bfk_{\bfm}
\label{Pwave}
\end{equation}
is the total wave momentum in the volume $L^3$ (we recover the conservation of the total momentum indicated at the beginning of this subsection). Equation (\ref{Ptot}) can also be written as
\begin{equation}
\frac{\rmd |A(t)|^2}{\rmd t} = 2 \gamma_{\rm L}(\bfk_{\bfm}) |A(t)|^2,
\label{Land}
\end{equation}
giving again Landau growth or damping to lowest order.

We now focus on the term of equation (\ref{Accavt}) involving $\frac{\sin (\Omega t)}{\Omega}$, calling it $\langle \ddot{\bfr}_{\rm{sec}1} \rangle$. As indicated after equation (\ref{Accavttotasym}), it corresponds to the contribution of particles close to being resonant with the wave. Since the derivative in $\Omega$ of $\frac{\sin (\Omega t)}{\Omega}$ is $B=[\Omega t \cos (\Omega t)- \sin (\Omega t)]\Omega^{-2}$, the component of $\langle \ddot{\bfr}_{\rm{sec}1} \rangle$ along $\bfk_{\bfm}$ has the sign of $B$. For $|\Omega|t \ll 1$, $B \simeq -\Omega t^3/3$ , which implies a stronger average synchronization when $|\Omega|$ grows at fixed $t$. This also shows this synchronization to vanish with $|\Omega|$, which rules out any role of trapped particles in Landau damping or growth\footnote{Actually, a possible role of trapping is a priori excluded since the bounce period is unbounded in the linear regime of Langmuir waves.}.
One easily sees that $B$ keeps the sign of $-\Omega$ for $\Omega t$ of the order of a few units.
Therefore, since Landau damping occurs over a time scale $\gamma_{\rm L}^{-1}(\bfk_{\bfm})$, synchronization occurs for $|\Omega|$ up to the order of $|\gamma_{\rm L}|$.
For $|\Omega| \gg |\gamma_{\rm L}|$, $B$ scales like $t/|\Omega|$ over the time scale $\gamma_{\rm L}^{-1}(\bfk_{\bfm})$. Consequently, the average synchronization is maximum for $|\Omega|$ of the order of $ |\gamma_{\rm L}|$.
More precisely, $B$ performs decreasing oscillations between positive and negative values when $|\Omega|$ grows at fixed $t$.

\begin{figure}
\begin{center}
\subfloat[]{\label{fig:B_vs_t_over_Omega_no_zoom}\includegraphics[width=0.49\textwidth]{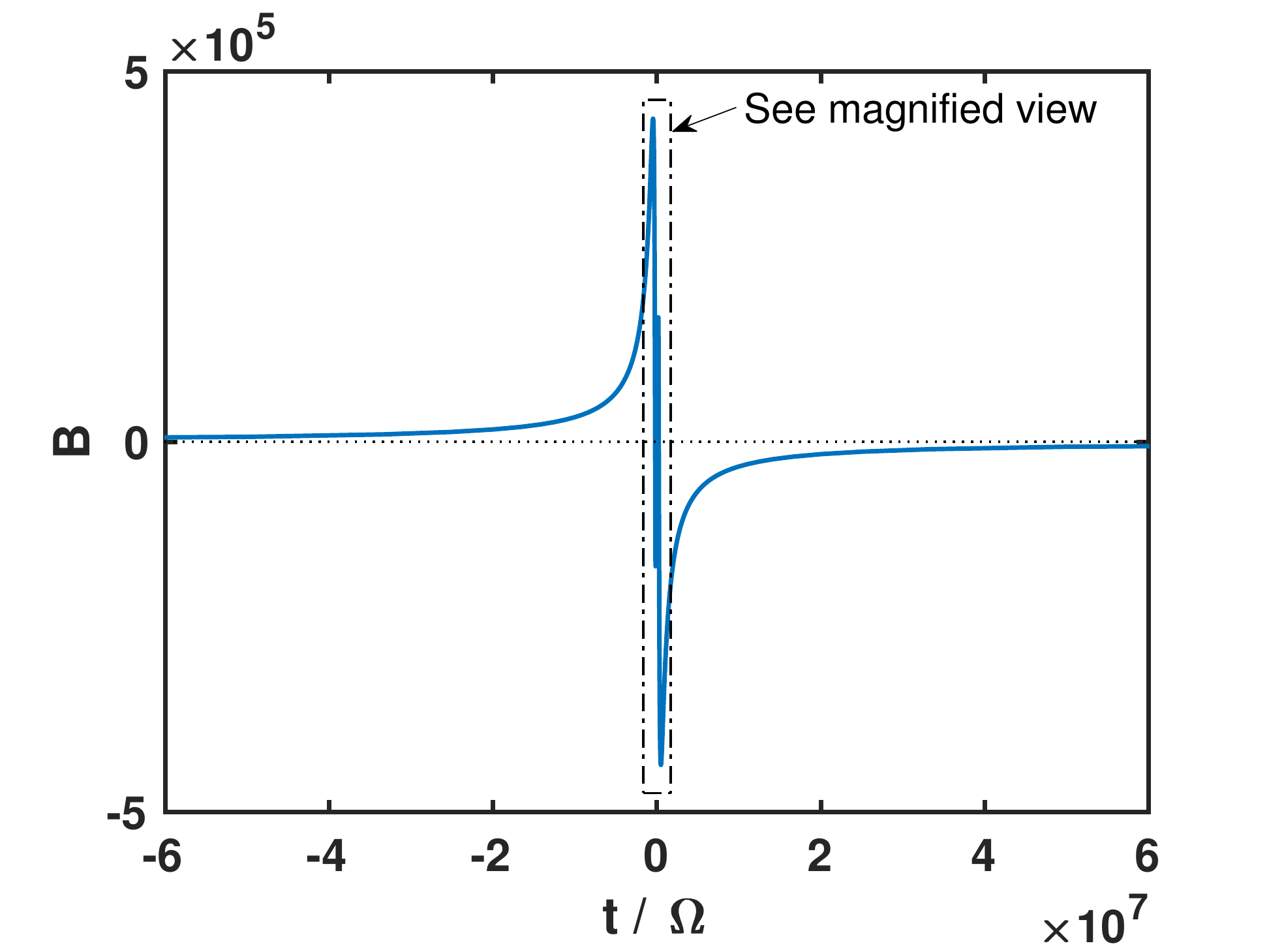}}
\subfloat[]{\label{fig:B_vs_t_over_Omega_zoom}\includegraphics[width=0.49\textwidth]{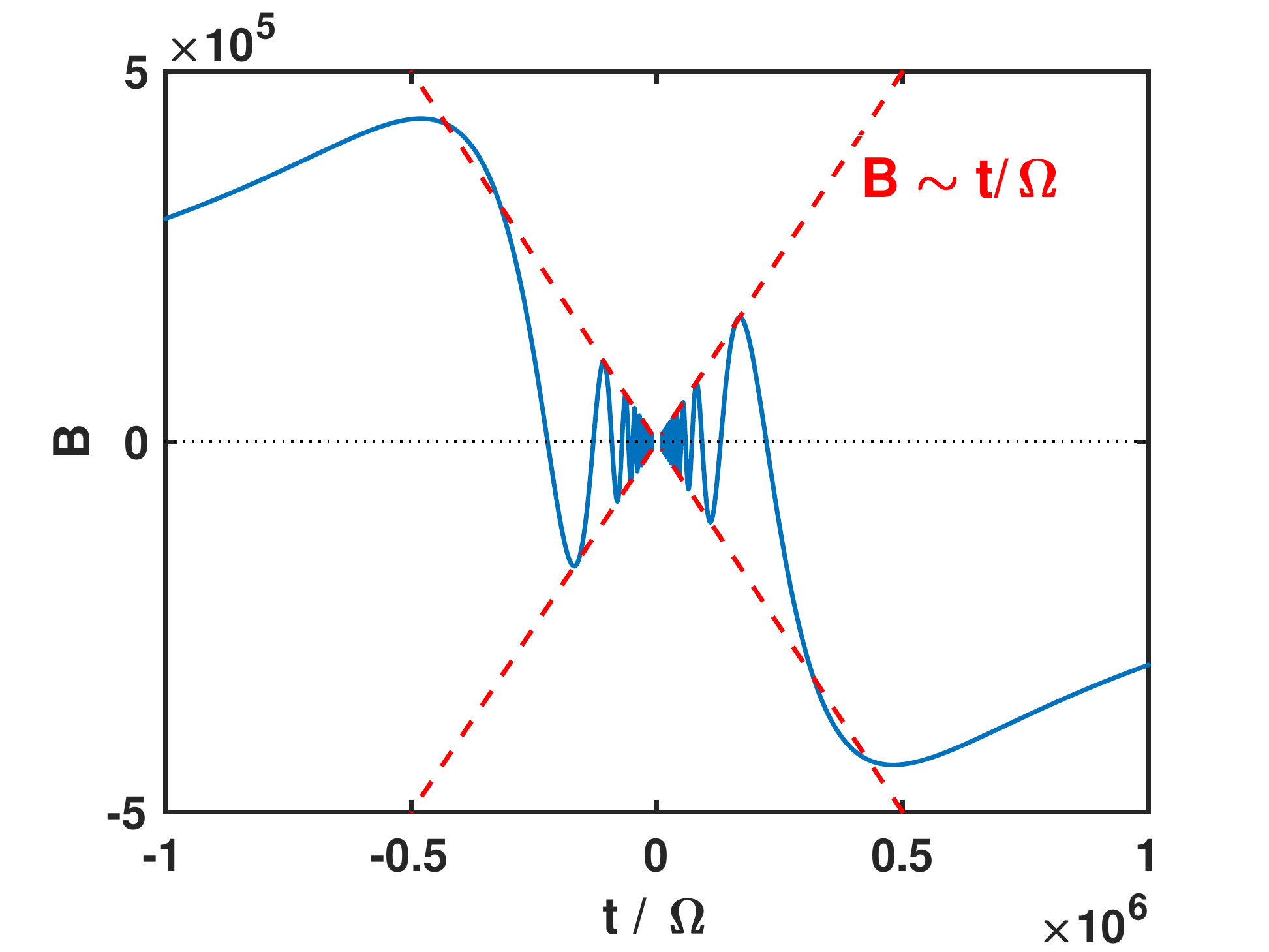}}
\caption{(Left) $B$ as a function of $t/\Omega$ for $t=10^3$, showing the decreasing amplitude
when $\vert\Omega\vert\rightarrow 0$. (Right) Magnified view of the rectangle indicated on the left panel,
showing the linear dependence on $t/\Omega$, as well as the decreasing oscillations when $\vert\Omega\vert$ grows.}
\label{fig:B_vs_t_over_Omega}
\end{center}
\end{figure}

The previous discussion is illustrated in figure \ref{fig:B_vs_t_over_Omega},
where the behaviour of $B$ as a function of $t/\Omega$ is plotted for a given $t$.
The left panel shows that $B$ indeed increases with $\vert\Omega\vert$ and therefore
also does the synchronization. For $\vert\Omega\vert$ large enough, the behaviour
$B\sim t/\Omega$ is also illustrated in the magnified view,
which highlights the oscillations with decreasing amplitude between positive and negative values of $B$.

\begin{figure}[t]
  \begin{center}
    \includegraphics[width=8cm,height=5cm]{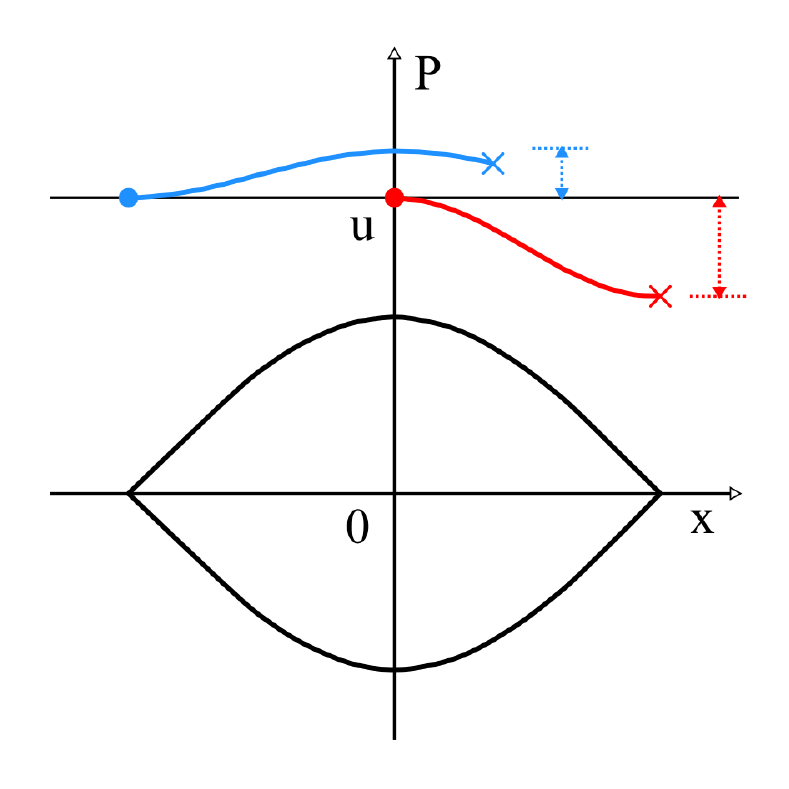}
    \caption{Phase space plot displaying the average synchronization of two particles with a wave, one starting at the position of the X-point of the separatrix (blue line), another one starting at the O-point of the trapping domain (red line).}
    \label{sync}
  \end{center}
\end{figure}

The average synchronization for small times can be intuitively understood by computing the velocity modulation of the particles having the same parallel velocity to the wave at $t=0$. Indeed, those whose absolute differential velocity to the wave is increasing at $t=0$ (blue line in figure \ref{sync}) have an average velocity which is higher than that of particles with an initially decreasing absolute differential velocity (red line in figure \ref{sync}). Since the former have a smaller velocity modulation than the latter, the desynchronization of the former is smaller than the synchronization of the latter, which yields an average synchronization.

As just shown, synchronization is mostly experienced by the particles whose absolute differential velocity to the wave scales as $ |\gamma_{\rm L}|/k_{\bfm}$. When $ |\gamma_{\rm L}|/k_{\bfm}$ is small enough, these particles are quasi-resonant. Consequently, there is a net loss of particles momentum when $g'(\frac{\omega_{\bfm \rmr}}{k_{\bfm}})>0$, and a net gain of momentum in the opposite case, in agreement with equation (\ref{Accav3}). Equation (\ref{Ptot}) shows that this momentum is exchanged with the wave momentum.

If the initial distribution of particles is meant to approximate a smooth distribution $g(v)$, this means that the length $L$ of the system is taken large enough to fulfill the condition ``there are many particles with parallel velocities in the range $(\omega - |\gamma_{{\rm L}}|)/k < v <(\omega + |\gamma_{{\rm L}}|)/k$".
However, whatever $L$, for $\frac{\omega}{k}$ large enough, this condition is unfulfilled. Therefore, one should not use $\epsilon_{\rm{r}}(\bfm,\omega)$, but its discrete counterpart computed with the $N$-body distribution function: Landau damping will not occur, but only the beam modes to be described in section \ref{Pm}.

Several textbooks, when trying to provide an intuitive explanation of Landau damping computed by a Vlasovian approach, make mechanical calculations similar to the above ones, in particular Nicholson's one in section 6.7 of \cite{Nicholson}. Here, the calculation is performed with the $N$-body mechanical model used to compute Landau damping itself. This calculation involves a smooth velocity distribution $g(v)$ at the end of the calculation only, while the Vlasovian approach deals with the dynamics of $g(v)$ from the outset.

The fact that both Landau damping and instability result from the same synchronization mechanism of particles with waves, is the \emph{fundamental reason why there is a single formula for the rates of Landau growth and damping}. While this comes naturally in Kaufman's derivation and in section \ref{LandamK}, growth and damping look very different in the van Kampen-Case \cite{vanKampen,vanKampen2,Case} approach and in the germane one of section \ref{LwKD}, since growth involves an eigenmode and damping only a phase-mixing instead; very different also in Landau's derivation where damping requires an analytic continuation, but not growth.

\section{Fundamental equation for the electrostatic potential}
\label{Feqpot}

This section derives a fundamental equation for the electrostatic potential \cite{DbCb,DbCbcor,DbCb2}, which is reminiscent of the one obtained by the calculation \`a la Landau starting with the Vlasov equation. This equation is introduced in a specific section, because it is the basis for the description of Langmuir waves of the next two sections. The derivation parallels completely the one in textbooks using the Vlasov-Poisson system: the motion is linearized with respect to the ballistic motion of particles of the system with a uniform density, as well as the electrostatic potential, and the Laplace transform in time is applied to the linearized equations.

We consider a given multi-beam-multi-array defined in section \ref{Aud}. Let $\bfr_{j0}$ be the initial position of the unperturbed beam particle with index $j$, and $\bfv_{j}$ be its velocity, and let $\Delta {\bfr}_j (t)= \bfr_j (t) - \bfr_{j0} - \bfv_j t$ be the mismatch of the actual position of particle $j$ with respect to the unperturbed beam particle with the same index. We define the ballistic approximation $\tilde{\varphi}^{(\rm{bal})}(\bfm,t)$
to $\tilde{\varphi}(\bfm,t)$ which is computed from equation (\ref{phitildetotM})
on setting $\bfr_j(t) = \bfr_{j0} + \Delta {\bfr}_j(0)
+ [\bfv_j +\Delta{\dot {\bfr}}_j(0)]t$
for all $j$'s in the latter
\begin{equation}
  \tilde{\varphi}^{(\rm{bal})}(\bfm,t)
  = -\frac{e}{\varepsilon_0 k_{\bfm}^2} \sum_{j = 1}^N
     \exp\Bigl(- \rmi \bfk_{\bfm} \cdot [\bfr_{j0} + \Delta {\bfr}_j(0)
+ (\bfv_j +\Delta{\dot {\bfr}}_j(0)\,) \, t] \Bigr).
\label{phitildetotball}
\end{equation}
If all $\Delta {\bfr}_j(0)$'s and $\Delta{\dot {\bfr}}_j(0)$'s are small, so is $\tilde{\varphi}^{(\rm{bal})}(\bfm,t)$ since $\tilde{\varphi}(\bfm,t)$ vanishes identically for the particles of a multi-beam-multi-array. We define the two mismatches to ballistic values
\begin{eqnarray}
  \delta {\bfr}_j(t)
  & = &
  \Delta {\bfr}_j (t) - \Delta {\bfr}_j(0) - \Delta{\dot {\bfr}}_j(0) t ,
  \label{Delta_xi}
  \\
  \delta \tilde{\varphi} (\bfm,t)
  & = &
  \tilde{\varphi}(\bfm,t) - \tilde{\varphi}^{(\rm{bal})}(\bfm,t) .
  \label{Delta_phi}
\end{eqnarray}

We now compute a linearized solution of the full $N$-body dynamics about the chosen multi-beam-multi-array\footnote{This derivation is close to that in \cite{DbCb}, but takes advantage of the simplification of \cite{DbCbcor}. The derivation of \cite{DbCb2} was convenient only when taking right away the singular limit of section \ref{LwDsL}.}, and the resulting Fourier components of the potential. We assume $\| \Delta {\bfr}_j (t) \| \ll b_{\mathrm{smooth}}$ for all $j$'s. To this end,
we replace $\delta \tilde{\varphi}$ with its expansion to first order in the $\Delta {\bfr}_j (t)$'s
\begin{equation}
\delta \tilde{\varphi} (\bfm,t)
  =
  \rmi \sum_{j = 1}^N
  \frac{ e}{\varepsilon_0 k_{\bfm}^2} \exp [- \rmi \bfk_{\bfm} \cdot (\bfr_{j0} + \bfv_{j} t)] \ \bfk_{\bfm} \cdot \delta {\bfr}_j(t),
\label{phitildnj}
\end{equation}
whose Laplace transform in time\footnote{The Laplace transform in time maps a function $ g(t)$
to $\widehat{g}(\omega) = \int_0^{\infty}  g(t) \exp(\rmi \omega t) \rmd t$
(with $\omega$ complex).} is\footnote{Since the arguments of functions are spelled explicitly, from now on we omit diacritics for the Laplace (or Fourier) transformed quantities.}
\begin{equation}
\delta \varphi (\bfm,\omega)
  =
  \rmi \sum_{j = 1}^N
  \frac{ e}{\varepsilon_0 k_{\bfm}^2} \exp [- \rmi \bfk_{\bfm} \cdot \bfr_{j0} ] \
  \bfk_{\bfm} \cdot \delta {\bfr}_j (\omega - \bfk_{\bfm}  \cdot \bfv_j) ],
\label{Lphitildnj}
\end{equation}
where the Doppler shift $- \bfk_{\bfm} \cdot \bfv_j$ comes from the linear dependence on $t$ of the exponent of equation (\ref{phitildnj}).

To compute $\delta {\bfr}_j (\omega - \bfk_{\bfm}  \cdot \bfv_j) $, we use Newton's equation (\ref{rsectot}) for the particles.
Since $\bfm$ runs over the finite domain such that $k_{\bfm} b_{\mathrm{smooth}} \leq 1$,
the self-field due to $\varphi$ vanishes, and it is not necessary to exclude self-interactions.
Therefore, one may use the harmonics of the electrostatic potential due to all particles defined by equation (\ref{phitildetotM}),
which yields
\begin{equation}
\varphi(\bfr_j,t) = \frac{1}{L^3}\sum_{\bfm, \, k_{\bfm} b_{\mathrm{smooth}} \leq 1} \tilde{\varphi} (\bfm,t) \exp(\rmi \bfk_{\bfm} \cdot \bfr_j) ,
\label{phiInvT}
\end{equation}
where $\tilde{\varphi} (\bfm,t) = \tilde{\varphi}^{(\rm{bal})}(\bfm,t) + \delta \tilde{\varphi} (\bfm,t)$. Using equation (\ref{phitildnj}), the linearized particles dynamics defined by equation (\ref{rsectot}) is then given by
\begin{equation}
\delta \ddot{\bfr}_j
 =
  \frac{\rmi e}{L^3 m} \sum_{\bfn, \, k_{\bfn} b_{\mathrm{smooth}} \leq 1} \bfk_{\bfn} \
    \tilde{\varphi}(\bfn,t) \exp[\rmi \bfk_{\bfn} \cdot (\bfr_{j0} + \bfv_{j} t)].
    \label{delrsec}
\end{equation}
The Laplace transform in time of this equation is
\begin{equation}
 - \omega^2 \delta {\bfr}_j(\omega)
 = \frac{\rmi e}{L^3 m} \sum_{\bfn, \, k_{\bfn} b_{\mathrm{smooth}} \leq 1}
                \bfk_{\bfn} \exp(\rmi \bfk_{\bfn} \cdot \bfr_{j0})
                  \ \varphi(\bfn,\omega + \bfk_{\bfn} \cdot \bfv_{j})
                  \nonumber \\ ,
\label{rLapl}
\end{equation}
where the Doppler shift $\bfk_{\bfn} \cdot \bfv_{j}$
comes from the linear dependence on $t$ in the exponent of equation (\ref{delrsec}), and where we take into account that $\delta {\bfr}_j(0) = \delta \dot{\bfr}_j(0)=0$. Computing $\delta {\bfr}_j (\omega - \bfk  \cdot \bfv) $ in equation (\ref{Lphitildnj}) from the expression of $\delta {\bfr}(\omega)$ given by equation (\ref{rLapl}) yields
\begin{eqnarray}
&&
   k_{\bfm}^2 \varphi(\bfm,\omega)
 - \frac{\omega_{\rmpp}^2}{N}
 \sum_{\bfn, \, k_{\bfn} b_{\mathrm{smooth}} \leq 1} \bfk_{\bfm} \cdot \bfk_{\bfn}
 \nonumber \\
 && \qquad \times
  \ \sum_{j = 1}^N \frac{\varphi(\bfn,\omega + \bfk_{\bfn} \cdot \bfv_{j} - \bfk_{\bfm} \cdot \bfv_{j})}{(\omega - \bfk_{\bfm} \cdot \bfv_{j})^2}
                             \exp[\rmi (\bfk_{\bfn}-\bfk_{\bfm}) \cdot \bfr_{j0}]
  \nonumber \\
& = &
  k_{\bfm}^2 \varphi^{(\rm{bal})}(\bfm,\omega) ,
\label{*phihat}
\end{eqnarray}
with
\begin{equation}
  \varphi^{(\rm{bal})}(\bfm,\omega) =
  \sum_{j = 1}^N \varphi_{j}^{(\rm{bal})}(\bfm,\omega)
  ,
\label{*phi0hat}
\end{equation}
where
\begin{equation}
  \varphi_{j}^{(\rm{bal})}(\bfm,\omega)
  = - \frac{\rmi e}{\varepsilon_0 k_{\bfm}^2}
      \frac{\exp[- \rmi \bfk_{\bfm}  \cdot \bfr_j(0)]}
             {\omega -\bfk_{\bfm}  \cdot \dot \bfr_j(0)} ,
\label{*phij0hat}
\end{equation}
is the ballistic potential related to particle $j$. Note that  assumption $k_{\bfm} b_{\mathrm{smooth}} \leq 1$ (equation (\ref{phiInv})), with $b_{\mathrm{smooth}} \ll \lambda_\rmD$ (equation (\ref{bsmlD})) excludes scales which are irrelevant to Debye shielding and Landau damping,
since these phenomena involve scales larger than, or of the order of $\lambda_\rmD$.

Because of equation (\ref{bsmnedge}), if $\bfn \neq \bfm$, for each beam the corresponding values of $\exp[\rmi (\bfk_{\bfn}-\bfk_{\bfm}) \cdot \bfr_{j0}]$ are uniformly distributed on the unit circle, and their global contribution to the non-diagonal term in equation (\ref{*phihat}) vanishes. Therefore, equation (\ref{*phihat}) becomes
\begin{equation}
 \epsilon_\rmd(\bfm,\omega) \, \varphi(\bfm,\omega)
  = \varphi^{(\rm{bal})}(\bfm,\omega) ,
\label{epsphibaldiscr}
\end{equation}
where
\begin{equation}
  \epsilon_\rmd(\bfm,\omega)
  = 1 - \frac{\omega_{\rmpp}^2}{N}
  \sum_{j = 1}^N  \frac{1}{(\omega - \bfk_{\bfm} \cdot \bfv_{j})^2}.
\label{epsdiscr}
\end{equation}
This is the \emph{fundamental equation for the potential}, which is going to be used for the description of the Langmuir waves \`a la van Kampen-Dawson and \`a la Landau.
Since $b_{\mathrm{smooth}} \ll \lambda_\rmD$, $b_{\mathrm{smooth}}$ is an intermediate scale between $\lid $ and $\lambda_\rmD$, which exists provided $\Lambda$ is large enough (see section \ref{Bs}). Then, equation (\ref{epsphibaldiscr}) describes the response of the plasma to the initial perturbation defined by $\varphi^{(\rm{bal})}(\bfm,\omega)$, and $\epsilon_\rmd(\bfm,\omega)$ is the dielectric function of the granular plasma.

The truncated Coulomb potential cannot correctly describe the encounters between particles with impact parameters smaller than $b_{\mathrm{smooth}}$,
which makes our description of the dynamics relevant for times shorter than the collision time
$\tau_{\rm{coll}} = 3(2 \pi)^{3/2} \Lambda /(\omega_\mathrm{p} \ln \Lambda) $,
as happens for the Vlasovian description.
Both descriptions are relevant for Langmuir waves,
since conditions $\Lambda \gg 1 $ and $\omega_{\rmpp} \tau_{\rm{coll}} \gg 1$ are equivalent.

As a result of equations (\ref{*phi0hat}-\ref{epsdiscr}), the part of $\varphi (\bfm,\omega)$ generated by particle $j$
is
\begin{equation}
\delta \varphi_{j} (\bfm,\omega)
= \varphi_{j}^{(\rm{bal})}(\bfm,\omega)/\epsilon_\rmd(\bfm,\omega).
\label{epsphibaldiscrj}
\end{equation}
By inverse Fourier-Laplace transform, after some transient discussed later,
the potential due to particle $j$ is the sum of two parts~:
one due to the zeros of $\epsilon_\rmd(\bfm,\omega)$ and one to the pole $\omega = \bfk_{\bfm}  \cdot \dot \bfr_j(0)$. These two contributions are computed in the next two sections: ``Landau damping \`a la van Kampen-Dawson" and ``Langmuir waves and Debye shielding \`a la Landau".

\section{Landau damping \`a la van Kampen-Dawson}
\label{LwKD}

This section deals anew with Langmuir waves by keeping the discrete summation in the expression of $\epsilon_\rmd(\bfm,\omega)$. Such a dielectric function was considered by Dawson in 1960 for a one-dimensional plasma made up of many fluid monokinetic beams \cite{Dawson60} whose velocities are successive multiples of a small velocity $\delta$. He showed that  $\epsilon_\rmd(\bfm,\omega)$ brings two beam modes per beam.
Their eigenfrequencies are pairs of complex conjugate values for $\omega$,
whose imaginary parts tend to vanish when $\delta$ decreases~:
this makes these modes analogous to the van Kampen modes.

Here, Landau damping is recovered by phase mixing of these modes by following a procedure similar to that used in section 3.8 of \cite{EEbook}. The spacing of the beam velocities is kept finite, but the limit of a vanishing spacing is used at the end of the calculation to approximate finite sums by integrals. Most of this section deals with $\Phi_{j \, \bfm}(\bfr, t)$, the part of the potential due to particle $j$, provided by the zeros of $\epsilon_\rmd(\bfm,\omega)$ in equation (\ref{epsphibaldiscrj}). A final subsection shows that the part of the potential coming from the pole $\omega = \bfk_{\bfm}  \cdot \dot \bfr_j(0)$ vanishes.

\subsection{Granular dielectric function}
\label{Gdf}

The calculation of the zeros of $\epsilon_\rmd(\bfm,\omega)$ is simpler when the cubic array has an edge parallel to $\bfk_{\bfm}$ and one of its points is at the origin of velocities. In this section, we make this choice, and show that, in this limit, the contribution of these zeros to $\Phi_{j \, \bfm}(\bfr, t)$ corresponds to two Langmuir waves, and that the chosen discretization of the beams is absent in the final expression of $\Phi_{j \, \bfm}(\bfr, t)$. One of these waves propagates in the direction of $\bfk_{ \bfm}$ and the other one in the direction of $\bfk_{- \bfm} = - \bfk_{\bfm}$. In the following, we consider only the first type of wave, and we associate the other one to $\bfk_{- \bfm}$.

Let $\delta$ be the mesh size of the grid, $\hat{1}$ be the unit vector parallel to $\bfk_{\bfm}$, and $\hat{2}$ and $\hat{3}$, be the other two unit vectors. We index a beam by three integers $(\sigma,\tau,\upsilon)$ corresponding to its position on the grid, so that its velocity is $\bfw_{\sigma,\tau,\upsilon} =(\sigma \hat{1}+\tau \hat{2}+\upsilon
 \hat{3}) \delta$. Let $N_{\sigma,\tau,\upsilon}$ be its number of particles. Then equation (\ref{epsdiscr}) becomes
\begin{eqnarray}
\epsilon_{\rmd}(\bfm,\omega)
 &=& 1 - \frac{\omega_{\rmpp}^2}{N}
 \sum_{\sigma,\tau,\upsilon}  \frac{N_{\sigma,\tau,\upsilon}}{(\omega - \sigma k_{\bfm} \delta)^2}
 \nonumber \\
&=& 1 - \frac{\omega_{\rmpp}^2}{N}
 \sum_{\sigma}  \frac{M_{\sigma}}{(\omega - \sigma k_{\bfm} \delta)^2},
\label{epsbeam}
\end{eqnarray}
where $M_{\sigma} = \sum_{\tau,\upsilon} N_{\sigma,\tau,\upsilon}$.

We consider again the beam velocity distribution to be a granular approximation to a spatially uniform smooth velocity distribution whose integral perpendicular to the $\hat{1}$ axis is $g(v)$. Then $M_{\sigma}$ is taken equal to the integer part of $N g(\sigma \delta) \delta$. When the number of particles in the Debye sphere increases, $g(v)$ is split over an increasing number of beams ($\sim \delta^{-1}$) whose velocities lie on grids with mesh size $\delta$ going to 0. The dielectric function of the granular plasma is now expressed with the one-dimensional distribution function $g(v)$ and reads
\begin{equation}
 \epsilon_{\rmd 1}(\bfm,\omega)
= 1 - \omega_{\rmpp}^2
 \sum_{\sigma}  \frac{g(\sigma \delta) \delta}{(\omega - \sigma k_{\bfm} \delta)^2}.
\label{epsbeampul}
\end{equation}

Subsection \ref{Zgdf} shows that at most two zeros of $\epsilon_{\rmd 1}(\bfm,\omega)$ have a finite imaginary part when $\delta$ goes to zero. The other zeros have an imaginary part vanishing in the limit where $\delta$ goes to zero. Dawson introduced in 1960 a very clever technique to compute them, when he considered a one-dimensional plasma made up of many fluid beams \cite{Dawson60}. He decomposed the sum in equation (\ref{epsbeampul}) into a regular part converging to
\begin{eqnarray}
\epsilon_{1}(\bfm,\omega)
 &=& 1 - \omega_{\rmpp}^2
 \int \frac{g(v) }{(\omega - k_{\bfm} v)^2} \ \rmd v
 \nonumber \\
&=& 1 + \frac{\omega_{\rmpp}^2}{k_{\bfm}}
 \int \frac{
 g'(v) }{(\omega - k_{\bfm} v)} \ \rmd v ,
\label{eps}
\end{eqnarray}
and a singular part, which is summed exactly using classical summation formulas for trigonometric functions (see \ref{App:InfNbBeams}). The conjugate zeros of $\epsilon_{\rmd 1}(\bfm,\omega)$ have real parts between the frequencies of the beams $\sigma k_{\bfm} \delta$, and their imaginary part scales like $\delta \, |\ln (\delta/v_{\rmT})\, |$ (see equations (\ref{epsaddsubtsol1})-(\ref{epsaddsubtsol2}) of \ref{App:InfNbBeams}). We write these zeros as $\nu_{\sigma,\mu} = \alpha_\sigma + \mu \rmi \beta_\sigma$, where $\mu = \pm 1$.

\subsection{Phase mixing}
\label{Pm}

We now follow a procedure similar to that used in section 3.8.3 of \cite{EEbook} (see also \cite{EFields}). The zero of $\epsilon_{\rmd 1}(\bfm,\omega)$ with index $(\sigma,\mu)$ brings to $\Phi_{j \, \bfm}(\bfr, t)$ a contribution
\begin{equation}
  \Phi_{j \, \bfm \, \sigma \, \mu}(\bfr, t)
  =
  - \frac{e}{\varepsilon_0 k_{\bfm}^2 L^3 \, \epsilon'_{\sigma \, \mu}}
      \frac{\exp[\rmi (\bfk_{\bfm} \cdot (\bfr - \bfr_j(0)) - \nu_{\sigma,\mu} t)]}
             {\nu_{\sigma,\mu} -\bfk_{\bfm}  \cdot \dot \bfr_j(0)} + \rmcc \, ,
\label{Phijt}
\end{equation}
where
\begin{equation}
\epsilon'_{\sigma,\mu} =
 \frac{\partial \epsilon_{\rmd 1}}{\partial \omega} (\bfm, \nu_{\sigma,\mu})
= 2 \omega_{\rmpp}^2
 \sum_{\sigma}  \frac{g(\sigma k_{\bfm} \delta) \delta}{(\nu_{\sigma,\mu} - \sigma k_{\bfm} \delta)^3}.
\label{epsbeampulpr}
\end{equation}
The sum over $j$ of the $\Phi_{j \, \bfm \, \sigma \, \mu}(\bfr, t)$'s yields
\begin{equation}
\Phi_{\bfm,\sigma,\mu }(\bfr, t)
 = - \frac{ \exp\rmi \bfk_{\bfm} \cdot \bfr}{\varepsilon_0 k_{\bfm}^2 L^3}
e N \int \frac{f(\bfm,\bfv)}
 {\nu_{\sigma,\mu}-\bfk_{\bfm} \cdot \bfv} \
 \rmd^3 \bfv \
 \frac{ \exp(- \rmi \nu_{\sigma,\mu} t)}{\epsilon'_{\sigma,\mu}}
 + \rmcc \, ,
 \label{Phiftbeam}
\end{equation}
where $f(\bfr,\bfv) = N^{-1}\sum_{j} \delta(\bfr - \bfr_j(0)) \  \delta(\bfv -  \dot \bfr_j(0))$, and $f(\bfm,\bfv)$ is its spatial Fourier transform ($\delta(\bullet)$ stands for the Dirac distribution).

Using again a decomposition into a regular part and a singular one, for small $\delta$, the expression of $\epsilon'_{\sigma,\mu}$ is shown in \ref{App:InfNbBeams} to converge toward
\begin{eqnarray}
 \epsilon'_{\sigma,\mu} \simeq - \rmi \mu \frac{2 \pi}{k_{\bfm} \delta} \left(1 + \rmP \! \! \! \int_{- \infty}^{\infty}  \frac{\omega_{\rmpp}^2 g'(v)}{ k_{\bfm}(\alpha_\sigma - k_{\bfm} v)} {\rmd v} - \rmi  \frac{\pi \mu \omega_{\rmpp}^2}{k_{\bfm}^2}g'(\frac{\alpha_\sigma}{k_{\bfm}})\right).
\label{epsbeampulprlim}
\end{eqnarray}
The sum of the first two terms in the bracket of this equation is the real part of $\epsilon_{1}(\bfm,\nu_{\sigma,\mu})$ defined by equation (\ref{epsr}). Since $\epsilon_{1}(\bfm,\omega_{\bfm})= 0$, this implies that the $\nu_{\sigma,\mu}$'s bring a contribution to $\Phi_{\bfm,\sigma,\mu }(\bfr, t)$, which is resonant in the vicinity of $\omega_{\bfm}$. Then, in the summation over $(\sigma,\mu)$, we may take
\begin{equation}
\epsilon'_{\sigma,\mu} \simeq -
 \frac{2 \pi}{k_{\bfm} \delta} \epsilon'_\bfm (\gamma_{\rm L}(\bfk_{\bfm}) + \rmi \mu (\alpha_\sigma - \omega _{{\bf{m}}  {\mathrm{r}}}
)),
\label{epsbeampulprlimapp}
\end{equation}
where $\epsilon'_\bfm =
  \frac{\partial \epsilon_1}{\partial \omega} (\bfm, \omega_{\bfm})$, and where the definition (\ref{gammaL}) of $\gamma_{\rm L}(\bfk_{\bfm})$ was used\footnote{We stress that the derivation of Landau damping in this section is completely independent of that in section \ref{LwaK} though.}. This yields
\begin{equation}
\Phi_{\bfm }(\bfr, t)
 =   \frac{\exp\rmi \bfk_{\bfm} \cdot \bfr}{\varepsilon_0 k_{\bfm}^2 L^3 \epsilon'_\bfm}
\, e N \int
 \frac{f(\bfm,\bfv)}
 {\omega _{{\bf{m}}  {\mathrm{r}}}
 -\bfk_{\bfm} \cdot \bfv} \
 \rmd^3 \bfv \ S
 + \rmcc \, ,
 \label{PhiftTot}
\end{equation}
where
\begin{eqnarray}
S &=&
\frac{k_{\bfm} \delta}{2 \pi} \Sigma_{\sigma,\mu}  \frac{\exp(- \rmi \nu_{\sigma,\mu} t)}{\gamma_{\rm L}(\bfk_{\bfm}) + \rmi \mu (\alpha_\sigma - \omega _{{\bf{m}}  {\mathrm{r}}}
)}
 \nonumber \\
&=& \frac{k_{\bfm} \delta}{ \pi} \Sigma_{\sigma} \exp(- \rmi \alpha_{\sigma} t) \frac{\gamma_{\rm L}(\bfk_{\bfm}) \cosh(\beta_\sigma t)}{\gamma_{\rm L}^2(\bfk_{\bfm}) + (\alpha_\sigma - \omega _{{\bf{m}}  {\mathrm{r}}}
)^2}
\nonumber \\
&\simeq& {{\rm{sgn}}}(\gamma_{\rm L}(\bfk_{\bfm}))\exp[- (|\gamma_{\rm L}(\bfk_{\bfm})| + \rmi \omega _{{\bf{m}}  {\mathrm{r}}}
)t] ,
\label{STot}
\end{eqnarray}
where the last expression makes $\cosh(\beta_\sigma t) = 1$, since $\beta_\sigma$ goes to zero in the continuous limit, and uses a Fourier transform identity in this limit. Equations (\ref{PhiftTot}) and (\ref{STot}) yields a contribution to the wave
\begin{eqnarray}
\Phi_{\bfm}(\bfr, t)
  &=& {\rm{sgn}}(\gamma_{\rm L}(k_{\bfm}))  \frac{ \exp[\rmi (\bfk_{\bfm} \cdot \bfr - \omega_{\bfm} t) - |\gamma_{\rm L}(\bfk_{\bfm})|t]}{\varepsilon_0 k_{\bfm}^2 L^3 \, \epsilon'_\bfm}
\times
\nonumber \\ 
& & \qquad \qquad
   e N \int
    \frac{f(\bfm,\bfv)}
         {\omega_{\bfm} -\bfk_{\bfm} \cdot \bfv} \
     \rmd^3 \bfv
     + \rmcc \, .
     \label{Phift}
\end{eqnarray}
We notice that the initially chosen discretization of the beams is absent in this expression.

\subsection{Vlasovian zero of the granular dielectric function}
\label{Zgdf}

When $\delta$ goes to 0, $\epsilon_{\rmd 1}(\bfm,\omega)$ as defined by equation (\ref{epsbeampul}) converges toward its continuous limit $\epsilon_{1}(\bfm,\omega)$ defined by equation (\ref{eps}), provided that $\rm{Im} \, \omega \neq 0$ or $g'(v)$ vanishes in a finite domain about $\omega/k_{\bfm}$. In (\ref{eps}) one recognizes the Vlasovian expression of the dielectric function.

We first look for the zeros $\omega _{{\bf{m}}  {\mathrm{r}}}
+\rmi \, \omega_{{\bf{m}}  {\mathrm{i}}}$ of $\epsilon_{\rmd 1}(\bfm,\omega)$ having a non vanishing $\omega_{{\bf{m}}  {\mathrm{i}}}$ when $\delta$ goes to zero. Since $|\omega_{{\bf{m}}  {\mathrm{i}}}|$ does not vanish, the convergence of $\epsilon_{\rmd 1}(\bfm,\omega)$ toward $\epsilon_{1}(\bfm,\omega)$ implies that the considered zero of the former converges to a zero of the latter. If the zero of the former has a positive $\omega_{{\bf{m}}  {\mathrm{i}}}$ when $\delta$ goes to zero, its limit value corresponds to a Vlasovian zero with the same property, which requires $g'(\omega _{{\bf{m}}  {\mathrm{r}}}
/k_{\bfm})>0$. Indeed, the corresponding Vlasovian growth rate is Landau's one and is given by equation (\ref{gammaL}). Since $\epsilon_{\rmd 1}(\bfm,\omega)$ has zeros coming in conjugate pairs, there is also a zero with the damping rate $- \gamma_{\rm L}(\bfk_{\bfm})$. This zero will be important to recover the correct Landau instability when $g'(\omega _{{\bf{m}}  {\mathrm{r}}}
/k_{\bfm})>0$. This is surprising for people used to the Vlasovian setting, where only an unstable root is present.

If $g'(\omega _{{\bf{m}}  {\mathrm{r}}}
/k_{\bfm})<0$, $\epsilon_{\rmd 1}(\bfm,\omega)$ cannot have the corresponding Vlasovian root when $\delta$ goes to zero, since its existence would imply that of the complex conjugate zero, which itself would reduce this case to the previous one, implying a contradiction. In the usual Vlasovian approach \cite{LL}, finding this zero results from continuing analytically\footnote{Actually, the analysis of Cauchy integrals, a more advanced topic in mathematics (section VIII.12 of \cite{Gode}), allows wilder $g(v)$'s. There, the analyticity of the integrals with respect to $\omega$, not $v$, in the upper or lower complex half-planes mirrors the use of Laplace transform. The relevant $g(v)$'s are such that their absolute values are integrable, as well as that of their Fourier transforms.} $g'(v)$ outside the real axis, which is nonsense for a derivation involving the sum of Dirac distributions corresponding to the many-beam distribution. Therefore, there is no zero of $\epsilon_{\rmd 1}(\bfm,\omega)$ having a non-vanishing imaginary part when $\delta$ goes to zero. We now show that the missing Vlasovian Landau damping is compensated by the damping resulting from the phase mixing of the many beam modes defined by the zeros of $\epsilon_{\rmd 1}(\bfm,\omega)$.

\subsection{Landau damping}
\label{Ldam}

As just explained, if $\gamma_{\rm L}(\bfk_{\bfm}) <0$, all zeros of $\epsilon_{\rmd 1}(\bfm,\omega)$ have an imaginary part vanishing in the limit where $\delta$ goes to zero. Therefore, the only contribution to the potential of the wave is provided by equation (\ref{Phift}), which is the Vlasovian result for the Langmuir wave with wavevector $\bfk_{\bfm}$ propagating in the direction of the wavevector, except for the granular character of the initial distribution in space and velocity.
$\Phi_{\bfm}(\bfr, t)$ is exponentially damped, because of the phase mixing of the beam modes, which makes these modes analogous to the van Kampen modes \cite{vanKampen,vanKampen2,Case} (see also section 6.14 of \cite{Nicholson}). However, while van Kampen modes and those in section 3.8.3 of \cite{EZE,EEbook} are eigenmodes, the Laplace transformed quantities of the present derivation are not. Finally, we stress that the phase mixing of this derivation is different from that in section \ref{LwaK}. Indeed the latter corresponds to the transients related to the terms complementing those of Landau poles when inverting the Laplace transform in the present derivation.

\subsection{Landau instability}
\label{Li}

For $\gamma_{\rm L}(\bfk_{\bfm}) > 0$, one must add to the expression of $\Phi_{\bfm }(\bfr, t)$ provided by equation (\ref{Phift}) the contributions of the two roots with the finite imaginary parts\footnote{In the limit where $\delta$ vanishes, the absence of these two roots does not modify equation (\ref{Phift}), since they bring an infinitesimal contribution to the result.}  $\pm \gamma_{\rm L}(\bfk_{\bfm})$. This yields the sum of three exponentials
\begin{equation}
 \rme^{- \rmi \omega _{{\bf{m}}  {\mathrm{r}}}
 t } \Bigl(\rme^{\gamma_{\rm L}(\bfk_{\bfm}) t}
          + \rme^{- \gamma_{\rm L}(\bfk_{\bfm}) t} -	\rme^{- \gamma_{\rm L}(\bfk_{\bfm}) t}
\Bigr) = \rme^{ - \rmi \omega _{{\bf{m}}  {\mathrm{r}}}
 t  + \gamma_{\rm L}(\bfk_{\bfm}) t}    ,
      \label{vKLand}
    \end{equation}
where the first two come from the two roots with finite imaginary parts, and the last one comes from equation (\ref{Phift}). Then, the above phase mixing term cancels the contribution of the damped root and one finally obtains again the Vlasovian expression (\ref{Phift})\footnote{Without the phase mixing term, a wave with an initial amplitude 1 would have only half the amplitude $\rme^{\gamma_{\rm L}(\bfk_{\bfm}) t}$ after a long enough time. This remark by Dr A. Samain was at the origin of the development of the van Kampen-like calculation in section 3.8.3 of \cite{EEbook}.}. It is therefore of paramount importance that the single root of $\epsilon_{1}(\bfm,\omega)$ yields two conjugate roots of $\epsilon_{\rmd 1}(\bfm,\omega)$. In section 3.8.3 of \cite{EEbook}, a similar calculation was done by using normal modes, and not the Laplace transform. An equation similar to equation (\ref{vKLand}) was obtained where the third exponential was $\rme^{- \gamma_{\rm L}(\bfk_{\bfm}) |t|}$ instead; this brings the cancellation of the second exponential for positive times, and of the first one for negative ones, because of the time-reversible character of the dynamics.

\begin{figure}
\begin{center}
\subfloat[]{\label{fig:Msigma_delta0p5_v0_4_frac_0p15B}\includegraphics[width=0.5\textwidth]{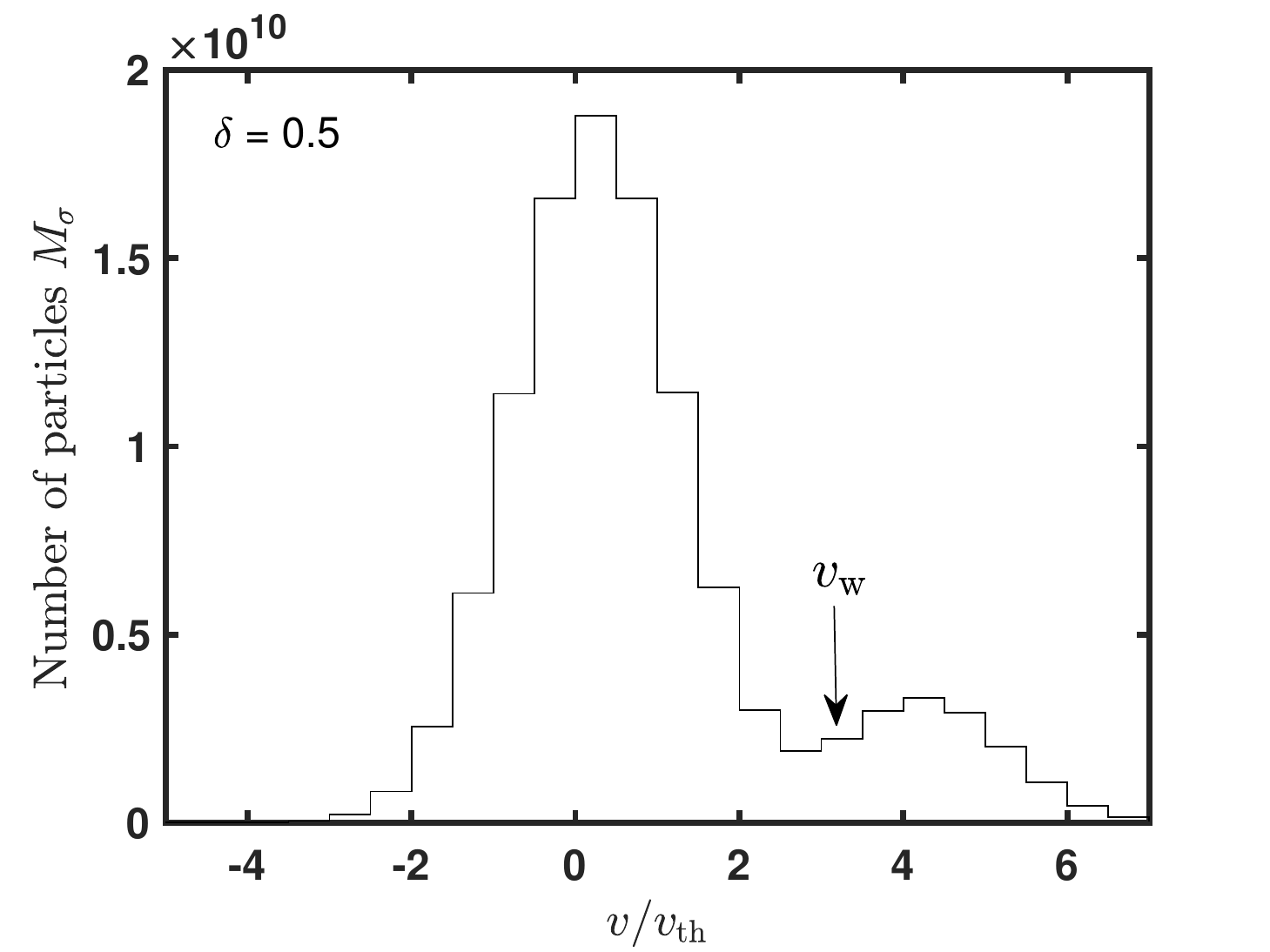}}
\subfloat[]{\label{fig:Zeros_epsilond_delta_0p5_v0_4_frac_0p15B}\includegraphics[width=0.5\textwidth]{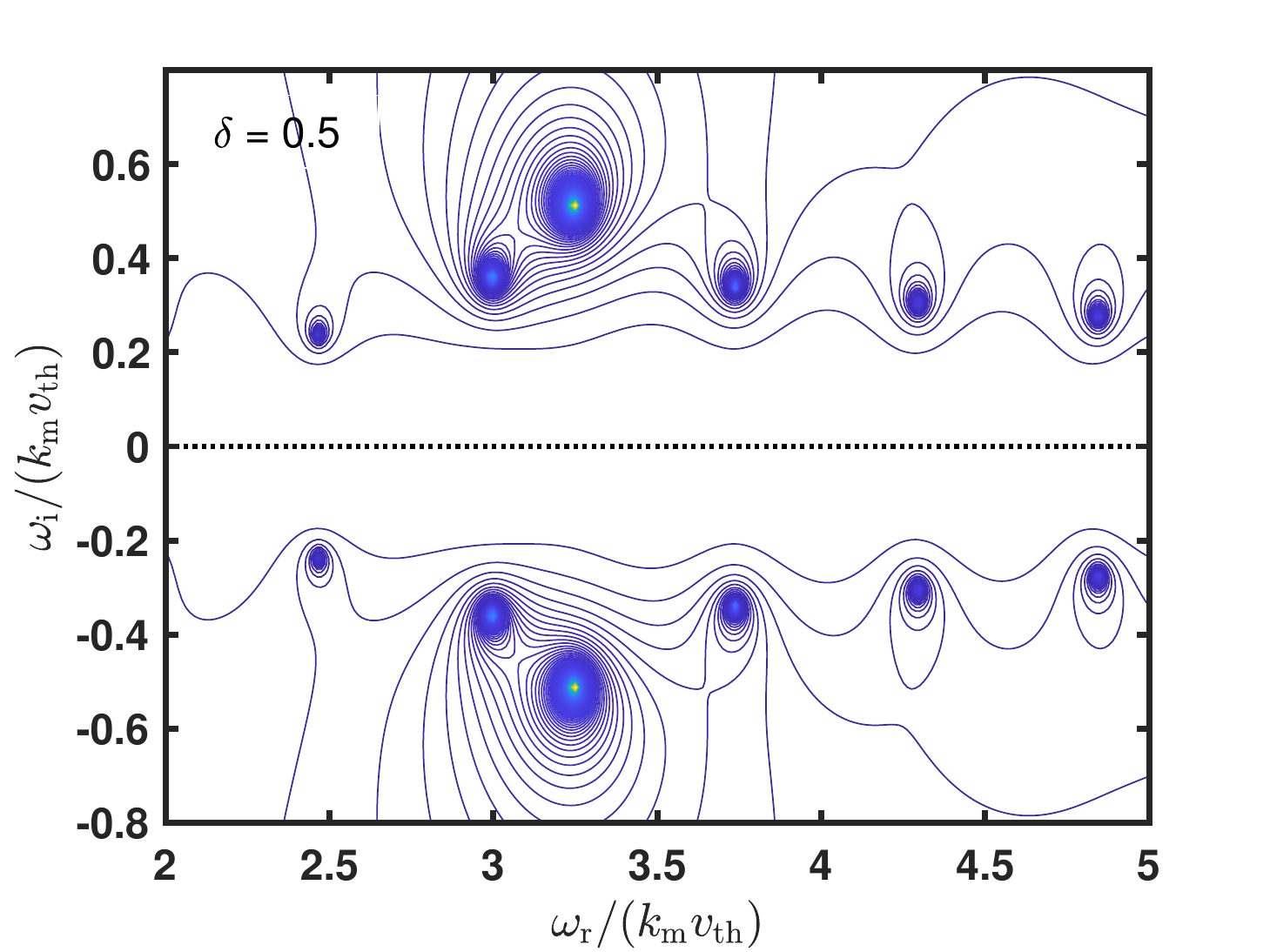}}\\
\subfloat[]{\label{fig:Msigma_delta0p1_v0_4_frac_0p15B}\includegraphics[width=0.5\textwidth]{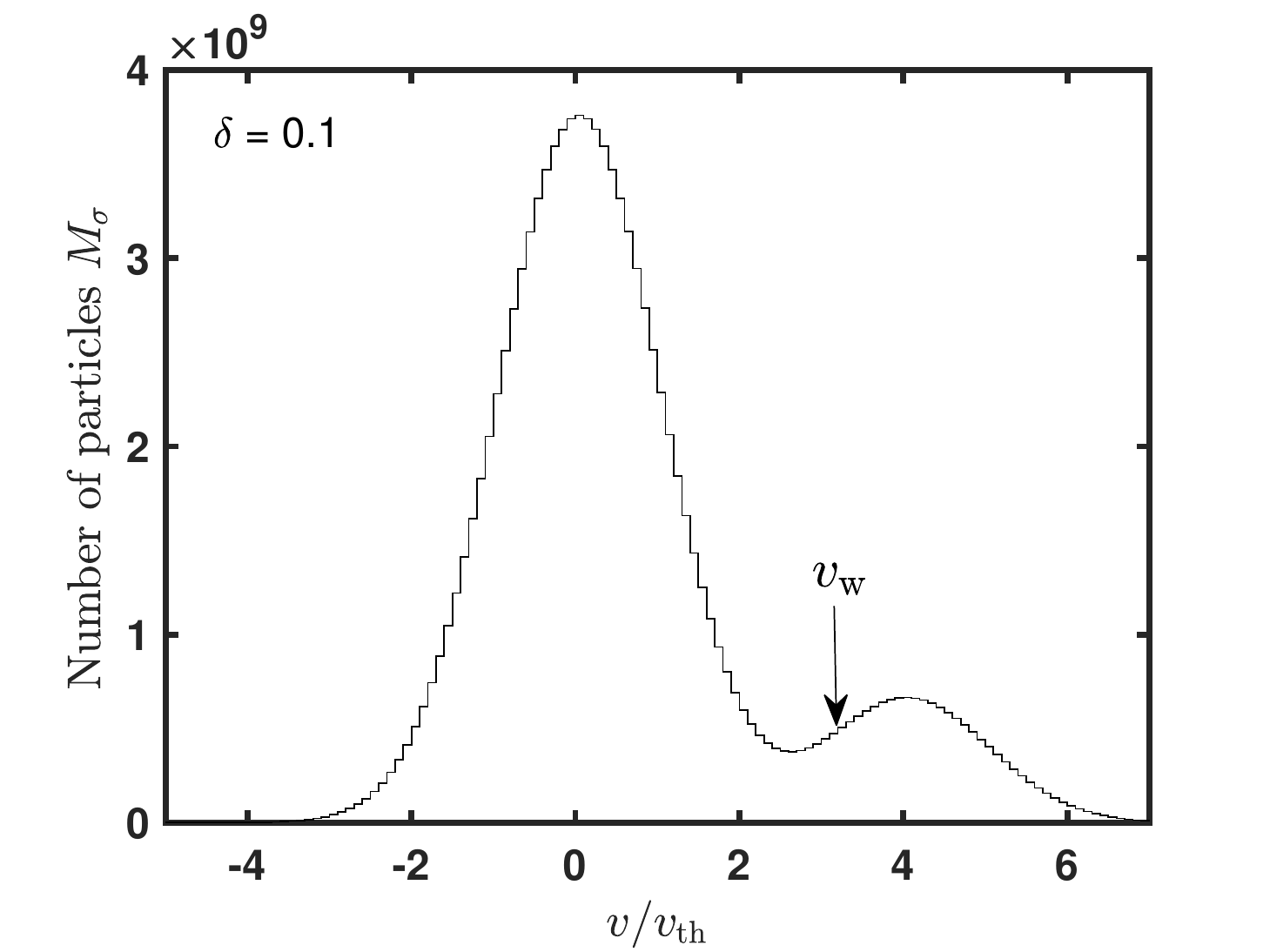}}
\subfloat[]{\label{fig:Zeros_epsilond_delta_0p1_v0_4_frac_0p15B}\includegraphics[width=0.5\textwidth]{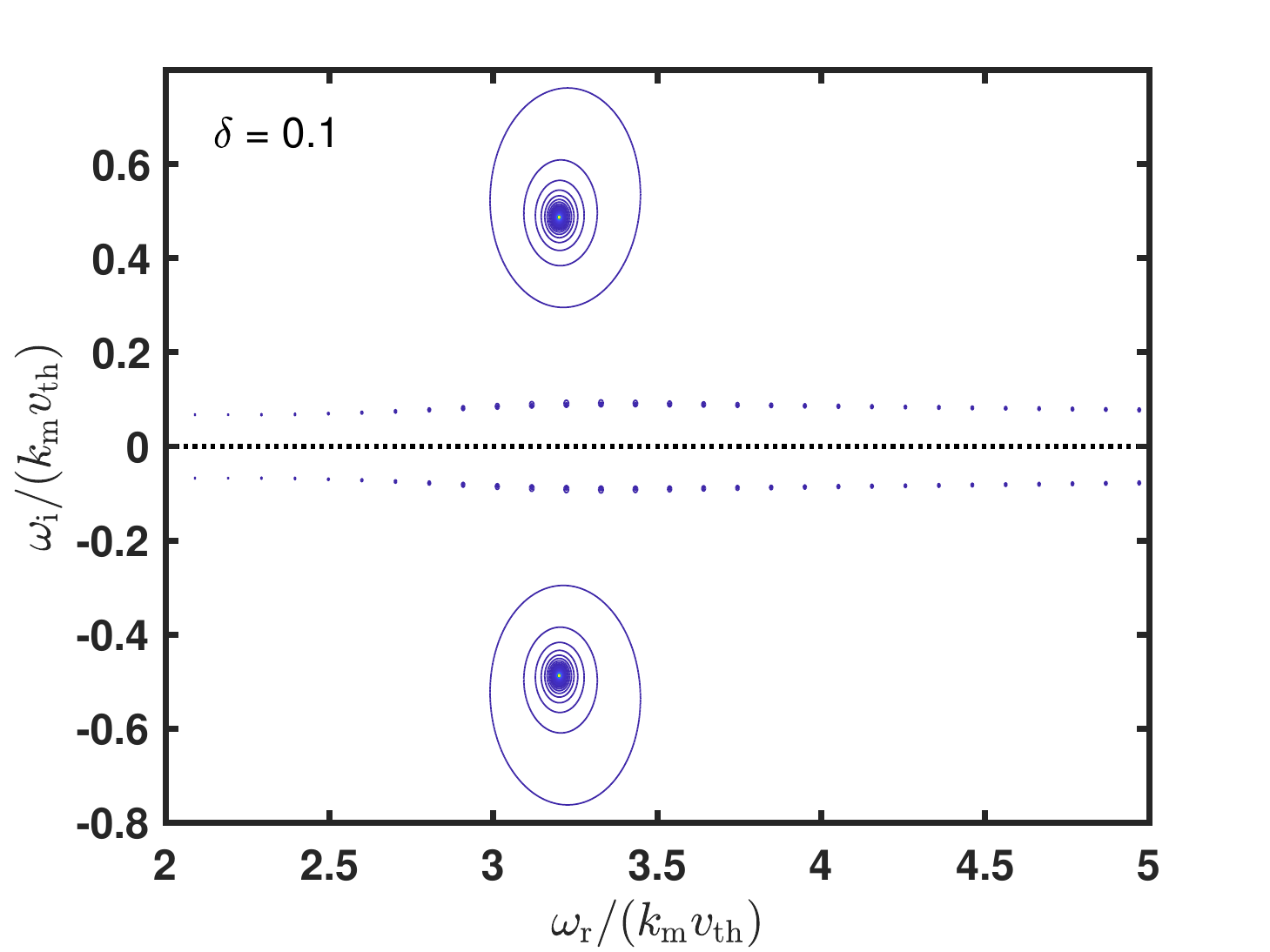}}
\caption{(Left) Distribution function. The position of the phase velocity is indicated by a vertical arrow. (Right) Contour plots of the modulus of the granular dielectric function. The calculations
are done for $\delta = 0.5$ (top panels) and $\delta = 0.1$ (bottom panels). The poles identified
as \textit{Van Kampen poles} get closer to the real
axis as $\delta$ decreases. Only two poles, identified as \textit{Landau poles} and responsible for the Landau instability,
exhibit a finite imaginary part when $\delta\rightarrow 0$.}
\label{fig:Msigma_Zeros_epsilon}
\end{center}
\end{figure}

To illustrate the previous discussion,
we have numerically calculated the zeros of the granular dielectric function (\ref{epsbeam}).
This was done for the following realistic plasma parameters: $n_e=10^{19}\,{\rm m}^{-3}$,
$T = 1$ keV and $L = 30\lambda_D$. We performed the calculation for the
mode number $m=1$. To model the distribution of particles in
velocity space, we considered a thermal population modelled
by a Maxwellian distribution and an energetic tail modelled by
a shifted Maxwellian
\begin{equation}
M_\sigma \sim f_{\rm th}\rme^{-\frac{v_\sigma^2}{2}} + f_{\rm t}\rme^{-\frac{\left(v_\sigma-v_0\right)^2}{2}}
\end{equation}
where, for educational reasons
in order to give a clear example of the Landau instability, we have chosen the
values $f_{\rm t} = 0.15$, $f_{\rm th} = 0.85$ and $v_0 = 4$. Velocities
are normalised to the thermal velocity $v_{\rm th}$ and frequencies in
equation (\ref{epsbeam}) are normalised to $(2\pi/L) v_{\rm th}$.
The velocity space is discretised to form the grid $v_{\rm min}\le v_\sigma=\delta\sigma\le v_{\rm max}$,
where $v_{\rm min} = -5$ and $v_{\rm max} = 7$, both normalised to the thermal velocity,
and $\sigma$ is used to parametrise the distribution of particles as in the previous discussion.
The zeros of the granular dielectric function are plotted in figures
\ref{fig:Zeros_epsilond_delta_0p5_v0_4_frac_0p15B}
and \ref{fig:Zeros_epsilond_delta_0p1_v0_4_frac_0p15B}
for two values
of $\delta$, namely $\delta = 0.5$ and $\delta = 0.1$, respectively\footnote{These two figures represent actually the poles of $1/\vert\epsilon_{\rmd}(\bfm,\omega)\vert$,
which are nothing else but the zeros of $\epsilon_{\rmd}(\bfm,\omega)$.
Identifying the poles of a function of complex variable is
easily done by identifying the closed contours in the complex plane.}.
Figures \ref{fig:Msigma_delta0p5_v0_4_frac_0p15B}
and \ref{fig:Msigma_delta0p1_v0_4_frac_0p15B} represent the distribution of particles
in velocity space, where the inversion of the slope is clearly visible
around $v\approx 3<v_0=4$. It can also be observed that the imaginary part of the zeros
decreases when $\delta$ decreases, and only two zeros emerge from the whole set of
zeros, leading to the Landau instability.

The wave echo experiment \cite{Baker} proved beam modes to be the actual support of Landau damped Langmuir waves excited by grids in a magnetized plasma column. In such a plasma, they would exist in a Landau unstable case too, but the above result shows their phase mixing contribution would be cancelled by a damped eigenmode.

For a wave with phase velocity $v_\rmw$, as in section \ref{Aspw}, the above passage to the smooth velocity distribution $g(v)$ implicitly assumes that there are many particles with parallel velocities in the typical range $[v_\rmw - |\gamma_{{\rm L}}|/k, v_\rmw + |\gamma_{{\rm L}}|/k]$, where $\gamma_{{\rm L}}$ is the Landau growth or damping rate of the wave, and $k$ the modulus of its wavevector. Indeed, this range is the one where the phase mixing \`a la van Kampen is occurring, as shown in section \ref{Pm}, and where the synchronization of particles with the wave brings the change of particle momentum inducing Landau damping or growth of this wave, as shown in section \ref{Aspw}. If the number of particles in the range $[v_\rmw - |\gamma_{{\rm L}}|/k, v_\rmw + |\gamma_{{\rm L}}|/k]$ is not large enough, the system behaves as a multi-beam in this range.

\subsection{Collisions at work}
\label{Caw}

We now comment on the nature of the unstable modes of the above many-beam system. When a cold beam with a high enough density is added to a Maxwellian plasma, it triggers the cold beam-plasma instability whose growth rate is defined by the beam and plasma densities (see for instance \cite{ONeil2}). When the beam becomes warmer and warmer, the beam-plasma instability becomes kinetic and its growth rate is given by Landau's formula \cite{ONeil2}. At some moment, the velocity spreading of the beam becomes so large that the slope of the global velocity distribution of the beam-plasma system becomes negative everywhere: the instability is quenched. Therefore, in the case of a kinetically stable distribution, the unstable modes of the above many-beam system cannot be interpreted as beam-plasma ones. Indeed, their growth rate (\ref{epsaddsubtsol2}) is mainly defined by global features of the plasma like the slope of $g(v)$, and depends weakly on the beam density through a logarithm of $g(n \delta)/\delta$.

Actually, the instability is due to what is traditionally called collisions, and will be described in section \ref{Coltrsp}. Indeed, section \ref{Aud} showed the distribution of the pure many-beam system is invariant in time. Therefore, it is insensitive to collisions. However, a typical perturbation to this distribution evolves because of collisions, and exponentially diverges from the many-beam distribution with the many-beam growth rates. What is traditionally called collisions involves scales up to the Debye length. This explains why the present derivation keeps a ``collisional" contribution, while using a smoothed version of the Coulomb potential. Because of the cutoff at $b_{\mathrm{smooth}}$, the collision frequency\footnote{It was recently noted that the collisional damping rate of Langmuir waves is much smaller than the one provided by the collision frequency \cite{Yoon1,Yoon2}.} involves a Coulomb logarithm $\ln (\lamD / b_{\mathrm{smooth}})$ instead of the usual one $\ln (\lamD / \lma)$.

The whole derivation of section \ref{Pm} heavily relies on the fact that the $\beta_\sigma$'s are larger than $k_{\bfm} \delta$, i.e.\  that collisions are strong enough. Therefore, not only \emph{collisions} make Landau damping irreversible \cite{Callen}, but they \emph{are necessary for the effect to exist}. Furthermore, in the second equality of equation (\ref{STot}), for a finite $\delta$, the $\exp(- \rmi \alpha_{\sigma} t)$'s might produce a recurrence of $S$ after a time $(k_{\bfm} \delta)^{-1}$. However, the $\cosh(\beta_\sigma t)$'s grow on the smaller time-scale $[k_{\bfm} \delta \, |\ln (\delta/v_{\rmT})\,|]^{-1}$, which prevents the recurrence\footnote{This property is the signature of the Floquet exponents mentioned in \ref{fdfe}.}. The last equality of equation (\ref{STot}) makes sense over a time $|\gamma_{\rm L}(\bfk_{\bfm})|^{-1}$ only if condition
\begin{equation}
\bfk_{\bfm} \delta \, |\ln (\delta/v_{\rmT}) \gamma_{\rm L}(\bfk_{\bfm}) \, |^{-1} \ll 1
\label{CondL}
\end{equation}
is fulfilled.

\subsection{Vanishing contribution of the ballistic poles}
\label{Vcbp}

We now consider the part of the potential due to particle $j$ coming from the pole $\omega = \bfk_{\bfm}  \cdot \dot \bfr_j(0)$. It is
\begin{equation}
  \Phi_{j \,\rmd}(\bfr, t)
  = \Phi_{\rmd}(\bfr - \bfr_j(0) - \dot{\bfr}_j(0) t,\dot{\bfr}_j(0)),
\label{phijd}
\end{equation}
where
\begin{equation}
  \Phi_{\rmd} (\bfr,\bfu)
  = - \frac{e}{L^3 \varepsilon_0} \sum_{{\bfm} \neq {\mathbf{0}}}
      \frac{\exp(\rmi \bfk_{\bfm} \cdot \bfr)}
           { k_{\bfm}^2 \, \epsilon_\rmd(\bfm,\bfk_{\bfm} \cdot \bfu + \rmi \varepsilon)}
\label{phid}
\end{equation}
with the usual $\rmi \varepsilon$ prescription resulting from inverting the Laplace transform, as $\epsilon_\rmd(\bfm,\omega)$ diverges for some real $\omega$'s. We again compute the limit of the contribution of component $\bfk_{\bfm}$ in this expression by using the convenient grid of the previous subsection. By taking the limit $\varepsilon \rightarrow 0$ for a fixed $\delta$, the first term in the bracket of equation (\ref{epsaddsubt}) diverges when $\delta$ goes to zero. This implies that $\epsilon_\rmd(\bfm,\bfk_{\bfm} \cdot \bfv + \rmi \varepsilon)$ becomes infinite, which cancels the contribution of the pole $\omega = \bfk_{\bfm}  \cdot \dot \bfr_j(0)$ in equation (\ref{epsphibaldiscrj}). Therefore, apart from transients, $\delta \varphi_{j} (\bfr,t)$ is a superposition of waves only.

\section{Langmuir waves and Debye shielding \`a la Landau}
\label{LwDsL}

The Vlasovian limit recovers the Landau damping/growth predicted by the $N$-body dynamics, by substituting $\epsilon_\rmd(\bfm,\omega)$ in equation (\ref{epsphibaldiscr}) with $\epsilon(\bfm,\omega)$ defined by
\begin{eqnarray}
\epsilon(\bfm,\omega)
  &=& 1 - \omega_{\rmpp}^2
     \int \frac{f_0(\bfv) }{(\omega - \bfk_{\bfm}  \cdot \bfv)^2} \ \rmd^3 \bfv
 \nonumber \\
&=& 1 + \frac{\omega_{\rmpp}^2}{k_{\bfm}^2}
     \int \frac{\bfk_{\bfm}
 \cdot \nabla_\bfv f_0(\bfv) }{(\omega - \bfk_{\bfm}  \cdot \bfv)} \ \rmd^3 \bfv ,
\label{epscont}
\end{eqnarray}
where $f_0(\bfv)$ is the smooth approximation of the velocity distributions of the  multi-beam-multi-arrays defined in section \ref{Aud}. This section considers the solutions of equation (\ref{epsphibaldiscr}) when this substitution is made, which corresponds to a \emph{singular limit of the problem}. It becomes
\begin{equation}
 \epsilon(\bfm,\omega) \, \varphi(\bfm,\omega)
  = \varphi^{(\rm{bal})}(\bfm,\omega).
\label{epsphibal}
\end{equation}
This singular limit was taken in \cite{DbCb,DbCbcor,DbCb2}. Then, $\Phi_{j \, \bfm}(\bfr, t)$, the part of the potential due to particle $j$ obtained by inverse Fourier-Laplace transform, was shown to comprise of two parts: one provided by the zeros of $\epsilon(\bfm,\omega)$ corresponding to the usual Vlasovian expression of Langmuir waves, and one provided by the pole $\omega = \bfk_{\bfm}  \cdot \dot \bfr_j(0)$ corresponding to the shielded Coulomb potential of the particle \cite{Gasio,Bal,Rost} (see for instance section 9.2 of \cite{Nicholson})\footnote{Therefore, taking this singular limit is a way to bring to students a rapid derivation of both Landau damping and Debye shielding.}. For particle $j$, this potential is
\begin{equation}
\Phi_{j \, \bfm} (\bfr, t)
  = \Phi(\bfr - \bfr_j(0) - \dot{\bfr}_j(0) t,\dot{\bfr}_j(0)),
\label{phij}
\end{equation}
where
\begin{equation}
\Phi (\bfr,\bfv)
  = - \frac{e}{L^3 \varepsilon_0} \sum_{{\bfm} \neq {\mathbf{0}}}
      \frac{\exp(\rmi \bfk_{\bfm} \cdot \bfr)}
           { k_{\bfm}^2 \, \epsilon(\bfm,\bfk_{\bfm} \cdot \bfv + \rmi \varepsilon)}
\label{phi}
\end{equation}
with the usual $\rmi \varepsilon$ prescription resulting from inverting the Laplace transform, as
the integral in equation (\ref{eps}) is undefined for the real-valued
$\omega = \bfk_{\bfm} \cdot \bfv$.

Using the smoothed version of Coulomb potential recovers the shielded Coulomb potential of equations (\ref{phij})-(\ref{phi})
down to a distance $\sim b_{\mathrm{smooth}}$ from the particle of interest.
Since $b_{\mathrm{smooth}} \ll \lambda_\rmD$, a part of the $1/r$ dependence of the genuine Coulomb potential is recovered,
and can be matched with the central $1/r$ dependence for closer distances.

The Vlasovian derivation of shielding requires a test particle to be added to the Vlasovian plasma (see for instance section 9.2 of \cite{Nicholson}). Similarly, the $N$-body approach requires $\epsilon_\rmd(\bfm,\omega)$ to be substituted with its Vlasovian analog. In the standard Vlasovian calculation, $\varphi^{(\rm{bal})}(\bfm,\omega) $ is replaced by its continuous limit for ${\rm{Im}} \, \omega$ finite
\begin{equation}
  \Phi^{(\rm{bal})}(\bfm,\omega)
  = - \frac{\rmi e}{\varepsilon_0 k_{\bfm}^2} \int
    \frac{f(\bfm,\bfv)}
         {\omega -\bfk_{\bfm} \cdot \bfv} \
     \rmd^3 \bfv,
\label{phi0hatcg}
\end{equation}
which is the smoothed version of the actual shielded potential in the plasma. Unfortunately, this quantity has no obvious interpretation. Therefore, generally textbooks do not dwell upon it, and just use it as the \emph{term of initial condition} for the calculation of Langmuir waves. The $N$-body description reveals that it is the continuous limit of a granular source term bringing not only the excitation of Langmuir waves, but also the Debye-shielded potential of the particles.

The Debye shielded potential is one of the simplest examples of a renormalized potential \cite{Mccomb}. This potential is a mean-field potential produced by the Coulomb deflections of the particles, as will be explained in section \ref{MIIDS}. \emph{Vlasov equation deals also with a mean-field potential}, as obvious from its mean-field derivation recalled in the introduction. This is also the case for the BBGKY derivation, because of its statistical aspect. Therefore, $\epsilon(\bfm,\omega)$ may be interpreted as a renormalized version of $\epsilon_\rmd(\bfm,\omega)$. When using $\epsilon(\bfm,\omega) = 0$ as the dispersion relation, one Langmuir wave is actually the renormalized version of a set of beam modes of the $N$-body system. As will be shown in section \ref{MIIDS}, similarly the Debye shielded potential of a particle $j$ is its Coulomb potential dressed with all the modifications of the potential of the other particles due to their deflections by $j$. Analytically, it is the renormalized potential provided by equations (\ref{phij}) and (\ref{phi}) by using the renormalized dielectric function $\epsilon(\bfm,\omega)$. Therefore, it is natural to recover both the Vlasovian Langmuir waves and the Debye shielded potentials of the particles, when using $\epsilon(\bfm,\omega) = 0$ as dispersion relation.

Equation (\ref{epsphibal}) can be introduced intuitively by the following simple argument. An almost uniform distribution of particles, written as a perturbation of a multi-beam-multi-array distribution related to a given grid, may be also written as a perturbation for many other nearby multi-beam-multi-array distributions; actually a continuum set. While $\epsilon_\rmd(\bfm,\omega)$ depends on the choice of the grid, the time evolution of the un-linearized dynamics is independent of it, and all the linearized dynamics corresponding to these various grids stay close to it for some time. This almost invariance of the linearized $N$-body  dynamics under the above continuous set of distributions suggests approximating this dynamics by a coarse-grained version of it, where $\epsilon_\rmd(\bfm,\omega)$ is substituted with $\epsilon(\bfm,\omega)$, the Vlasovian expression of the dielectric function, which is present in equation (\ref{epsphibal}). \ref{Aslcg} describes a coarse-graining leading to this equation. As shown in \ref{SCP}, the shielded Coulomb potential can also be recovered by a singular limit of the many-beam description.

\section{The Vlasovian limit: a singular one}
\label{Vlso}

The singularity of the Vlasovian limit is mathematically visible in the dielectric function. Indeed, finding the zeros of $\epsilon_\rmd(\bfm,\omega)$ corresponds to finding those of a polynomial of degree $2 n_\rmb$ with real coefficients ($n_\rmb$ is the number of beams). While when increasing $n_\rmb$, the number of zeros of $\epsilon_\rmd(\bfm,\omega)$ keeps increasing, this number is fixed for $\epsilon(\bfm,\omega)$. Even more singular is the transition from a finite increasing number of poles for $\epsilon_\rmd(\bfm,\omega)$ to a cut for $\epsilon(\bfm,\omega)$\footnote{However, this has no practical consequence, since the time scale over which the system is observed, or measured, introduces a natural granularity in frequency space below which the actual continuum (or singular) limit and the discretized system should not be distinguishable.}.

The results of sections \ref{LwaK} and \ref{LwKD} show that the Vlasovian limit is also singular from a physical point of view. Indeed, the mechanical description of the plasma as a granular system shows that, as to the electrostatic potential, Landau damping is due to a phase mixing, and that Landau growth involves a phase mixing too. As already mentioned, the wave echo experiment \cite{Baker} proved the existence of the beam modes in a genuine (granular) plasma. In contrast, in the usual Vlasovian setting, the derivation of Landau damping requires an analytic continuation, which is, though powerful, far from physically intuitive. While section \ref{Aspw} shows that Landau growth and damping are due to the same average synchronization of particles nearly-resonant with the wave, the fate of individual particles is usually eluded in a Vlasovian setting\footnote{In contrast, the global exchange of energy and momentum between waves and particles is easily available in a Vlasovian setting.}. As recalled in the introduction of this paper, this lack of physical intuition made it hard for the plasma community to accept the reality of this damping, and is the origin of still discordant physical interpretations of the effect, some invoking trapping or surfing of the particles, which is wrong mechanically.

Adding a test particle to the $N$-body system, induces a mere modification of the calculations of section \ref{Feqpot}: one more term is added to the ballistic potential (\ref{phitildetotball}), but the remaining of section \ref{Feqpot} stays the same. Therefore, equation (\ref{epsphibaldiscr}) stays the same except for the contribution of the test particle in the ballistic term. The calculation of section \ref{Vcbp} also applies to this contribution, which does not provide the shielded potential of this test particle. This is in contrast with the Vlasovian case, another symptom of the singularity of the Vlasovian limit.

The reason, why the derivation of Landau damping in a Vlasovian setting does not correspond to an eigenmode, is deeply rooted in mechanics. Indeed, imagine Landau damping would correspond to a zero of $\epsilon_\rmd(\bfm,\omega)$: since the above polynomial has real coefficients, its complex solutions come in conjugate pairs, and the damped eigenmode would come with a growing companion, which would dominate for a typical initial condition of the system. This contradicts observation. Therefore, Landau damping cannot correspond to a zero of $\epsilon_\rmd(\bfm,\omega)$, and section \ref{Pm} showed that, for the genuine granular system, Landau damping occurs by phase mixing of beam modes.

As a matter of fact, this reductio ad absurdum can be done without any calculation. Indeed, since the $N$-body system is a Hamiltonian one, its Lyapunov exponents come in pairs: consistently with the preservation of phase space areas, a damped eigenmode comes with a growing companion. Therefore, while Landau damping corresponds to a zero of $\epsilon(\bfm,\omega)$, it cannot for $\epsilon_\rmd(\bfm,\omega)$. This issue does not exist for the unstable case.

Furthermore, while introducing the van Kampen modes was a creative contribution in the Vlasovian case, they are spontaneously present in the $N$-body approach, because the multi-beam-multi-array distribution is a natural way to deal with a ``uniform" plasma. The van Kampen modes are beam modes recovered with $\epsilon(\bfm,\omega)$, which is a renormalized dielectric function, as was shown in section \ref{LwDsL}. As was explained in section \ref{Caw}, the instability of the beam modes is the signature of Coulomb collisions. If a collision operator is added in Vlasov equation, the continuous spectrum of the Van Kampen-Case modes is eliminated and replaced by a discrete spectrum, even in the limit of zero collision, and these discrete eigenmodes form a complete set of solutions \cite{Ng1,Ng2}. This is similar to the $N$-body case. However, the Landau-damped solutions are recovered as true eigenmodes, which they are neither in the collisionless Vlasovian theory, nor in the $N$-body approach. Finally, we notice that in the $N$-body approach, Landau damping occurs without requiring $g(v)$ to be analytically continuable like in the usual Vlasovian approach, which is better for the experimental observability of the phenomenon.

\section{Mediated interactions imply Debye shielding}
\label{MIIDS}

In this section\footnote{For completeness, most of this section follows closely section 5 of \cite{DbCb}.}, Picard's iteration technique\footnote{Picard's iteration technique is one of the standard methods to prove the existence and uniqueness of solutions to first-order equations with given initial conditions. It uses the fact that the exact solution of equation $\frac{\rmd \bfX}{\rmd t} = f(\bfX)$ is the fixed point of the iterative process starting from $n = 0$, and providing $\bfX_{n+1}$ from $\bfX_{n}$ by equation $\frac{\rmd \bfX_{n+1}}{\rmd t} = f(\bfX_n)$ with any choice of $\bfX_{0}(t)$. This iteration technique is very convenient, in particular to alleviate the algebra of perturbation calculations.}
is applied to the \emph{full equation of motion of a particle $j$ due to the Coulomb forces of all other ones} for any $\Lambda$, i.e.\ by using equation (\ref{phiInv}) without the restriction $k_{\bfm} b_{\mathrm{smooth}} \leq 1$. It stresses that, for $\Lambda \gg 1$, a part of the effect on particle $j$ of another particle $j'$
is mediated by all other particles (equation~(\ref{rsecnAccDev2})) and reduces the direct part.
Indeed, particle $j'$ modifies the motion of all other particles,
implying that the action of the latter ones on particle $j$ is affected by particle $j'$. This will be shown to provide an intuitive picture of Debye shielding.

We consider the dynamics in real time of the particle with index $l$ defined by equation (\ref{rsectot}) with the full OCP Coulomb potential of equation (\ref{phiInv}). To this end we use Picard's iteration technique.
From equation (\ref{rsectot}), $\bfr_l^{(n)}$, the $n$-th iterate for $\bfr_l$, is computed from
\begin{equation}
  \ddot{\bfr}^{(n)}_l
  = \frac{e}{m} \nabla \varphi^{(n-1)}(\bfr^{(n-1)}_l),
\label{rsecn}
\end{equation}
where $\varphi^{(n-1)}$ is obtained by the inverse Fourier transform of equation (\ref{phitildetotM}) with the $\bfr_j$'s substituted with the $\bfr_j^{(n-1)}$'s. The iteration starts with the ballistic approximation of the dynamics $\bfr_m^{(0)} = \bfr_{m0} + \bfv_{m} t$, where $\bfr_{m0}$ and $\bfv_{m}$ are respectively the initial position and velocity of particle $m$. The actual orbit of equation (\ref{rsectot}) corresponds to $n \rightarrow \infty$. Let $\bm{\xi}_m^{(n)} = \bfr_m^{(n)} - \bfr_m^{(0)}$ be the mismatch of the position of particle $m$ with respect to the ballistic one at the $n$-th iterate, viz.\ the effect of Coulomb interactions to that order of iterations~; because of the initial conditions, $\bm{\xi}_m^{(0)}$'s and $\dot{\bm{\xi}}^{(0)}_m$'s vanish identically for all $m$'s. It is convenient to write equation (\ref{rsecn}) as
\begin{equation}
 \ddot{\bm{\xi}}^{(n)}_l
  = \sum_{j \in S;j \neq l} \ddot{\bm{\xi}}^{(n)}_{lj},
\label{rsecnAccl}
\end{equation}
with $S$ denoting the set of integers from 1 to $N$ labeling particles, and
\begin{equation}
  \ddot{\bm{\xi}}^{(n)}_{lj}
  = \bfa_{\rm{C}}(\bfr_l^{(n-1)}-\bfr_j^{(n-1)})
\label{rsecnAcclj}
\end{equation}
 and
\begin{equation}
  \bfa_{\rm{C}}(\bfr)
  =
  \frac{\rmi e^2}{\varepsilon_0 m L^3}
  \sum_{{\bfm} \neq {\mathbf{0}}}
      k_{\bfm}^{-2} \, \bfk_{\bfm}
      \exp(\rmi \bfk_{\bfm} \cdot \bfr).
\label{phiCbPer}
\end{equation}
Let $\bm{\xi}^{(n)}_{lj}
    = \int_0^t \int_0^{t'} \ddot {\bm{\xi}}^{(n)}_{lj} (t'') \, \rmd t'' \rmd t' = \int_0^t (t-t'') \, \ddot {\bm{\xi}}^{(n)}_{lj} (t'') \, \rmd t''$.
For $n \geq 2$, one finds
\begin{equation}
  \ddot{\bm{\xi}}^{(n)}_l
  = \sum_{j \in S;j \neq l}
       [\ddot{\bm{\xi}}^{(1)}_{lj} + \bfM_{lj}^{(n-1)}
        + 2 \nabla \bfa_{\rm{C}}(\bfr_l^{(0)}-\bfr_j^{(0)}) \cdot \bm{\xi}_{lj}^{(n-1)}] + O(a^3),
\label{rsecnAccDev2}
\end{equation}
where $a$ is the order of magnitude of the total Coulombian acceleration, and
\begin{eqnarray}
\bfM_{lj}^{(n-1)}
  & = & \nabla \bfa_{\rm{C}}(\bfr_l^{(0)}-\bfr_j^{(0)})
      \cdot [\bm{\xi}_{l}^{(n-1)} - \bm{\xi}_{j}^{(n-1)} - 2 \bm{\xi}_{lj}^{(n-1)}]
  \\
  & = & \nabla \bfa_{\rm{C}}(\bfr_l^{(0)}-\bfr_j^{(0)})
      \cdot \sum_{i \in S; i \neq l,j} (\bm{\xi}_{li}^{(n-1)} + \bm{\xi}_{ij}^{(n-1)})   ,
\label{Mlj}
\end{eqnarray}
where the second expression takes into account that $\bfa_{\rm{C}}(\bfr)$ is anti-symmetrical in $\bfr$.
The latter expression displays $\bm{\xi}_{ij}^{(n-1)}$ which is the deflection of particle $i$ by particle $j$.
It shows how the bare Coulomb acceleration of particle $l$ due to particle $j$
is modified by the following process~:
particle $j$ modifies the motion of all other particles,
so that the action of the latter ones on particle $l$ is modified by particle $j$.
Therefore $\bfM_{lj}^{(n-1)} $ \emph{is the acceleration of particle $l$ due to particle $j$ mediated by all other particles}.
The last term in the bracket in equation (\ref{rsecnAccDev2}) accounts for the fact
that both particles $j$ and $l$ are shifted with respect to their ballistic positions.
Both $\bfM_{lj}^{(n-1)} $ and this last term are anti-symmetrical with respect to the labels $j$ and $l$, since $\nabla \bfa_{\rm{C}}(\bfr)$ is an even function of $\bfr$.

The full Coulombian dynamics of the plasma includes the relaxation to a thermal state. If the corresponding temperature is low enough, the particles have a vanishing mean velocity. In order to describe such an equilibrium, it is advisable to take $\bfv_{m}=0$ for all $m$'s for a faster convergence of the iterative process toward the equilibrium solution.

Since the shielded potential of section \ref{LwDsL} was found by first order perturbation theory,
it is felt in the acceleration of particles computed to second order.
This acceleration is provided by equation (\ref{rsecnAccDev2}) for $n=2$.
Therefore, its term in brackets is the shielded acceleration of particle $l$ due to particle $j$.
As a result, though the summation runs over all particles, its effective part is only due
to particles $j$ typically inside the Debye sphere (with radius $\lambda_\rmD$) about particle $l$.
Starting from the third iterate of the Picard scheme, the effective part of the summation in equation (\ref{rsecnAccDev2})
ranges inside this Debye sphere, since the $\bm{\xi}_{lj}^{(n-1)}$'s are then computed with a shielded acceleration. In equation (\ref{Mlj}), the compound effect of the $\bm{\xi}_{ij}^{(n-1)}$'s,
the deflections of particle $i$ by particle $j$, is to diminish the negative charge inside a sphere centered on particle $j$.

This calculation yields the following interpretation of shielding.
At $t = 0$, consider a set of uniformly distributed particles, and especially particle $j$.
At a later time $t$, the latter has deflected all particles which made a closest approach to it
with a typical impact parameter
$b \lesssim  v_{\rmT} t$, where $v_{\rmT}$ is the thermal velocity.
This \emph{part of their global deflection} due to particle $j$ reduces the number of particles
inside the sphere $S_j(t)$ of radius $v_{\rmT} t$ about it.
Therefore, according to Gauss' theorem, the effective charge of particle $j$ as seen out of $S_j(t)$ is reduced~:
the charge of particle $j$ is shielded due to these deflections.
This shielding effect increases with $t$, and thus with the distance to particle $j$.
It becomes complete at a distance on the order of $\lambda_\rmD$.
Since the global deflection of particles includes the contributions of many other ones, the density of the electrons does not change,
at variance with the shielding at work next to a probe (see e.g.\ section 2.2.1 of \cite{APiel}). This interpretation explains how \emph{a given particle can be shielded by all other particles, while contributing to their individual shieldings}. The Debye shielding of a test particle can be computed by using explicitly the just described Coulombian deflections \cite{Meyer1993}.

When starting from random particle positions, the typical time-scale for shielding to set in is the time for a thermal particle to cross a Debye sphere, i.e.\ $\omega_{\rmpp}^{-1}$,  where $\omega_{\rmpp}$ is the plasma frequency.
Furthermore, shielding, though very fast a process, is a cooperative dynamical one, not a collective (viz.\ coherent) one~:
it results from the accumulation of almost independent repulsive deflections with the same qualitative impact on the effective electric field of particle $j$ (if point-like ions were present, the attractive deflection of charges with opposite signs would have the same effect).
So, shielding and Coulomb scattering are two aspects of the same two-body repulsive process. For $\Lambda \gg 1$, this transport is negligible on fairly long time scales. However, collisions are of paramount importance to provide shielding over the plasma period $\omega_{\rmpp}^{-1}$. We now understand that Debye shielding cannot work for a multi-beam-multi-array, because it does not experience collisions, as anticipated in section \ref{Aud}. Finally, in contrast to what occurs if electron $j$ is substituted with a Langmuir probe, this does not change the density of charges of the plasma, because the deflections due to $j$ are compensated by those of the other electrons.

This approach clarifies the mechanical background of the calculation of shielding using the equilibrium pair correlation function,
which shows shielding to result from the correlation of two particles occurring through the action of all the other ones
(see e.g.\ section 12.3 of \cite{BoydSan}).
As discussed in section \ref{LwDsL}, the derivation of Debye shielding using Vlasov equation plus a test particle takes advantage of the renormalization involved in the Vlasovian dielectric function. However, rigorously speaking, the derivation of Debye shielding using fluid equations is not justified in a collisionless plasma, since these equations are not justified, especially in a non magnetized plasma (see chapter 3 of \cite{HazelW}).

\section{Wave-particle interaction: linear and nonlinear effects}
\label{Wpilnle}

This section introduces a wave-particle description of Langmuir waves. This is done by splitting the set of particles into a bulk, which cannot resonate with Langmuir waves, and a tail. Then, amplitude equations are derived for the Fourier components of the potential where tail particles provide a source term. These equations, together with the equations of motion of the tail particles, provide a Hamiltonian description of the dynamics of tail particles coupled with Langmuir waves. This description is well fitted to statistical descriptions unifying Landau damping and spontaneous emission, or showing that the transition from Landau damping to O'Neil's damping with trapping is a second order phase transition. This description is also amenable for studying the nonlinear saturation of various regimes of the beam-plasma instability, and to prove the nonexistence of Bernstein-Greene-Kruskal (BGK) modes corresponding to traveling-wave solutions. Finally, this description enables the use of modern tools of nonlinear dynamics and chaos available for finite dimensional systems.

This section first derives the Hamiltonian ruling the self-consistent dynamics of tail particles with Langmuir waves. On this basis, a statistical analysis shows Langmuir waves evolve under the compound action of Landau damping and of spontaneous emission. Then, these two effects are shown to have their counterpart in the diffusion and friction coefficients of the quasilinear Fokker-Planck equation ruling the dynamics of particles in a broad spectrum of waves. This sets the ground to face the controversial issue of the saturation of the weak warm beam instability. The remaining of the section deals with the single wave Hamiltonian both in the unstable and the damped case.

\subsection{Self-consistent Hamiltonian}
\label{AmplitudeEqn}

Till now, we described Langmuir waves by a fully linear theory.
Following the analysis of section 6 of \cite{DbCb}, we now generalize the analysis of section \ref{Feqpot} to afford the description of both linear and nonlinear effects in wave-particle dynamics.
Indeed, resonant particles may experience trapping or chaotic dynamics,
which imply $\bfk_{\bfm} \cdot \Delta {\bfr}_j$'s of the order of $2 \pi$ or larger for wave $\bfk_{\bfm}$'s. For such particles,
it is not appropriate to make the linearizations leading to equations (\ref{phitildnj}) and (\ref{delrsec}).
However, these linearizations may still be justified for non-resonant particles over times
where trapping and chaos show up for resonant ones.
In order to keep the capability to describe the latter effects, we now split the set of $N$ particles into bulk and tail,
in the spirit of references \cite{OWM,OLMSS,AEE,EZE,EEbook}.
The bulk is defined as the set of particles which are not resonant with Langmuir waves.
We then perform the analysis of section \ref{Feqpot} for these $N_{\mathrm{bulk}}$ particles,
while keeping the exact contribution of the remaining $N_{\mathrm{tail}}$ particles to the electrostatic potential.
To this end, we number the tail particles from 1 to $N_{\mathrm{tail}}$,
the bulk ones from $N_{\mathrm{tail}}+1$ to $N = N_{\mathrm{bulk}} + N_{\mathrm{tail}}$,
and we call these respective sets of integers $S_{\mathrm{tail}}$ and $S_{\mathrm{bulk}}$.
Then, equation (\ref{*phihat}) is substituted with
\begin{equation}
 \epsilon_\rmd(\bfm,\omega) \, \varphi(\bfm,\omega)
  = \varphi^{(\rm{bal})}(\bfm,\omega) + U(\bfm,\omega) ,
\label{*phihattail}
\end{equation}
where $\epsilon_\rmd(\bfm,\omega)$ is defined by equation (\ref{epsdiscr}), and $\varphi^{(\rm{bal})}_{\mathrm{bulk}}(\bfm,\omega)$ is defined by equation (\ref{*phi0hat}), where the sums over $j$ runs over $S_{\mathrm{bulk}}$ only, and where
\begin{equation}
U(\bfm, t)
= -\frac{e}{\varepsilon_0 k_{\bfm}^2}
   \sum_{j \in S_{\mathrm{tail}}} \exp(- \rmi \bfk_{\bfm} \cdot \bfr_j(t)),
\label{U}
\end{equation}
for $N_{\mathrm{bulk}} \gg 1$.

Since Langmuir waves are not resonant with the bulk, the singularities of $\epsilon_\rmd(\bfm,\omega)$ do not show up, and in the limit of large numbers of particles in the Debye sphere, this quantity may be approximated by $\epsilon(\bfm,\omega)$, or more precisely by $\epsilon_{\mathrm{bulk}}(\bfm,\omega)$ defined by
\begin{eqnarray}
\epsilon_{\mathrm{bulk}}(\bfm,\omega)
  &=& 1 - \omega_{\rmpp}^2
     \int \frac{f_0(\bfv) }{(\omega - \bfk_{\bfm}  \cdot \bfv)^2} \ \rmd^3 \bfv
 \nonumber \\
&=& 1 + \frac{\omega_{\rmpp}^2}{k_{\bfm}^2}
     \int \frac{\bfk_{\bfm}
 \cdot \nabla_\bfv f_0(\bfv) }{(\omega - \bfk_{\bfm}  \cdot \bfv)} \ \rmd^3 \bfv ,
\label{epscontZ}
\end{eqnarray}
where $f_0(\bfv)$ is the bulk distribution only, and $\omega_{\rmpp}$ is computed with the bulk particles only.
Then, equation (\ref{*phihattail}) becomes
\begin{equation}
 \epsilon_{\mathrm{bulk}}(\bfm,\omega) \, \varphi(\bfm,\omega)
  = \varphi^{(\rm{bal})}_{\mathrm{bulk}}(\bfm,\omega) + U(\bfm,\omega) .
\label{epsphibaltail}
\end{equation}

For the scales much larger than $\lambda_\rmD$, the electric potential for the bulk is a superposition of Langmuir waves. The presence of tail particles slightly modifies these waves. Therefore, as shown in section 6.1 of \cite{DbCb}, one can derive an amplitude equation for the potential $\varphi(\bfm, t)$ of the wave with wavevector $\bfk_{\bfm}$ in a way similar to references \cite{OWM,OLMSS}
\begin{equation}
\frac {\rmd \varphi(\bfm, t)} {\rmd t}  + \rmi \omega_{\bfm} \varphi(\bfm, t)
  =
  \frac{\rmi e}
         {\varepsilon_0 k_{\bfm}^2
                      \frac{\partial \epsilon_{\mathrm{bulk}}}{\partial \omega}(\bfm,\omega_{\bfm})
         }
  \sum_{j \in S_{\mathrm{tail}}} \exp(- \rmi \bfk_{\bfm} \cdot \bfr_j),
\label{eqampl}
\end{equation}
where $\omega_{\bfm}$ is the eigenfrequency solving $\epsilon_{\mathrm{bulk}} (\bfm,\omega_{\bfm}) = 0$ corresponding to the wave propagating in the direction of $\bfk_\bfm$~; this frequency is real, since it is not resonant with the support of the bulk distribution function (indeed, we assumed $f_0 (\bfv) = 0$ for all $\bfv$'s such that $\bfk_\bfm  \cdot \bfv = \omega_{\bfm}$). We notice that $\epsilon_{\mathrm{bulk}} (\bfm,\omega_{\bfm}) = 0$ is a Bohm-Gross type dispersion relation associated with plasma oscillations of the bulk.

The self-consistent dynamics of $M$ Langmuir waves and of the tail particles is ruled by equation (\ref{eqampl}) written for each wave
and by the equation of motion of these particles due to the $M$ waves,
\begin{equation}
  \ddot{\bfr}_j
  = \frac{\rmi e}{L^3 m} \sum_{\bfn \in \mathcal{M}}
         \bfk_{\bfn} \varphi(\bfn,t)
            \exp(\rmi \bfk_{\bfn} \cdot \bfr_j) ,
\label{delrsecwv}
\end{equation}
where $\mathcal{M}$ is the set of the indices of the $M$ waves,
and the tail-tail interactions were neglected owing to the low density of the tail particles. These two sets of equations generalize to three dimensions the self-consistent dynamics defined in references \cite{MK,AEE,EEbook} for the one-dimensional $N$-body dynamics.

We now write $\varphi(\bfn,t)$ as $a_{\bfn} \sqrt{2 I_{\bfn}} \exp (- \rmi \theta_{\bfn})$, where $I_{\bfn}$ and $\theta_{\bfn}$ are real, and where
\begin{equation}
a_{\bfn} = \Bigl[
\frac{L^3}
         {2 \varepsilon_0 k_{\bfn}^2
                      \frac{\partial \epsilon_{\mathrm{bulk}}}{\partial \omega}(\bfn,\omega_{\bfn})}\Bigr]^{1/2}.
\label{an}
\end{equation}
Then equations (\ref{eqampl})-(\ref{delrsecwv}) can be cast as the canonical equations related to the self-consistent Hamiltonian
\begin{eqnarray}
  H_{\rm sc}
   =& & \sum_{j=1}^{N_{\mathrm{tail}}} {\frac {\bfp_j^2} {2m}}
        + \sum_{\bfn \in \mathcal{M}} \omega_{\bfn} I_{\bfn}
        \nonumber\\
        &-& \varepsilon \sum_{j=1}^{N_{\mathrm{tail}}} \sum_{\bfn \in \mathcal{M}}
           k_{\bfn}^{-1} \kappa_{\bfn} \sqrt{2 I_{\bfn}} \cos (\bfk_{\bfn} \cdot \bfr_j- \theta_{\bfn})
\label{eqHItheta3D}
\end{eqnarray}
where $\kappa_{\bfn} = [\partial \epsilon_{\mathrm{bulk}} (\bfn,\omega_{\bfn}) / \partial \omega]^{-1/2}$, and $\varepsilon = \omega_{\rm p} [2 m / N]^{1/2}$ is the coupling parameter ruling the intensity of the wave-particle interaction. The conjugate variables for $H_{\rm sc}$ are $(\bfp_j,\bfr_j)$ for the particles and $(I_{\bfn},\theta_{\bfn})$ for the waves. $H_{\rm sc}$ is the sum of the kinetic energy of particles, of the energy of waves (harmonic oscillators described in action-angle variables), and of a coupling term. It may be useful to write $H_{\rm sc}$ by using the ``Cartesian" coordinates of the harmonic oscillators instead of their intensity-phase (or action-angle) components. To this end we write $\varphi(\bfn,t)$ as $a_{\bfn}(X_{\bfn} + \rmi Y_{\bfn})$, where $X_{\bfn}$ and $Y_{\bfn}$ are real. This yields
\begin{eqnarray}
  H_{\rm sc}
   =& & \sum_{j=1}^{N_{\mathrm{tail}}} {\frac {\bfp_j^2} {2m}}
        + \sum_{\bfn \in \mathcal{M}} \omega_{\bfn} \frac{X_{\bfn}^2 + Y_{\bfn}^2}{2}
        \nonumber\\
        &-& \varepsilon \sum_{j=1}^{N_{\mathrm{tail}}} \sum_{\bfn \in \mathcal{M}}
           k_{\bfn}^{-1} \kappa_{\bfn} [X_{\bfn} \cos (\bfk_{\bfn} \cdot \bfr_j)
           - Y_{\bfn} \sin (\bfk_{\bfn} \cdot \bfr_j)].
\label{eqHItheta3DC}
\end{eqnarray}

For $N_{\mathrm{tail}}/N$ fixed, the coupling parameter $\varepsilon$ scales like $N_{\mathrm{tail}}^{-1/2}$ when $N$ goes to infinity. This keeps constant the typical size of the coupling term for random phases and/or particle positions in this limit. It can be checked that, on top of the total energy $E_{\rm sc} = H_{\rm sc}$, the total momentum
\begin{equation}
P_{\rm sc} =  \sum_{j=1}^{N_{\mathrm{tail}}}  \bfp_j +  \sum_{\bfn \in \mathcal{M}} I_{\bfn} \bfk_{\bfn},
\label{Totmom}
\end{equation}
is conserved, which is natural since the system is isolated. The total momentum is made of a wave contribution and of a particle one, while the total energy has a coupling contribution on top of these two. This invalidates the interpretations of Landau damping using energy conservation, while neglecting this coupling. We further notice that waves are now explicit degrees of freedom, while the electric potential is slaved to the particles in the One Component Plasma.

In references \cite{AEE,EEbook} the one-dimensional analog of $H_{\rm sc}$ was obtained by a direct mechanical reduction of degrees of freedom starting with the one-dimensional $N$-body problem\footnote{This was the starting point of the one-dimensional $N$-body approach.}. This derivation had been introduced in \cite{DescrLd89,LargeSc91}. In chapter 2 of reference \cite{EEbook}, this derivation was performed rigorously by making error estimates involving three small parameters. These parameters control the slow evolution of the wave amplitudes, the slow evolution of the average velocity of a given bulk particle in presence of the waves, and ensure that the bulk particles are in the linear regime of oscillation. Such a Hamiltonian was first introduced by Mynick and Kaufman \cite{MK}, and derived in a consistently Hamiltonian way from the Vlasov--Poisson system of equations by Tennyson\emph{ \textit{et al.}} \cite{Tennyson1994}\footnote{We notice that the derivations of the self-consistent dynamics starting with a Vlasovian description \cite{OWM,OLMSS,Tennyson1994} perform kind of a zigzag with respect to the $N$-body description, since they go back to a finite number of degrees of freedom after going through the continuous Vlasovian description.}.

\subsection{Spontaneous emission and Landau damping}
\label{SpontEmission}

For the sake of brevity, we do not develop here the full generalization of the one-dimensional analysis in Refs \cite{EZE,EEbook}~;
it is lengthy, but straightforward. However, since this analysis unifies spontaneous emission with Landau growth and damping,
we provide the generalization to three dimensions of the result of \cite{EZE} and of section 4.1.4 of \cite{EEbook} ruling the evolution of the amplitude of a Langmuir wave. It is provided by perturbation calculations
where the right hand sides of equations~(\ref{eqampl})-(\ref{delrsecwv}) are considered small (of first order).
This is natural for equation~(\ref{eqampl}) since $N_{\mathrm{tail}} \ll N_{\mathrm{bulk}}$,
and for equation~(\ref{delrsecwv}) if the Langmuir waves have a low amplitude.
Let $J(\bfm,t) = \langle \varphi(\bfm,t) \varphi(- \bfm,t) \rangle$,
where the average (mathematical expectation) is over the random initial positions of the tail particles (their distribution being spatially uniform and pairwise independent).
Then a calculation to second order in $\varphi$ yields \begin{equation}
    \frac {\rmd J(\bfm,t)} {\rmd t}
    = 2 \gamma_{\rm L}(\bfk_{\bfm}) J(\bfm,t) + S_{\bfm \, \mathrm{spont} }  ,
   \label{LandSpont}
\end{equation}
where $\gamma_{\rm L}(\bfk_{\bfm})$ is the Landau growth or damping rate defined by equation (\ref{gammaL}), and $S_{\bfm \, \mathrm {spont} }$ is given by\footnote{This expression corrects equation (46) of \cite{DbCb}.}
\begin{equation}
  S_{\bfm \, \mathrm{spont} }
  =
  \frac{2 \pi N e^2 }{\varepsilon_0^2 [\frac{\partial \epsilon_{\mathrm{bulk}}}{\partial \omega}(\bfm,\omega_{\bfm})]^2 k_{\bfm}^5}
     g \Bigl( \frac{\omega _{{\bf{m}}  {\mathrm{r}}}
}{k_{\bfm}} \Bigr).
  \label{Spont}
\end{equation}
$S_{\bfm \, \mathrm {\rm{spont}} }$ corresponds to the spontaneous emission of waves by particles and induces an exponential relaxation of the waves to the thermal level in the case of Landau damping (the three-dimensional analogue of what was found in \cite{EZE,EEbook}). This spontaneous emission was already present for a given realization of the $N$-body system in the beginning of section \ref{LandamK}, by terms enabling the excitation of a wave with an initially vanishing amplitude. It was implicitly present in the sum of the shielded potentials of equation (\ref{phij}). Indeed, the space-time average of the square of the corresponding electric field provides the estimate of spontaneous emission \cite{Bekefi}.

\subsection{Diffusion and friction coefficients}
\label{Diffric}

This subsection and the next ones deal with the one-dimensional case and review, in particular, the main results of \cite{EZE} and of chapters 4, 8, and 9 of \cite{EEbook}. The corresponding Hamiltonians are given by equations (\ref{eqHItheta3D}) and  (\ref{eqHItheta3DC}), where $\bfp_j,\bfr_j,\bfn,\bfk_{\bfn}$ are scalars. In order to go to the limit of a continuous wave spectrum, we define an
interpolating function $J(k)$ such that
\begin{equation}
   J(k_n)
   =
   I_n \frac L {2 \pi}  .
\label{ch4spectrumJk}
\end{equation}
Then the continuous spectrum limit reads
\begin{equation}
   \sum_j I_{j0} \bullet
   \to
   \int  \bullet I_j {\frac L {2 \pi}} \rmd k
   =
   \int  \bullet J(k) \rmd k.
\label{SZ136}
\end{equation}
In the case of a broad spectrum of Langmuir waves, a second order perturbation calculation for the particles (see \cite{EZE} and sections 4.1.3 of \cite{EEbook}), similar to the above one for the waves, yields the diffusion and friction coefficients of a Fokker-Planck equation ruling the particle dynamics. This equation can be written in the compact way
\begin{equation}
  \frac {\partial f} {\partial t}
  =
  \frac \partial {\partial p}
  \Bigl( D_{\rm QL} (p) \frac {\partial f} {\partial p} \Bigr)
  - \frac \partial {\partial p} ( F_{\rm c} f ),
  \label{SZ142}
\end{equation}
with
\begin{eqnarray}
   D_{\rm QL}(p)
   &=&
   \frac {\pi k(p) \varepsilon^2 \kappa^2(k(p))} {\omega_{\rm p}}
   J(k(p)),
   \label{SZ137}
   \\
   F_{\rm c}(p)
   &=&
   - \frac { k^2(p) L \varepsilon^2 \kappa^2(k(p))} {4 \omega_{\rm p}},
   \label{SZ138}
\end{eqnarray}
where $\kappa(k)$ and $k(p)$, are respectively $\kappa_n$ and
$k_n$ with $n$ chosen such that respectively $|k-k_n|$ and $|p -
\omega_n/k_n|$ are minimum. The Fokker-Planck equation displays two terms on the right
hand side. The first one involves only the wave amplitude through
the diffusion coefficient, and the second one is a friction term due
to the previously introduced spontaneous, or Cherenkov, emission. The former
is the so-called quasilinear diffusive term \cite{Vedenov,Drumm}. The latter accounts for the dynamical friction due to the
spontaneous emission responsible for $S_j$ in equation (\ref{Spont}).

Equation (\ref{SZ142}) is coupled to a wave evolution equation, which is the analogous of equation (\ref{LandSpont}) in one dimension
\begin{equation}
    \frac {\rmd J(k,t)} {\rmd t}
    = 2 \gamma_{\rm L}(k) J(k,t) + S_{\mathrm{spont}}(k)  ,
   \label{LandSpont1D}
\end{equation}
with
\begin{equation}
  S_{\mathrm{spont}}(k)
  =
  \frac{ N e^2 \kappa^2(k)}{ \varepsilon_0 L^2 k^3}
     g \Bigl( \frac{\omega_{r}(k)}{k} \Bigr)  .
  \label{Spont1D}
\end{equation}

The second order perturbation calculations leading to the quasilinear results make sense only over time scales $\Delta t$ such that $\tau_{\rm c} \ll \Delta t \ll \tau_{\rm spread}$, where $\tau_{\rm c}$ and $\tau_{\rm spread}$ are respectively the correlation time of the wave potential and the time over which the particle positions spread over $k^{-1}$, where $k$ is a typical wave number, because of the diffusion of velocities. These two times are defined as
\begin{equation}
  \tau_{\rm c}
  =
  (k \Delta u)^{-1}
\label{deftaucSC}
\end{equation}
with $\Delta u$ the spread in particle and phase velocities, and
\begin{equation}
  \tau_{\rm spread}
  =
  4 \Bigl( k^{2} D_{\rm QL 0} \Bigr)^{-1/3},
\label{tausp'}
\end{equation}
with $D_{\rm QL 0}$ a typical value of $D_{\rm QL}(p)$.

When particles diffuse, the local momentum conservation underlying Landau damping and growth no longer corresponds to the synchronization with a single wave. Therefore, \emph{this effect corresponds} to a coherent mechanism for isolated waves and \emph{to a (quasilinear) diffusive one for a broad spectrum}.

The coupled equations (\ref{LandSpont}) and (\ref{SZ142}) may be used to
describe the nonlinear evolution of the kinetic beam-plasma
instability, as long as the above perturbation theory remains
correct. They are called the quasilinear equations in the
literature when the Cherenkov terms are omitted, which occurs naturally in a Vlasovian setting.

The latter equations imply that the initial bump (corresponding to the
beam) in the distribution function flattens into a plateau (so
that $\partial f/\partial v = 0$ over the corresponding velocity interval). The presence of
the Cherenkov term implies a relaxation toward a distribution
function with a negative slope on a longer time scale. This is the
prelude to the relaxation toward
a new thermal equilibrium for the whole plasma.\footnote{This further relaxation cannot be described by the self-consistent Hamiltonian, since the latter corresponds to a given bulk.}

The system (\ref{SZ142}), (\ref{LandSpont1D}), without the terms due to spontaneous emission, was first introduced in 1962 in the framework of the Vlasov-Poisson description of the saturation of the weak warm beam instability, by Vedenov, Velikhov, and Sagdeev \cite{Vedenov}, and by Drummond and Pines \cite{Drumm}. These two works dealt with the plasma in a quasilinear way, as they neglected mode coupling for the wave growth and considered the particle evolution to be close to a ballistic one\footnote{Rigorously speaking, the quasilinear equations had already been mentioned in 1961 by Romanov and Filippov \cite{Romanov}. This reference makes the Ansatz of a Fokker-Planck equation for particle evolution and computes the corresponding diffusion coefficient; it describes the evolution of the Langmuir wave amplitude as the result of spontaneous and stimulated emission of quanta and estimates the corresponding coefficients.}. This enabled the introduction of the quasilinear diffusion coefficient long before chaos became fashionable in physics. Furthermore, this introduction was first done for the self-consistent problem.

\subsection{Saturation of the weak warm beam instability}
\label{Swwb}

The theory of the saturation of the weak warm beam instability is at the origin of the $N$-body approach of this review paper. This is why we briefly recall the corresponding historical background in the next paragraph. The following one summarizes results about the chaotic transport of a particle in a prescribed spectrum of waves, which are needed to tackle the full self-consistent problem. The next one shows there is a depletion of nonlinearity when the distribution is a plateau, and the last one summarizes the results of numerical simulations of the instability.

\paragraph{Historical background}

Quasilinear theory shows that the weak warm beam instability saturates by the formation of a plateau in the distribution function \cite{Vedenov,Drumm}. This agrees with experimental observation \cite{Roberson}.

During the above defined spreading time $\tau_{\rm spread}$, the velocity diffuses by the amount $\Delta v_{\rm spread} = 4 (D_{\rm QL 0}/k)^{1/3}$, whose size is defined by the turbulent wave spectrum, which contains an energy bounded by the weak beam energy. Therefore, $\Delta v_{\rm spread}$ is typically a quantity much smaller than the width of the plateau. This implies that the saturation time is much longer than $\tau_{\rm spread}$, which turns out to be also the (Lyapunov) time of separation of nearby orbits in the chaos induced by the waves (see for instance section 6.8.2 of \cite{EEbook}). Therefore, the \emph{formation of the plateau} can be proved without quasilinear theory: it \emph{comes from the chaos induced by the unstable Langmuir waves among the resonant particles}, whatever be the precise description of the corresponding chaotic transport.

When the motion of particles is chaotic, the perturbative approach used in the usual derivations, in particular in section \ref{Diffric}, of the quasilinear equations cannot be justified. Indeed, waves scatter the particle positions far away from their ballistic value. At time $\tau_{\rm spread}$, when the corresponding spreading of positions becomes on the order of the wavelength, the perturbative approach fails. Since leveling out the beam distribution function needs a time much longer than $\tau_{\rm spread}$, one may doubt the validity of quasilinear equations to describe the saturation of the instability.

The validity of quasilinear theory was first questioned in 1979 by Adam, Laval, and Pesme \cite{Adam} when accounting for nonlinear wave coupling, but the importance of this coupling was denied by Galeev  \textit{et al.} in 1980 \cite{Galeev}. In 1981, Krivoruchko  \textit{et al.} performed a beam-plasma experiment in a plasma column, which confirmed the quasilinear predictions to be correct, but displaying the non-quasilinear onset of field correlations (formation of coherent packets) and particle-motion correlation (formation of tails of accelerated particles and acceleration of the energy exchange between the waves and the particles) \cite{Krivoruchko}. However, in 1983, Laval and Pesme proved the existence of a renormalization of the growth rate and of the diffusion coefficient during the growth of the instability \cite{LPren}, and the inconsistency of quasilinear theory due to mode coupling \cite{LPinc}. In 1984, they proposed a new Ansatz to substitute the quasilinear one, and predicted that whenever $\gamma_{\rm Landau} \tau_{\rm spread} \ll 1$ both the wave growth rate and the velocity diffusion coefficient should be renormalized by a factor 2.2 \cite{LPtt}\footnote{More information about this controversy can be found in \cite{Doxas,LPcontr}.}.

This motivated Tsunoda, Doveil, and Malmberg to perform a new experiment with a traveling wave tube in order to have a much lower noise than in a magnetized plasma column \cite{Tsunoda1987a,Tsunoda1987b,TSU}. It brought a surprising result: quasilinear predictions looked right, while quasilinear assumptions were completely violated. Indeed no renormalization was measured, but mode-mode coupling was not negligible at all, while it is neglected in the quasilinear approach. These results were further documented in an extension of the experiment \cite{Hartmann}. This sets the issue: is there a rigorous way to justify quasilinear estimates in the chaotic regime of the beam-plasma instability?

Tackling this issue in a Vlasovian setting sounded formidable. However, the theory of chaos for finite number of degrees of freedom Hamiltonian systems had been developing in the plasma physics community for more than a decade \cite{EscContr,EscFromThermo}, and this was an incentive to address the weak warm beam-plasma instability by generalizing \cite{LargeSc91,Tennyson1994} a model originally introduced for the numerical simulation of the cold beam-plasma instability \cite{OLMSS,OWM}. There the beam was described as a set of particles, while the wave was present as a harmonic oscillator. If one considers a wave-particle interaction occurring in a finite range of velocities $[v_{\rm{min}},v_{\rm{max}}]$, then it is sufficient to include in the Hamiltonian the waves with phase velocities in this interval, which defines their number $M$ when the length $L$ of the system is defined. This leads to the self-consistent Hamiltonian (\ref{eqHItheta3D}). In retrospect, in a system where the transport due to short range interactions (``collisions") is weak, it is natural to think about plasma dynamics by working directly with classical mechanics, and by taking into account that the collective field dominates over the graininess field.

Before embarking in the study of the chaotic dynamics of the self-consistent Hamiltonian corresponding to the weak warm beam instability, three preliminary investigations sounded useful: deriving the self-consistent Hamiltonian from the underlying $N$-body dynamics, recovering Landau damping/growth from this Hamiltonian, and studying the chaotic transport of a particle in a prescribed spectrum of waves, which was propaedeutic to the self-consistent case. The first investigation led to the derivations of the one-dimensional version of the self-consistent Hamiltonian recalled at the end of section \ref{AmplitudeEqn}. The second one led to the one-dimensional version of the van Kampen-Dawson approach presented in section \ref{LwKD}. The third one brought a series of results concerning chaotic transport in a prescribed spectrum of waves, which are now briefly recalled\footnote{Experimentally, studying the chaotic transport of particles in a prescribed spectrum of waves was propaedeutic to the self-consistent case too. This led to the experimental observation of resonance overlap \cite{Doveil2005a}, and of the transition from stochastic diffusion in a large set
of waves to slow chaos associated to a pulsating separatrix \cite{Doveil2011}. Nonlinear resonances excited by injected waves were both observed as a ``devil's staircase" \cite{Macor2005} and cancelled to build a barrier to transport \cite{Chandre2005}.}.

\paragraph{Chaotic transport}

While for uncorrelated phases, it is natural to expect the diffusion coefficient to converge to its quasilinear estimate from below when the resonance overlap of the waves increases, for intermediate values of the overlap, unexpectedly the diffusion coefficient turns out to exceed its quasilinear value by a factor about 2.5 \cite{Cary1990}. This further suggested the possibility of a renormalization for the self-consistent case. Nevertheless, the diffusive picture for the chaos due to waves needed to be substantiated. It was shown to be right, provided adequate averages are performed on the dynamics; however, this picture is wrong if one averages only over the initial positions of particles with the same initial velocity \cite{Benisti1997} (see also section 6.2 of \cite{EEbook}, and \cite{Escande2007,Escande2008}). These results were completed by mathematical results: individual diffusion and particle decorrelation were proved for the
dynamics of a particle in a set of waves with the same wavenumber and integer frequencies if their electric field is gaussian \cite{Elskens2010}, or if their phases have enough randomness \cite{Elskens2012}. The randomness of amplitudes also affects significantly the overal diffusion \cite{Elskens2010nql}.

The intuitive reason for the validity of the diffusive picture is given in \cite{Benisti1997}: it is due to the locality in velocity of the wave-particle interaction, which makes the particle to be acted upon by a series of uncorrelated dynamics when experiencing large scale chaos. This locality of the wave-particle interaction was rigorously proved by B\'enisti in \cite{Benisti1998a}. On taking into account that the effect of two phases on the dynamics is felt only after a long time when there is strong resonance overlap, it can be approximately proved that the diffusion coefficient is larger than quasilinear, but converges to this value when the resonance overlap goes to infinity \cite{Benisti1997,Escande2002b} (see also \cite{Escande2002a,Escande2003}, and section 6.8.2 of \cite{EEbook}).

\paragraph{Depletion of nonlinearity when the distribution is a plateau}

When the plateau forms in velocity during the saturation of the weak warm beam instability, density becomes also almost uniform spatially in this range of velocities. Indeed, chaos tends at equidistributing particles all over the chaotic domain in $(r,v)$-space. Actually, KAM tori, bounding the chaotic domain
defined by a prescribed spectrum of waves, experience a sloshing motion due to
the waves. This brings a small spatial modulation to the particle density
which provides a source term for the Langmuir waves.
However, if the plateau is broad, the source term in equation (\ref{eqampl}) almost vanishes, since the particles are equidistributed spatially, and the waves keep a fixed amplitude~:
the self-consistency of equations (\ref{eqampl})-(\ref{delrsecwv}) is quenched and the wave spectrum is frozen,
even when particle dynamics is strongly chaotic in the plateau domain\footnote{Equations (\ref{eqampl})-(\ref{delrsecwv}) are used here to avoid writing their one-dimensional analog, which is the relevant one for this discussion.} (see sec.\ 2.2 of \cite{BEEB}). Clumps of particles may experience a strong
temporary trapping, but the distribution function stays flat.
As a result, the plateau dynamics is almost the
same as in a prescribed field of Langmuir waves\footnote{For a plateau with a finite width, the small remaining source brings a further evolution of the wave-particle system toward a state where the wave spectrum collapses toward small wavelengths together with the escape of initially resonant particles towards low bulk plasma thermal speeds \cite{FLA}. This corresponds to a further step toward a new thermal equilibrium of the $N$-body system corresponding to the initial beam-plasma system. The description of the subsequent steps toward thermal equilibration require to use a full $N$-body model.}. This is an instance where nonlinear effects increase the symmetry of the system, and lead to a depletion of nonlinearity\footnote{This phenomenon, also called depression of nonlinearity, was introduced in fluid mechanics \cite{KP}. In Navier-Stokes turbulence, the mean-square value of the nonlinear term of the equation was found significantly depressed, i.e.\ smaller than the same quantity in the Gaussian field with the same energy distribution. This was identified to result from the emergence of long-lived vortices where the enstrophy cascade is inhibited. It also exists in systems with quadratic nonlinearities  \cite{KP,BRF}.}. Then, it is possible to use the tools of 1.5 degree-of-freedom Hamiltonian chaos mentioned in the above paragraph ``Chaotic transport" to compute the diffusion of particle velocities.

In a Vlasovian description, the bump-on-tail instability saturates with the previous plateau substituted with a very jagged distribution in both space and velocity, resulting from the chaotic stretching and bending of the initial beam-plasma distribution. Indeed, since the initial distribution $f_0(v)$ is conserved along particle motion, there is no finite range in velocity where $f_0(v)$ has the amplitude of the plateau. This plateau can be obtained only by coarse-graining (local averaging) of the Vlasovian distribution of the saturated state.

\newpage

In the above paragraph ``Chaotic transport", we recalled that, in the chaotic regime, $D/D_{\rm QL}$ may cover a large range of values
\cite{Cary1990,EEbook}. In particular $D \simeq D_{\rm QL}$ is obtained for
random phases of the waves and strong resonance overlap
\cite{Cary1990,EEbook,Elskens2007,Elskens2010nql,Elskens2010,Elskens2012}. The plateau regime corresponds to $\gamL = 0$ and therefore to
$\gamL \tau_{\rm spread} = 0$. Since $D/D_{\rm QL}$ may cover a
large range of values in this regime, $\gamL \tau_{\rm spread}
\ll 1$ does not imply per se any renormalization or non-renormalization of
$D/D_{\rm QL}$ (nor of $\gamma /
\gamL$ by wave-particle momentum conservation). This contradicts previous works
using $\gamL \tau_{\rm spread}
\ll 1$ to try and prove the validity of quasilinear theory
\cite{LD1,LD2,EE2,EEbook} along with
the ``turbulent trapping'' Ansatz aiming at the contrary \cite{LPtt}. The value
of $D/D_{\rm QL}$ in
the plateau regime of the bump-on-tail instability depends on the kind of wave
spectrum the beam--plasma system reaches during the saturation of the
instability, and not only on condition $\gamL\tau_{\rm spread}
\ll 1$, as assumed by these works.

\paragraph{Numerical simulations}

The difficulty of dealing analytically with the strongly nonlinear
regime of the Vlasov-Poisson system led from 1989 to the
development of the finite number of degrees of freedom approach using the self-consistent Hamiltonian (\ref{eqHItheta3D}). This enabled numerical simulations of the self-consistent dynamics to be performed \cite{Cary,Doxas}.

Let $\tau_{\rm w}$ be the typical time for the growth of the wave amplitudes. In the regime $\tau_{\rm w}
\ll \tau_{\rm spread}$ the system (\ref{SZ138})-(\ref{Spont1D})
was found to be correct, but for one realization
the wave spectrum appeared to be jagged with respect to the
average one, each wave having a temporal behaviour strongly marked
by nonlinear wave coupling. This confirmed the behaviour
found by Theilhaber, Laval and Pesme \cite{Theilhaber}, and Berndtson \cite{Berndtson} with Vlasovian codes. The simulation of Doxas and Cary \cite{Doxas} indicated a possible renormalization, but the wave spectrum was not dense enough for this to be completely convincing. Furthermore, the renormalization factor was much smaller than the one proposed by \cite{LPtt}.

In 2011, taking advantage of the increased power of computers, more precise numerical simulations were performed using a semi-Lagrangian code for the Vlasov--wave model \cite{BEEB}. This model is the mean-field limit of the granular dynamics defined by the self-consistent Hamiltonian: waves are still present as $M$ harmonic oscillators, but particles are described by a continuous distribution function (which is discretized in the numerical scheme, though). The simulations were benchmarked in various ways. In particular, the conservation laws were checked, as well as the above depletion of nonlinearity when the distribution is a plateau. They were repeated for a large number of random realizations of the initial wave phases for a fixed initial spectrum of amplitudes. As shown by previous simulations, the final wave spectrum was found to be quite jaggy, and not smooth as that predicted by QL theory \cite{Vedenov,Drumm}.

For each of the realizations, one computed the spreading of the velocities of test particles when acted upon by the final set of waves. The first four even moments of this spreading were compared with those of the solution to the quasilinear Fokker-Planck equation for velocity diffusion, using the velocity-dependent diffusion constant $D_{QL}$ computed with the final wave spectrum. The agreement was found to be excellent: the plateau verified the predictions of QL theory. However, as found in previous numerical simulations and experimentally, mode-mode coupling was found to be very strong during the saturation, which invalidated the QL assumptions. Similar results were obtained in \cite{Volokitin} with a code using the dynamics of the self-consistent Hamiltonian.

At this point, the validity of QL predictions while QL assumptions are wrong sounded still like a mystery. However, the simulations brought an unexpected clue to elucidate it: the variation $\Delta \phi(t) = \phi(t) - \phi(0)$ of the phase $\phi(t)$ of a given wave, was found to be almost non fluctuating with the random realizations of the initial $\phi(0)$'s of the waves \cite{BEEBEPS}\footnote{This is reminiscent of the fact that the initial correlations were not disturbed in the course of the relaxation in the beam-plasma experiment \cite{Krivoruchko}.}. Since $\Delta \phi(t)$ does not depend on $\phi(0)$, the randomness of the final wave phases is the same as that of initial phases. As a result, the self-consistent dynamics was shown to display an important ingredient for the validity of a quasilinear diffusion coefficient for the dynamics in a prescribed spectrum. Some analytic support was brought to this finding by using a third order Picard iterate of the dynamics \cite{BEEBEPS2}.

\subsection{Dynamics with a single wave}
\label{C8l}

The single wave case is relevant to two kinds of situation.
First, in the unstable beam-plasma system, one wave may have a
larger growth rate than the others, so that it soon dominates over
these and the single wave approximation is reasonable~: this is the case
for a cold beam. Second, the physical device of interest, because
of its finite length, may have a single wave being resonant with a
beam, even a warm one.

If a single wave interacts with all particles,
the locality in
velocity of the wave-particle interaction is particularly
important. In the nonlinear regime of the beam-wave system, it appears that only particles with velocities close to the wave phase velocity, up to the resonance width,
interact significantly with the wave, and that the relevant
velocities thus depend on the instantaneous wave intensity (see chapter 8 of \cite{EEbook}).

Numerical simulations considered the growth and saturation of a single wave whose phase velocity lies in the range with positive slope of the velocity distribution function of a warm beam \cite{Dov1w,Firpo}. They showed that the dynamics with a finite number $N$ of particles leads to a long-time evolution of a single wave qualitatively distinct from the numerical integration of the kinetic Vlasov model. Indeed, while the latter integration leads the amplitude to saturate with an amplitude $A$, the dynamics with a finite number $N$ of particles brings a slow second growth of the amplitude after an apparent saturation at the value $A$.

This occurs despite the fact that the limit $N \to \infty$
formally reduces the many-particle evolution equation to the
Vlasov kinetic equation (see section G.1.2 of \cite{EEbook}), but there is
no contradiction.
Kinetic limit theorems show that
the limit $N \to \infty$ commutes with the dynamics
over any finite time interval $0 \leq t \leq t_*$, but the discrepancy
between both evolutions may well diverge exponentially
as $t_* \to \infty$, especially because the wave-particle
system has dynamical instabilities \cite{FirpoE1998}. Then, in the
long term, given a finite number of particles, there is a time
beyond which the smooth solution to the kinetic model
(approximating the finite $N$ system initially) may evolve
significantly differently from the physical finite-$N$ system~:
the limits $t \to \infty$ and $N \to \infty$ need not commute
\cite{DescrLd89}.

In short, in its evolution, the plasma
eventually reminds the physicist of its microscopic granular
nature, so that the Vlasov equation cannot accurately
describe it over long times \cite{DescrLd89}, another symptom of the singular Vlasovian limit. In the above single wave simulations, the granular nature of the plasma induces fluctuations of the width of the separatrix of the single wave in the apparent saturated state. This enables the wave to exchange almost trapped particles with average velocities smaller and higher than the phase velocity of the wave. However, when the wave reaches amplitude $A$, the beam distribution is not globally flattened, and there are still more faster particles than slower particles. Therefore, the exchange of almost trapped particles provides momentum to the wave, i.e.\ further growth.

Stating this result in a different way, the BGK equilibrium corresponding to the saturated Vlasovian mode does not exist because of the fluctuations induced by the granular nature of the plasma. In reality, this problem is for all Vlasovian travelling-wave BGK equilibria, as shown in \cite{ElskBGK} : they cannot exist for the dynamics ruled by the single wave finite-$N_{\mathrm{tail}}$ self-consistent Hamiltonian, because they would come with singularities in complex time, which are inconsistent with the travelling-wave assumption.

\subsection{Damping with trapping results from a phase transition}
\label{DwT}

As was shown by O'Neil in 1965, when the initial amplitude of a Landau damped Langmuir wave is increased, there is a threshold above which the wave amplitude enters an oscillatory regime after a small initial damping \cite{O'N}. This oscillatory regime results from the trapping of particles inside the wave. The description of this phenomenon by the self-consistent dynamics enables proving that this transition is a second order phase transition \cite{FirEl2}. This is shown in a series of steps: (i) In the many-particle limit $N \to \infty$, the self-consistent system is described
by an invariant distribution over $(r,v)$-space, and explicit forms
can be found for several physical observables in a Gibbsian setting. (ii) For a
single wave, the partition function can be computed analytically, and one distinguishes two
regimes separated by a phase transition. The two thermodynamic phases are linked
to the different dynamical evolutions of the system. In
particular, the two regimes of wave damping (Landau damping and damping with trapping) are recovered in
connection with the two phases\footnote{The envelope equation of an electron plasma wave has a sudden variation when going from the linear to the trapping regimes, in a way similar to a first order phase transition \cite{benisti2017}. Furthermore, there are other aspects of non collisional damping for a wave having trapped electrons \cite{benisti2012}.}.

A striking property
of the one-dimensional One Component Plasma
is that it undergoes no phase transition. A phase transition in
the wave-particle
model was thus unexpected. However, the two systems are very different thermodynamically.
Indeed, the thermodynamic treatment of the binary interaction assumes that
all particles reach a ``global'' equilibrium, whereas the
thermodynamic treatment of the wave-particle system applies to a
reduced system, in which only the faster, more efficient
interactions are taken into account.

\subsection{Dynamics with two waves}
\label{D2w}

The dynamics of $H_{\rm sc}$ defined by equation (\ref{eqHItheta3DC}) are extremely rich, and do not at all reduce to those of the single or of the many waves cases described above. In particular, a work with the one-dimensional analog of this Hamiltonian involving two waves coupled to many particles displays the emergence of long-lived quasi-stationary states (QSS) \cite{Carlev}. Motivated by the problem of $\alpha$-particle thermalization in a burning thermonuclear plasma, this work focuses on the case where $\omega_{\bfn} \ll 1$ for the two waves, which endows them with vanishing phase velocities, and yields a dramatic importance to the coupling term. Spatially, one of the two waves is the second harmonic of the other one. In numerical simulations, at $t=0$ particles are spread uniformly spatially and in velocity in the interval $[- p_0,p_0]$. A threshold value of $p_0$ is found, such that below this threshold the system responds to the beam injection by the emergence of long-lived QSS: the clustering of the beam particles into resonant rotating clumps. This threshold is also present for the case of a single wave. The values of the two coupling constants select a leading harmonic, and the passage from a leading one to the other one corresponds to a first-order phase transition. This transition can be described analytically by applying Lynden-Bell's technique to the Vlasovian limit of the problem \cite{Carlev}.

\section{Coulomb scattering}
\label{Coltrsp}

This section reviews the theory developed in \cite{CbCollTrsp}, which shows that an $N$-body model where electrons interact through their Debye shielded Coulomb potential, enables the calculation of Coulomb scattering describing correctly all impact parameters $b$, with a convergent expression reducing to Rutherford scattering for small $b$. This derivation explains why a two-body calculation yields a correct estimate of Coulomb scattering, while most of this transport is due to the simultaneous action of many particles with impact parameters between the inter-particle distance and the Debye length.

The word ``collision" is borrowed from the physics of gases where it qualifies the close encounter of two particles. However, the interaction of particles in a plasma corresponds seldom to two-body collisions, even when taking into account Debye shielding~:
in the plasmas considered here, where the interparticle distance $\lid$ is much smaller than the Debye length $\lambda_\rmD$,
a particle $j$ feels the simultaneous unshielded short-range action of many particles.
Except for those particles very close to $j$, this action produces a slow and simultaneous deflection of $j$.
Rigorously speaking, one should speak about ``short range induced interactions",
``unshielded Coulomb interactions", or so, and not ``collisions". Such a designation is a result of the development of plasma physics after that of gases,
which made natural for the former to borrow concepts and tools from the latter.
In particular, the unshielded interactions of particles in kinetic plasmas were considered as collisions. These interactions, occurring at scales $\lambda_\rmD$ or smaller, set a bound to the Vlasovian description, and require another specific one.

About sixty years ago, two groups at UC Berkeley's Radiation Laboratory
simultaneously studied transport due to collisions in non-magnetized plasmas,
and they quoted each other's results in their respective papers~:
one in 1956 by Gasiorowicz, Neuman and Riddell \cite{Gasio} and,
in 1957, one by Rosenbluth, MacDonald and Judd \cite{RoMcDJu}.
The second group of authors used the Rutherford picture of two-body collisions, while
the first group of authors dealt with the mean-field part of the interaction by using perturbation theory in electric field amplitude\footnote{Their derivation suggested to the first author of the present review that a direct $N$-body approach might be possible.}.
Later on, within the same approximations as \cite{Gasio},
a more elegant derivation of the Coulomb scattering coefficients was provided,
in a ``post-Vlasovian'' approach,
by taking the limit ``infinite number of particles in the Debye sphere" of the Balescu--Lenard equation
(see section 8.4 of \cite{BalBk97} and sections 7.3 and 7.4 of \cite{HaWa04}),
though the rigorous foundation of this equation is still a challenge \cite{Spohn,Lancellotti}.

Each of the above works on Coulomb scattering
has a difficulty in describing the interactions at distances of the order of the typical interparticle distance $\lid$.
Indeed, the mean-field approach cannot describe the graininess of these scales,
and the Rutherford picture cannot describe the simultaneous collisions with several particles.
Consequently, the mean-field approach is suited to describing scales larger than $\lid$,
and should be used with a corresponding ultraviolet cutoff,
while the Rutherford picture holds for scales smaller than $\lid$, and should be used with a corresponding infrared cutoff.
Fortunately, in both approaches the transport coefficients depend only logarithmically on these cutoffs.
Furthermore, forgetting about the latter ones, and considering in both cases the scales
typically between $\lma$ and $\lambda_\rmD$, the two results are found to agree \cite{Gasio,RoMcDJu}.
This provided confidence in these complementary extrapolations, which were for long the basis
of the description of Coulomb scattering in plasmas, as presented in many plasma physics textbooks.

However, till 2015 a calculation of the contribution of scales about $\lid$ to Coulomb scattering had been missing,
and no theory provided a calculation of this transport covering all scales between $\lma$ and $\lambda_\rmD$.
This gap was filled by reference \cite{CbCollTrsp}, which computes the trace $T_D$ of the velocity diffusion tensor of a given particle by a convergent expression
including the particle deflections for all impact parameters.
The main ideas of the new derivation are (i) the substitution of the Coulomb potential of a particle with its Debye-shielded potential, i.e.\ the substitution of the bare potential with its ``dressed'' one defined by equation (\ref{phij}),
(ii) the computation of Coulombian deflections by first order perturbation theory in the total electric field, except for those due to close encounters,
(iii) the contribution to $T_D$ of the former ones is matched with that of the latter ones computed by \cite{RoMcDJu}.
The detailed matching procedure includes the scale of the inter-particle distance,
and is reminiscent of that of \cite{Hub}, however without invoking the cancellation of three infinite integrals.
It leads to the same expression as \cite{RoMcDJu},
except for the Coulomb logarithm which is modified by a velocity dependent quantity of the order of 1. More precisely, the computation of the deflection of particle $l$ is performed in four steps.

The first step uses first order perturbation theory in the electrostatic potential,
which shows the total deflection to be the sum of the individual deflections due to all other particles. Indeed, to this order
\begin{equation}
  \delta \dot{\bsr}_l(t)
  =
  \sum_{j \in S ; j \neq l} \delta \dot{\bsr}_{lj}(t) ,
\label{delrl1}
\end{equation}
where $S$ denotes the set of integers from 1 to $N$ labeling particles, and $\delta \dot{\bsr}_{lj}(t) $ is the contribution of particle $j$ to the change of velocity of particle $l$.  This yields
\begin{equation}
  \langle \norm{\delta \dot{\bsr}_l (t)}^2 \rangle
  =
  \sum_{j \in S ; j \neq l}  \langle \norm{\delta \dot{\bsr}_{lj} (t)}^2 \rangle,
\label{delrlj1^2}
\end{equation}
where the average is performed over the initial random positions of particles, which kills the terms involving the contributions of two different particles.
Therefore, though being due to the simultaneous scattering of particle $l$ with the many particles inside its Debye sphere,
$\langle \norm{\delta \dot{\bsr}_l (t)}^2 \rangle$ turns out to be the sum of individual two-body deflections
for impact parameters such that first order perturbation theory is correct.

For an impact parameter $b$ much smaller than $\lambda_\rmD$,
the main contribution of the acceleration due to particle $j$
 to the deflection of particle $l$
comes from times when this acceleration takes on its bare Coulombian value.
Therefore, $\delta \dot{\bsr}_{lj}(t) $ is a first order approximation
of the effect on particle $l$ of a Rutherford collision with particle $j$.
The perturbative calculation is seen to be correct for
$\lambda_\rmD \gg b \gg \lma$. This explains why the contribution to $\langle \norm{\delta \dot{\bsr}_l (t)}^2 \rangle$ of this range of $b$'s can be computed as if it would result from successive two-body collisions, as was done by \cite{RoMcDJu} and in many textbooks.

The second step of the computation of the deflection of particle $l$ proves that for a close encounter with particle $j$,
the deflection of particle $l$ is exactly the one it would undergo if the other $N-2$ particles were absent. The contribution of such collisions to $\langle \norm{\delta \dot \bsr_l (t)}^2 \rangle $ was calculated by \cite{RoMcDJu}. Now, since the deflection of particle $l$ due to particle $j$ as computed by the above perturbation theory
is an approximation to the Rutherford deflection for the same impact parameter,
one may conversely approximate the perturbative deflection with the full Rutherford one,
and obtain an obvious matching of the theories
for $\lid \gg b \sim \lma$ and for $\lambda_\rmD \gg b \sim \lid$~:
one may thus use the estimate of \cite{RoMcDJu} in the whole domain $b \ll \lambda_\rmD$.

The third step shows that the deflection for an impact parameter of the order of $\lambda_\rmD$ is given by
the Rutherford expression multiplied by some function of the impact parameter reflecting shielding. This enables a good matching of the deflections for large impact parameters with those for smaller ones.

The final and fourth step obtains an analytic expression for deflection whatever the impact parameter, by taking advantage of the fact that the individual deflections due to impact parameters $b$ exceeding $\lambda_\rmD$ decay rapidly with $b$. One finds that
the Coulomb logarithm $\ln (\lamD / \lma)$ of the second equation~(14) of \cite{RoMcDJu}
becomes $\ln (\lamD / \lma) + C$ where $C$ is of order unity. The same approach can provide the calculation of the other elements of the diffusion tensor and of dynamical friction, the latter requiring second order perturbation theory. The tensors corresponding to electron-ion collisions and to ion-ion collisions can be computed likewise.

As was shown in section \ref{MIIDS}, Debye shielding results from the Coulombian deflections of particles usually called ``collisions". In turn, the resulting Debye shielded effective potential
yields a description of pair interaction which provides a direct calculation of particle deflections,
viz.\ of Coulomb scattering. Shielding and collisions are thus intimately linked,
and the present ability of a thorough calculation of Coulomb scattering rests on this link.

A startling aspect of collisions in plasmas is that,
although each particle interacts simultaneously with many other ones on the Debye length scale
(suggesting the need for a collective description),
the transport effect of these interactions is well approximated by a sum of independent binary estimates,
because the deflections are so weak that they can be treated perturbatively.

\section{Discussion}
\label{Disc}

\subsection{New physical picture of microscopic plasma physics}
\label{Nppmpp}

The $N$-body approach brings a new physical picture of microscopic plasma physics. First, it shows that collisions play an essential role in collisionless plasmas. In particular, Debye shielding is a direct consequence of collisions (section \ref{MIIDS}). Indeed, the Coulombian deflections of electrons by a given electron P decrease the number of electrons about P, which decreases its apparent charge, according to Gauss' theorem. In contrast to what occurs if electron P is substituted with a Langmuir probe, this does not change the density of charges of the plasma, because the deflections due to P are compensated by those of the other electrons. Debye shielding results from a cooperative dynamical self-organization process, produced by the accumulation of almost independent Coulomb deflections over a time-scale $\omega_{\rmpp}^{-1}$. It is now clear how a given particle can be shielded by all other particles, while contributing to their individual shieldings. Unexpectedly on the basis of the implicit picture of most basic textbooks, shielding and Coulomb scattering are two aspects of the same two-body repulsive/attractive process. The $N$-body approach to Coulomb scattering shows there is a smooth connection between impact parameters where the two-body Rutherford picture is correct, and those where a collective description is mandatory (section \ref{Coltrsp}).

The essential role of collisions was also exemplified in section \ref{Aud} for distributions of particles corresponding to a set of monokinetic beams where each beam is a simple cubic array of particles: multi-beam-multi-arrays. Indeed, such distributions correspond to invariant states of the dynamics where neither collisions, nor waves are present. When perturbing the positions of particles of a multi-beam-multi-array with a dense enough set of velocities, they exponentially diverge from their initial positions, because of collisions (section \ref{Caw}). The finite value of the corresponding exponentiation rates is crucial for the equivalent of the van Kampen phase mixing to occur in the $N$-body system (section \ref{Pm}). Finally, Landau damping occurs after a phase mixing time (section \ref{LandamK}), which is longer than $\omega_{\rmpp}^{-1}$, the time necessary for collisions to establish Debye shielding. Therefore, this damping occurs in a plasma already organized by collisions.

The $N$-body approach shows unequivocally that Landau damping results from the simultaneous average synchronization of almost resonant passing particles with the wave (section \ref{Aspw}). It is because this synchronization is the same for Landau growth and damping, that a single formula applies to both phenomena (section \ref{Aspw}). The phase mixing of many beam modes produces Landau damping, which cannot correspond to a damped eigenmode because of Hamiltonian mechanics  (section \ref{Vlso}). This phase mixing is also active for Landau growth (section \ref{Li}). If there is a too low number of particles in a range of velocities proportional to the damping rate, about the phase velocity of a wave, the system behaves as a multi-beam in this range, and no Landau damping occurs (section \ref{Li}). When particles diffuse, Landau damping and growth no longer correspond to the synchronization with a single wave, but result from this diffusion (see section \ref{Diffric}). If there is a plateau in the distribution function and the waves with phase velocities in the range of this plateau make particle dynamics chaotic in this range, these waves have a stationary amplitude because of a depletion of nonlinearity due to the chaotic dynamics of particles: there is no component of the electron density resonating with these waves (section \ref{Swwb}). This implies that the case with a plateau is very different from one where there is a vanishing number of particles in the same velocity interval, while Landau damping vanishes in both case in a Vlasovian setting.

The $N$-body approach incorporates spontaneous emission naturally (sections \ref{LandamK} and \ref{SpontEmission}). When this emission is taken into account, Landau damping is nothing but a relaxation mechanism to the thermal level of Langmuir waves (section \ref{SpontEmission}). This emission prevents a plateau on the tail of the velocity distribution to be stationary, and triggers the walk toward the eventual thermal distribution (section \ref{Diffric}).

The travelling-wave BGK equilibria of the Vlasovian case do not exist in the $N$-body one, because of the fluctuations induced by the granular nature of the plasma (section \ref{C8l}). In particular, a beam-plasma system with a single unstable wave, after reaching the equivalent of the Vlasovian saturation, evolves on a longer time-scale toward a saturated state where the wave has a larger amplitude (section \ref{C8l}). O'Neil's damping with trapping typical of initially large enough Langmuir waves results from a phase transition (section \ref{DwT}).

\subsection{New insight into the Vlasovian limit}
\label{Ta}

The $N$-body approach reveals two important features of the Vlasovian limit: it is singular and it corresponds to a renormalized description of the actual $N$-body dynamics.

The singularity of the Vlasovian limit shows up: (i) in the dielectric function: that of a multi-beam-multi-array does not converge toward the Vlasovian expression when the density of the beam velocities increases, since both zeros and poles do not match (section \ref{Vlso}), (ii) in the importance of phase mixing for Landau growth in the $N$-body approach (section \ref{Li}); van Kampen's theory displays a view of the dynamics of a plasma closer to the genuine one than Landau's, (iii) in the impossibility to give a physical interpretation to the term of initial conditions (\ref{phi0hatcg}) in Landau's calculation of Langmuir waves, which is nothing but the continuous limit of the sum of the ballistic potentials of the $N$ electrons, (iv) in the fact that adding a test particle to the $N$-body system does not provide the shielded potential of this particle (section \ref{Vcbp}), in contrast with the Vlasovian case, (v) in the usual requirement of an analytically continuable velocity distribution function too (see section \ref{Zgdf}; also its footnote, which shows that a softer requirement is possible).

One of the simplest examples of a renormalized potential is the Debye shielded potential \cite{Mccomb}. It is a mean-field potential produced by the Coulomb deflections of the particles (section \ref{MIIDS}). Both its mean-field and BBGKY derivations show that Vlasov equation deals with a mean-field potential (section \ref{LwDsL}). Therefore, the Vlasovian dielectric function is a renormalized version of that of a multi-beam-multi-array, and a Vlasovian Langmuir wave is the renormalized version of a set of beam modes of the $N$-body system (section \ref{LwDsL}). The renormalized dielectric function enables the calculation of the shielded potentials of the $N$ particles of the granular plasma considered here, or of a test particle added to a Vlasovian plasma (section \ref{LwDsL}).

\section{Conclusion}

Laplace's dream was not a mere utopia, since the calculation of classical orbits starting from prescribed initial conditions can genuinely describe and explain many phenomena of microscopic plasma physics : Landau damping and growth, Debye shielding, Coulomb scattering, etc... Many of the calculations can be done for a single realization of the plasma by using standard tools of elementary mechanics, calculus and no probabilistic setting. This provides a stronger ground to face the complexity of plasmas, a difficult issue \cite{escande2013face,escande2014complexity}. An alternative title of this review paper might be ``Mechanical foundations of microscopic plasma physics".

Vlasov equation and calculations \`a la Landau have proved to be very efficient tools of theoretical plasma physics. However, computing and understanding are very different, as shown by the almost two decades the plasma physics community took to accept the reality of Landau damping. To introduce basic microscopic plasma phenomena, the $N$-body approach is short, since it avoids the introduction of a kinetic equation, and provides a clear-cut interpretation of the derived phenomena. So, Landau damping can be taught or given as an exercise to students knowing Newton's second law of motion, Fourier series, but neither Vlasov equation, nor Laplace transform; in particular, to students studying Newtonian mechanics. The additional knowledge of Laplace transform makes Debye shielding accessible too. Since it is rigorous and starts from first principles, the $N$-body approach could make plasma physics more attractive to colleagues in other disciplines and to prospective students who are fundamentally minded --- even more so, because of the issues of singular limits, of cooperative self-organization, of renormalization, and of the depletion of nonlinearity. Furthermore, basic microscopic plasma physics has both a description corresponding to a singular limit, the Vlasovian one, and a non-singular
one, the $N$-body approach: exceptionally, reductionism works for this physics.

Plasma physicists might enjoy the intuitive mechanics, the unifications and simplifications of the $N$-body approach, and the new insight into the Vlasovian limit it provides. Retrospectively, Landau's derivation of his damping was the first, but is the less physically intuitive; indeed, the Vlasovian limit is a singular and renormalized one. The $N$-body approach is also the occasion to consider the many facets of what is usually called ``the distribution function". Most theoretical calculations do not specify which aspect of this function is considered. In particular, is it thought in a statistical setting describing an \emph{ensemble} of plasmas, or as an idealization of a \emph{single} granular distribution \cite{ElsVla}? In the $N$-body approach, this distribution is introduced at the end of the mechanical calculations. Finally, plasma physicists might enjoy the full description of the contribution of all impact parameters to Coulomb scattering (section \ref{Coltrsp}). In particular, most of this contribution corresponds to a sum of independent binary estimates, while each particle interacts simultaneously with many other ones on the Debye length scale.

It would be useless to rewrite the whole of plasma physics in an $N$-body setting. However, some extensions might turn useful. In particular, to clarify the issue of Debye shielding in a magnetized plasma. Indeed, in the direction perpendicular to the magnetic field, the repulsive electric field due to a particle provides an $\mathbf{E} \times \mathbf{B}$ drift of the ``colliding" ones. What is the corresponding shielding in the direction perpendicular to $\mathbf{B}$? Is it the same as in a non-magnetized plasma where Coulomb deflections are present\footnote{The conspicuous modification of relaxation processes by magnetic fields \cite{Ichimaru,Silin,Hassan1,Baldwin,Hassan2}, suggests that shielding might be modified too. They also suggest the importance of the relative ordering of spatial scales, such as Larmor radius versus the Debye length.}?

\begin{acknowledgements}
D. F. E. is grateful to the members of Equipe Turbulence Plasma in Marseilles, since the theory reviewed in this paper is the result of three decades of collaboration with them. He thanks Professor M. Kikuchi for suggesting him to write this review. He also thanks Professor A. Sen for many useful suggestions, and Professor P. Huneman for pointing out to him the book ``Reductionism, emergence and levels of reality" by Chibbaro  \textit{et al.} He thanks Drs F. Bonneau, M.-C. Firpo, and F. Sattin for helpful comments on the manuscript. Also D. F. G. Minenna who brought the precious views of a newcomer in the field. One of the authors (D. Z.) has been supported by the A*MIDEX project (no. ANR-11-IDEX-0001-02) funded by the ``Investissements d'Avenir" French Government program, managed by the French National Research Agency (ANR).
\end{acknowledgements}

\renewcommand*{\thesection}{Appendix \Alph{section}}
\setcounter{section}{0}
\section{Rigorous derivation \`a la Kaufman}
\label{fdfe}

\paragraph{$A(t)$ is entire}

The dynamics considered in section \ref{LwaK} corresponds to the linearized motion of the $N$ electrons with respect to a given multi-beam-multi-array, when a single wave with wavevector $\bfk_{\bfm}$ is excited at $t=0$. Let $\bfr_{j0}$ be the initial position of the unperturbed beam particle with index $j$, and $\bfv_{j}$ be its velocity, and let $\Delta {\bfr}_j (t)= \bfr_j (t) - \bfr_{j0} - \bfv_j t$ be the mismatch of the actual position of particle $j$ with respect to the unperturbed beam particle with the same index. Setting $\bfr_j = \bfr_{j0} + \bfv_j t + \Delta {\bfr}_j (t)$ in equation (\ref{phitildetotM}),
we replace $\tilde{\varphi}$ with its expansion to first order in the $\Delta {\bfr}_j (t)$'s
\begin{equation}
\tilde{\varphi} (\bfm,t)
  =
  - \sum_{l = 1}^N
  \frac{ e}{\varepsilon_0 k_{\bfm}^2} \exp [- \rmi \bfk_{\bfm} \cdot (\bfr_{l0} + \bfv_{l} t)] \ [1 - \rmi \bfk_{\bfm} \cdot \Delta {\bfr}_l(t)].
\label{phitildnj2}
\end{equation}
Using equations (\ref{phitildetotM}) and (\ref{phitildnj2}), the linearized particle dynamics defined by equation (\ref{rsectot}) is then given by
\begin{equation}
\Delta \ddot{\bfr}_j
 =
  \frac{\rmi e}{L^3 m} \bfk_{\bfm} \
    \tilde{\varphi}(\bfm,t) \exp[\rmi \bfk_{\bfm} \cdot (\bfr_{j0} + \bfv_{j} t)] + \rmcc \,.
    \label{delrsec2}
\end{equation}
Because of ordering (\ref{bsmnedge}), for each beam the corresponding values of $\exp[\rmi \bfk_{\bfm} \cdot \bfr_{l0}]$ are uniformly distributed on the unit circle, and their global contribution to the 1 factor in the last bracket of equation (\ref{phitildnj2}) vanishes. Therefore equation (\ref{delrsec2}) yields the compact expression
\begin{equation}
\Delta \ddot{\bfr}_j
 =
  - \frac{2 \omega_{\rmpp}^2}{N k_{\bfm}^2}  \bfk_{\bfm} \
    \sum_{l = 1}^N
    \cos[\bfk_{\bfm} \cdot [\bfr_{j0} - \bfr_{l0} + (\bfv_{j} - \bfv_{l}) \, t \, ]
    \, \bfk_{\bfm} \cdot \Delta {\bfr}_l(t).
    \label{delrsec3}
\end{equation}
This defines a system of $N$ linear differential equations whose coefficients are entire functions of $t$. Therefore, the $\Delta {\bfr}_l(t)$'s are entire functions. Through equation (\ref{phitildnj2}), this property is transferred to $\tilde{\varphi} (\bfm,t)$ 
and to the amplitude $A(t) = \tilde{\varphi} (\bfm,t) \exp (\rmi \omega t)$ for $\omega$ real~: $A(t)$ is an entire function.

If the $\bfk_{\bfm} \cdot \bfv_{j}$'s are multiples of a given number, equation (\ref{delrsec3}) has coefficients with some period $T$, and belongs to the Floquet class of differential equations. Then its solutions are of the type
\begin{equation}
    U(t)
  =
  V(t) \rme^{\alpha t}  ,
  \label{SZ43}
\end{equation}
where $V(t)$ is a vector of period $T$, and $\alpha$ a complex
number. The corresponding Floquet exponents are the $\beta_\sigma$'s introduced at the end of section \ref{Zgdf}. A similar equation was met in the self-consistent wave-particle approach introduced in section \ref{AmplitudeEqn} (section 3 of \cite{EEbook}). Its full solution turned out to be a superposition of wave-like and ballistic solutions. This is remarkable, since such equations are generally not
explicitly solvable with elementary functions, even for the simplest one, the Mathieu equation.

\paragraph{Solution to all orders}

By expanding $A(t-\tau)$ in Taylor series, the second term of equation (\ref{At1}) becomes (with $k = k_{\bfm}$ and $v_\phi = \omega/k)$)
\begin{eqnarray}
   S &=&
   - \rmi \omega_\rmpp^2 \sum_{n=0}^{+\infty} \frac{\rmd^n A(t)}{\rmd t^n}
       \int \frac{\partial}{\partial \Omega}
             \int_0^t (-1)^n\frac{\tau^n}{n!} \exp( - \rmi \Omega \tau) \, \rmd \tau
             \, g(v) \rmd v
   \nonumber
   \\
   &=&
   - \rmi \omega_\rmpp^2 \sum_{n=0}^{+\infty} \frac{\rmd^n A(t)}{\rmd t^n}
          \int  \frac{\partial}{\partial \Omega}
                 \int_0^t \frac{\rmi^{3n}}{n!} \frac{\partial^n}{\partial \Omega^n} \exp( - \rmi \Omega \tau) \, \rmd \tau
                 \, g(v) \rmd v
   \nonumber
   \\
   &=&
   \omega_\rmpp^2 \sum_{n=0}^{+\infty} \frac{\rmi^{3n}}{n!} \frac{\rmd^n A(t)}{\rmd t^n}
         \int  \frac{\partial^{n+1}}{\partial \Omega^{n+1}} \frac{\exp( - \rmi \Omega t)-1}{\Omega} \, g(v) \rmd v
   \nonumber
   \\
   &=&
   \omega_\rmpp^2 \sum_{n=0}^{+\infty} \frac{\rmi^{3n}}{n! k^{n+1}} \frac{\rmd^n A(t)}{\rmd t^n}
       \int (-1)^{n+1} g^{(n+1)}(v) \,  \frac{\exp( - \rmi \Omega t)-1}{\Omega} \rmd v
   \nonumber
   \\
   &=&
   - \omega_\rmpp^2 \sum_{n=0}^{+\infty} \frac{\rmi^{n}}{n! k^{n+1}} \frac{\rmd^n A(t)}{\rmd t^n}
       \int g^{(n+1)}(v) \frac{\exp( - \rmi \Omega t)-1}{\Omega} \rmd v
   \nonumber
   \\
   & \sim_{t\rightarrow +\infty} &
   \omega_\rmpp^2 \sum_{n=0}^{+\infty} \frac{\rmi^{n}}{n! k^{n+1}} \frac{\rmd^n A(t)}{\rmd t^n}
        \left[ {\rm{P}} \! \! \!  \int \frac{g^{(n+1)}(v)}{\Omega} \rmd v + \rmi \frac{\pi}{k}  g^{(n+1)}(v_\phi) \right].
   \label{eq:AllOrders1}
\end{eqnarray}
The time evolution of $A(t)$, and the frequency $\omega$, are then derived from
\begin{equation}
   1 = \frac{\omega_\rmpp^2}{A} \sum_{n=0}^{+\infty} \frac{\rmi^{n}}{n! k^{n+1}} \frac{\rmd^n A(t)}{\rmd t^n}
             \left[ {\rm{P}} \! \! \!  \int \frac{g^{(n+1)}(v)}{\Omega} \rmd v + \rmi \frac{\pi}{k}  g^{(n+1)}(v_\phi) \right].
  \label{D1}
\end{equation}
Looking for a solution $A(t)=A(0)e^{\nu t}$ with $\nu$ real, equation~(\ref{D1}) reads
\begin{equation}
  1 = \omega_\rmpp^2 \sum_{n=0}^{+\infty}\frac{\rmi^{n}}{n! k^{n+1}} \nu^n
          \left[ {\rm{P}} \! \! \!  \int \frac{g^{(n+1)}(v)}{\Omega} \rmd v + \rmi \frac{\pi}{k}  g^{(n+1)}(v_\phi) \right],
   \label{D2}
\end{equation}
whose real part is
\begin{equation}
   1 = \omega_\rmpp^2 \sum_{p=0}^{+\infty} \frac{(-1)^p}{(2p)! k^{2p+1}} \nu^{2p}
            \left[{\rm{P}} \! \! \!  \int \frac{g^{(2p+1)}(v)}{\Omega} \rmd v - \frac{\pi \nu}{(2p+1)k^2} g^{(2p+2)}(v_\phi)\right],
   \label{D3}
\end{equation}
and imaginary part is
\begin{equation}
   0 = \sum_{p=0}^{+\infty} \frac{(-1)^p\nu^{2p}}{(2p+1)!k^{2p}}
           \left[\nu {\rm{P}} \! \! \!  \int \frac{g^{(2p+2)}(v)}{\Omega} \rmd v +\pi(2p+1)g^{(2p+1)}(v_\phi) \right].
   \label{D4}
\end{equation}
Since $g^{(n+1)}(v) \sim v_{\rmT}^{-n}g'(v)$, equations~(\ref{D3})~and~(\ref{D4})~may be seen as expansions in the small parameter $\varepsilon \equiv \nu / (k \lambda_\rmD \omega_{\rmpp})$. Then, these equations may be solved order by order. The first two orders were used in section \ref{LandamK}.

It is also noteworthy that, when $g(v)$ is analytic, equation~(\ref{D2}) reads
\begin{equation}
   1 = \frac{\omega_\rmpp^2}{k}
          \left[{\rm{P}} \! \! \!  \int \frac{g'(v + \rmi \nu / k)}{\Omega} \rmd v + \rmi \frac{\pi}{k} g'(v_\phi+i\nu/k) \right],
   \label{D5}
\end{equation}
which is exactly the formula obtained by Landau.

\section{Relevance of transients before Landau damping}
\label{RtbLd}

In this Appendix, we discuss the relevance of transients that would occur before the self-consistent electrostatic field may experience Landau damping. To do so, we have to specify how the self-consistent electrostatic field has actually been generated, an issue that is usually eluded. This implies that we do account for the external drive (e.g. a laser, a polarized grid, electrodes\dots) used to induce the self-consistent field in the plasma, when calculating the electron motion. The calculation is, therefore, slightly more general than that leading to Eqs~(\ref{delrj})-(\ref{Atint}). Such a generalization has already been performed in Ref.~\cite{benisti2007} when the self-consistent field was slowly driven. An important point of Ref.~\cite{benisti2007} was to prove that a wave may be considered as slowly varying, provided that its complex amplitude did not change much during a time interval of the order of $\tau_{\rm{mix}}=(k v_\rmT)^{-1}$. More precisely, the typical wave growth rate, $\gamma$, had to be such that $\vert \gamma \vert \tau_{\rm{mix}}<0.1$. 

We now further discuss the importance of the product $\vert \gamma \vert \tau_{\rm{mix}}$ in terms of the transients. 
For the sake of simplicity, the discussion is restricted to the situation when the electrons only feel the effect of the drive during a finite time interval, 
namely when $0<t<\tau_\rmd$ (see Ref.~\cite{benisti2007} for a calculation that does not make use of this hypothesis). 
We also assume that the total force, including the effect of the self-consistent field and the drive, 
derives from an effective potential which reads $\varphi=A(t) \exp[\rmi(\mathbf{k_m}.\mathbf{r}-\omega t)] + \rmcc$ 
(which has been proved to be correct in Ref.~\cite{benisti2007}~when the plasma wave was laser driven). 
Because we include the effect of the drive, we can now integrate the electrons motion from the time when they are at equilibrium, $\bfr_j=\bfr_{j0}$ when $t=0$. 
Then, Eq.~(\ref{delrj}) is changed into
\begin{equation}
\Delta \bfr_{j1}(t)
  =  \alpha \bfk_{\bfm} \int_{0}^{t} \tau A(t-\tau) \exp [\rmi (\Omega_j (t-\tau) + \bfk_{\bfm} \cdot \bfr_{j0})] \rmd \tau + \rmcc
\label{delrj2}
\end{equation}
If the wave is slowly driven, $A(t)$ is a slowly-varying function, and one may stop the Taylor expansion of $A(t-\tau)$ at first order, which yields
\begin{eqnarray}
A(t) =&& \int  \frac{\omega_{\rmpp}^2}{k_{\bfm}} A(t) g'(v) \frac{1-\cos (\Omega t)+ \rmi \sin (\Omega t)}{\Omega}  \rmd v 
 \nonumber \\
&&- \, \rmi \frac{\omega_{\rmpp}^2}{k_{\bfm}^2}
\dot{A}(t)  \int g^{\prime \prime}(v) \frac{\cos (\Omega t)- \rmi \sin (\Omega t) -1}{\Omega} \rmd v,
\label{Atint2}
\end{eqnarray}
which is the same as Eq.~(\ref{Atint}) except that the term proportional to $h(v)$ no longer appears, since the calculation has been performed with $\delta \bfr_{ j}=\mathbf{0}$. Hence, unlike in Eq.~(\ref{Atint}), no transient is expected from this term. Now, when $t>\tau_\rmd$ it is clear that $A(t)$ is nothing but the amplitude of the self-consistent potential, since the electrons no longer feel the effect of the drive.  Moreover, because it is slowly driven, only when $\tau_\rmd \gg \tau_{\rm{mix}}$ may the self-consistent field reach a significant amplitude, and may effectively be Landau damped. Hence, when $t>\tau_\rmd$, Eq.~(\ref{Atint2}) leads to Eq.~(\ref{Atintasym}). This means that, if the wave is slowly driven, it is Landau damped just after the drive has been turned off, and there is no transient.

Let us now investigate the situation when the wave may no longer be considered as slowly-varying when $t < \tau_\rmd$. Then, Eq.~(\ref{Atint2}) is no longer valid, and we have to pay a specific attention to the electron motion during the driving phase ($t\leq\tau_\rmd)$. To do so, we still assume that the electrons are at equilibrium when $t=0$. Moreover, we use Eq.~(\ref{delrj2}), which is exact, to calculate the shift in their positions at $t=\tau_\rmd$, induced by the drive and the self-consistent electric fields. This yields
\begin{eqnarray}
\nonumber
\Delta \bfr_{j1}(t) &=&\alpha \bfk_{\bfm}  \int_{0}^{\tau_\rmd} \tau A(\tau_\rmd-\tau) \exp [\rmi (\Omega_j (\tau_\rmd-\tau) + \bfk_{\bfm} \cdot \bfr_{j0})] \rmd \tau  + \rmcc \\
\label{ben1}
&\equiv & \delta \bfr_{ j} \sin(\bfk_{\bfm} \cdot \bfr_{j0}+\psi_0).
\end{eqnarray}
As noted in Section~\ref{LandamK}, the phase $\psi_0$ has no importance in the derivation, and we henceforth drop it. Using the result of~Eq.~(\ref{ben1}), we now calculate the shift in position when $t>\tau_\rmd$, which will let us conclude about the evolution of the wave amplitude. Using again Eq.~(\ref{delrj2}), we find
\begin{eqnarray}
\Delta\mathbf{r}_{j_1}(t)
&=&  \delta \bfr_{ j} \sin(\bfk_{\bfm} \cdot \bfr_{j0})
        + \left\{ \alpha \mathbf{k_m}\int_{\tau_\rmd}^t \tau \mathcal{A}(t-\tau) \rmd \tau 
        \right.  \nonumber   \\
&& \qquad \qquad + \left. \alpha \mathbf{k_m} \int_0^{\tau_\rmd} \tau \left[\mathcal{A}(t-\tau)-\mathcal{A}(\tau_\rmd-\tau) \right] \rmd \tau + \rmcc \right\} \ ,
\label{ben2}
\end{eqnarray}
where we introduced
\begin{equation}
\mathcal{A}(t) \equiv A(t)\exp [\rmi (\Omega_j t + \bfk_{\bfm} \cdot \bfr_{j0})].
\label{ben3}
\end{equation}
In Section~\ref{LandamK}, the limit $\tau_\rmd \rightarrow 0$ was considered, which yields Eq.~(\ref{delrj}). 
In this Appendix, we specify how small $\tau_\rmd$ has to be for the results of Section~\ref{LandamK} to be valid. 
First, we want the third term in the right-hand side of Eq.~(\ref{ben2}) to be negligible, 
which is only true if $t\gg \tau_\rmd$ and if $A(t)$ does not abruptly vanish within a time interval smaller than $\tau_\rmd$. 
The first condition implies that the results of Section~\ref{LandamK} are only valid when $t\gg\tau_\rmd$. 
The second condition is only true provided that $\gamma_{\rm L} \tau_\rmd \ll 1$. 

Neglecting the third term of Eq.~(\ref{ben2}) and following the same steps as in Section~\ref{LandamK}, 
one finds that Eq.~(\ref{Atint}) is changed into
\begin{eqnarray}
A(t) =&& \int \left[\frac{Ne}{2\varepsilon_0 k_{\bfm}^2 }\, h(v)  \exp (- \rmi \Omega t) \right.
\nonumber    \\
&& \qquad \qquad  \left.
+ \frac{\omega_{\rmpp}^2}{k_{\bfm}} A(t) g'(v) \frac{\rme^{-\rmi\Omega\tau_\rmd}-\cos (\Omega t)+ \rmi \sin (\Omega t)}{\Omega} \right] \rmd v
 \nonumber \\  
&&- \, \rmi \frac{\omega_{\rmpp}^2}{k_{\bfm}^2}
\dot{A}(t)  \int g^{\prime \prime}(v) \frac{\cos (\Omega t)- \rmi \sin (\Omega t) - \rme^{-\rmi\Omega\tau_\rmd}}{\Omega} \rmd v.
\label{ben4}
\end{eqnarray}
Then, Eq.~(\ref{Atint}) is recovered only in the limit $\Omega\tau_\rmd \rightarrow 0$. Hence, this equation is only valid provided that the wave could reach a significant amplitude, due to the external drive, during a time much smaller than $\tau_{\rm{mix}}$ and the plasma period. When the latter condition is fulfilled, Eq.~(\ref{Atint}) implies that, once the drive is turned off, one has to wait for a time of the order of $\tau_{\rm{mix}}$ before Landau damping is effective.

Therefore, we recover here the usual difference in a system's response to a sudden excitation compared to an adiabatic one. When a system is subjected to a sudden change, it usually rapidly oscillates before entering a stationary, or slowly-varying, regime. These transient oscillations do not exist under an adiabatic-like external force, and the system keeps on varying in a smooth way.

Note that we only derived the wave evolution in two opposite limits, either when the driving time, $\tau_\rmd$, was much smaller than $\tau_{\rm{mix}}$ and the plasma period, or when it was much smaller than $\tau_{\rm{mix}}$. The situation when $\tau_\rmd$ is either of the order of the plasma period or of $\tau_{\rm{mix}}$ would deserve further investigation.

\section{Infinite number of beams}
\label{App:InfNbBeams}

This appendix provides calculations similar to those in \cite{Dawson60}, but reformulated in a form suitable for the derivation of section \ref{LwKD}, and with the correction of several errors. We first focus on $ \epsilon_{\rmd 1}(\bfm,\omega)$ defined by equation (\ref{epsbeampul}). In order to separate the regular and the singular parts of this quantity, we add and subtract to the right hand side of equation (\ref{epsbeampul}) the quantity
\begin{equation}
\omega_{\rmpp}^2[\frac{\pi^2 g(\omega/k)}{k^2 \delta \sin^2(\pi \omega/k \delta)}-\frac{2 \pi g'(\omega/k)}{k^2}\cot (\pi \omega/k \delta)],
\label{addsubt}
\end{equation}
where $k=k_{\bfm}$. This yields
\begin{eqnarray}
\epsilon_{\rmd 1}(\bfm,\omega)=&& 1 - \omega_{\rmpp}^2 \left[ \frac{\pi^2 g(\omega/k)}{k^2 \delta \sin^2(\pi \omega/k \delta)}-\frac{2 \pi g'(\omega/k)}{k^2} \cot (\pi \omega/k \delta) \right.
 \nonumber \\
 && \left. + \Sigma_{\sigma = - \infty}^{\infty}( \frac{[g(\sigma \delta) -g(\omega/k)]\delta}{ (\omega - \sigma k \delta)^2}+\frac{2 g'(\omega/k) \omega \delta}{k (\omega^2 - (\sigma k \delta)^2)} ) \right],
\label{epsaddsubt}
\end{eqnarray}
where use has been made of the relations (see \cite{Abramowitz})\footnote{Equation (\ref{idcirc2}) corrects a typo in equation (33) of \cite{Dawson60}.}
\begin{eqnarray}
\frac{\pi^2 }{\sin^2(\pi x)} &=& \Sigma_{\sigma = - \infty}^{\infty}\frac{1}{ (x - \sigma)^2}
\label{idcirc1}
 \\
\pi \cot (\pi x) &=& \Sigma_{\sigma = - \infty}^{\infty}\frac{x}{ (x^2 - \sigma^2)}.
\label{idcirc2}
\end{eqnarray}
We notice that the function of $\sigma \delta$ inside the summation in equation (\ref{epsaddsubt}) has no real poles. Therefore, the sum passes smoothly to an integral as $\delta$ goes to zero\footnote{We assume $g'(v)$ continuous, $|g(v)|$ to be integrable, and $g(v) \ge 0$.}. In this limit, equation (\ref{epsaddsubt}) becomes
\begin{eqnarray}
\epsilon_{\rmd 1}(\bfm,\omega)=&& 1 - \omega_{\rmpp}^2 \left[ \frac{\pi^2 g(\omega/k)}{k^2 \delta \sin^2(\pi \omega/k \delta)}-\frac{2 \pi g'(\omega/k)}{k^2} \cot (\pi \omega/k \delta) \right.
 \nonumber \\
 && \qquad \left. + \int_{- \infty}^{\infty}( \frac{g'(v)}{ k(\omega - k v)}+\frac{2 g'(\omega/k) \omega}{k (\omega^2 - k^2 v^2)} )\rmd v \right].
\label{epsaddsubt2}
\end{eqnarray}

We now compute the zeros of $\epsilon_{\rmd 1}(\bfm,\omega)$ and write $\omega = \alpha + \rmi \beta$. We first consider those with $\beta$ vanishing when $\delta$ goes to zero. If $\beta$ vanished like or faster than $\delta$, the first term in the bracket of equation (\ref{epsaddsubt2}) would diverge, while the second and third one would remain finite, which is impossible. Therefore, $\beta$ vanishes slower than $\delta$, which forces the cotangent to converge toward $- \mu \, \rmi$, where $\mu = \pm 1$ is the sign of $\beta$. Then, in order to stay finite, the first term requires $\beta$ to scale like $\delta \, |\ln (\delta/v_{\rmT})\,|$. With this in mind, and looking for solutions in the vicinity of $n k \delta$, equation (\ref{epsaddsubt2}) requires
\begin{eqnarray}
&&\omega_{\rmpp}^2 [- \frac{4 \pi^2 g(n \delta) \exp [2 \pi (\rmi  \mu \alpha_1 - |\beta|)/k \delta]}{k^2 \delta}+\frac{\rmi \mu \pi g'(n \delta)}{k^2}]
 \nonumber \\
 &=& 1 +  {\rm{P}} \! \! \! \int_{- \infty}^{\infty} \frac{\omega_{\rmpp}^2 g'(v)}{ k^2(n \delta - v)} \, \rmd v,
\label{epsaddsubt0}
\end{eqnarray}
for $n \delta$ in the support of $g$, with $\alpha_1 = \alpha - n k \delta$. Equation (\ref{epsaddsubt2}) provides two contributions to the term in $g'(n \delta)$: one from the term in cotangent, and one from the pole of the term in $g'(v)$ in the integral, while the two poles of the term in $g'(\omega/k)$ bring contributions cancelling each other. Solving for $\alpha_1$ and $\beta$ yields
\begin{eqnarray}
&&\tan \frac{2 \pi \alpha_1}{k \delta}  =  - \frac{\pi \omega_{\rmpp}^2}{k^2} g'(n \delta)/[1 + {\rm{P}} \! \! \! \int_{- \infty}^{\infty} \rmd v \frac{\omega_{\rmpp}^2 g'(v)}{ k^2(n \delta - v)}]    \,   ,
\label{epsaddsubtsol1}
\\
&& \beta  =  \mu \frac{k \delta}{2 \pi}  \ln \Biggl\{ \left[\frac{k^2 \delta}{4 \pi^2 \omega_{\rmpp}^2 g(n \delta)}\right] \Biggr.
\times \nonumber \\ 
&& \qquad \qquad \qquad \Biggl. \left[ \left(1 \! \! +  \rmP \! \! \! \int_{- \infty}^{\infty} \! \! \! \rmd v \frac{\omega_{\rmpp}^2 g'(v)}{ k^2(n \delta - v)} \right)^2 \! \! + \!
\left(\frac{\pi \omega_{\rmpp}^2 g'(n \delta)}{k^2} \right)^2 \right] \Biggr\}.
\label{epsaddsubtsol2}
\end{eqnarray}
Equation (\ref{epsaddsubtsol1}) yields $2 \pi \alpha_1/k \delta$ modulo $\pi$, and the right solution is obtained by requiring $\cos (2 \pi \mu \alpha_1/k \delta)$ to have the opposite sign to the denominator of this equation.

Like the natural frequencies of the beams, the roots are spaced $k \delta$ apart in $\alpha$. Therefore, the above zeros have real parts between these natural frequencies. There are two roots for each beam, since $\mu$ can be either positive or negative, and no root away form the support of $g$. Thus we obtain two modes for each beam, as required.

We now compute $ \frac{\partial \epsilon}{\partial \omega} (\bfm, \omega)$ defined in equation (\ref{epsbeampulpr}) for the case of a vanishing imaginary part of $\omega$ when $\delta$ goes to zero. Here again, we handle the singularity by adding and subtracting to the right hand side of equation (\ref{epsbeampulpr}) the quantity
\begin{equation}
2 \omega_{\rmpp}^2[- \frac{\pi^3 \cos (\pi \omega/k \delta) g(\omega/k)}{k^3 \delta^2 \sin^3(\pi \omega/k \delta)}
- \frac{\pi^2 g'(\omega/k)}{k^2 \delta \sin^2(\pi \omega/k \delta)}
+ \frac{\pi g^{\prime \prime}(\omega/k)}{2 k^3}\cot (\pi \omega/k \delta)].
\label{addsubtZ}
\end{equation}
These terms can also be written in the form of sums by using again equations (\ref{idcirc1}) and (\ref{idcirc2}), and (see \cite{Abramowitz})
\begin{eqnarray}
\pi^3 \frac{\cos(\pi x) }{\sin^3(\pi x)} &=& \Sigma_{\sigma = - \infty}^{\infty}\frac{1}{ (x - \sigma)^3}.
\label{idcirc3}
\end{eqnarray}

Using the latter expression, for $\delta$ small we find that $ \epsilon'_{\sigma,\mu}$ is given by equation (\ref{epsbeampulprlim}).

\section{Approaching the singular limit by coarse-graining}
\label{Aslcg}

In the introduction, we recalled the pulverization procedure for deriving Vlasov equation from the BBGKY hierarchy. The singular limit can be obtained by a coarse-graining procedure, which is germane to the pulverization procedure: each particle is substituted by a continuum of particles with velocities close to its velocity, with a mismatch in velocity $\Delta \bfv$ distributed with the continuous distribution $P(\Delta \bfv)$, instead of a discrete distribution in the case of the pulverization\footnote{More precisely, the pulverization leads to infinitesimal beamlets whose summation corresponds to a coarse-graining of the previous multi-beam-multi-arrays.}. This procedure may be viewed as the counterpart in velocity of the quantum regularization of small spatial scales for collisions recalled in the introduction. Indeed, the coarse-graining in velocity may appear as a way to account for the quantum uncertainty on the particle velocities.

The calculation leading to equation (\ref{epsphibaldiscr}) can be performed again, but the summation over particles now involves an integral over the nearby velocities of the coarse-grained system. Then $\varphi^{(\rm{bal})}(\bfm,\omega)$ is substituted with
\begin{equation}
 \varphi_{{\rm{cg}},j}^{(\rm{bal})}(\bfm,\omega)
 = - \frac{\rmi e}{\varepsilon_0 k_{\bfm}^2}
 \int \frac{\exp[- \rmi \bfk_{\bfm}  \cdot \bfr_j(0)]}
 {\omega -\bfk_{\bfm}  \cdot (\dot \bfr_j(0) + \bfu)} P(\bfu) \rmd^3 \bfu ,
\label{*phij0hatcg}
\end{equation}
and
\begin{equation}
 f_0(\bfv)
 = \sum_{\sigma = 1}^{n_\rmb} N_\sigma P(\bfv-\bfw_\sigma) ,
\label{f0}
\end{equation}
where $n_\rmb$ is the number of beams. If the width of $P$ is large with respect to the edge of an elementary cube of the velocity grid, $f_0$ is a smooth function, the grid may be taken as very tight, and for practical purposes $P(\bfu)$ may be taken as a Dirac distribution in equation (\ref{*phij0hatcg}), which then becomes equation (\ref{*phij0hat}).

\section{Shielded Coulomb potential by a singular limit of the many-beam description}
\label{SCP}

If in equation (\ref{phid}) we take first the limit $\delta \rightarrow 0$ and then the limit $\varepsilon \rightarrow 0^{+}$, according to equation (\ref{epsaddsubt2}), $\epsilon_\rmd(\bfm,\bfk_{\bfm} \cdot \bfv + \rmi \varepsilon)$ converges toward
\begin{equation}
\epsilon_{\rm{lim}} (\bfm,\bfk_{\bfm} \cdot \bfu) = 1 + \omega_{\rmpp}^2 \, {\rm{P}} \! \! \! \int \frac{\frac{\partial g}{\partial v}(v)}{k_{\bfm}(\bfk_{\bfm} \cdot \bfu - k_{\bfm} v)} \rmd v
 - \rmi \mu \frac{\pi \omega_{\rmpp}^2}{k_{\bfm}^2}
\frac{\partial g}{\partial v}(\frac{\bfk_{\bfm} \cdot \bfu}{k_{\bfm}}).
\label{epsbeamprlim}
\end{equation}
This is nothing but the contribution of $\epsilon(\bfm,\bfk_{\bfm} \cdot \bfu + \rmi \varepsilon)$ in the same limit (see for instance equation (9.12) of \cite{Nicholson}). Therefore, for $\delta$ small enough, equation (\ref{phijd}) becomes equation (\ref{phij}), and equation (\ref{phid}) becomes equation (\ref{phi}).

%


\end{document}